\documentstyle[floats,aps,epsf,amsfonts]{revtex}
\tighten
\begin{document}
\draft

\title{Perturbations of Schwarzschild black holes in the
Lorenz gauge: Formulation and numerical implementation}
\author{Leor Barack}
\address
{School of Mathematics, University of Southampton, Southampton,
SO17 1BJ, United Kingdom}
\author{Carlos O. Lousto}
\address
{Department of Physics and Astronomy,
and Center for Gravitational Wave Astronomy, The University of Texas
at Brownsville, Brownsville, Texas 78520, USA}
\date{\today}
\maketitle

\begin{abstract}
We reformulate the theory of Schwarzschild black hole perturbations in terms
of the metric perturbation in the Lorenz gauge. In this formulation, each
tensor-harmonic mode of the perturbation is constructed algebraically from
10 scalar functions, satisfying a set of 10 wavelike equations, which are
decoupled at their principal parts. We solve these equations using numerical
evolution in the time domain, for the case of a pointlike test particle set in
a circular geodesic orbit around the black hole. Our code uses characteristic
coordinates, and incorporates a constraint damping scheme. The axially-symmetric,
odd-parity modes of the perturbation are obtained analytically. The approach
developed here is especially advantageous in applications requiring knowledge
of the local metric perturbation near a point particle; in particular,
it offers a useful framework for calculations of the gravitational self force.
\end{abstract}



\section{introduction} \label{SecI}

There has been much progress, over the past few years, on the problem of
finding the ``self force'' (SF) correction to the geodesic motion of point
particles in curved geometries \cite{PoissonRev,CQGRev}. This had strong
motivation from the need to accurately model the orbital evolution in
astrophysical binaries with extreme mass ratios---a key type of sources
for the planned gravitational-wave detector LISA (the Laser Interferometer
Space Antenna) \cite{BarCut}. Yet, despite the fact that a formal framework for
calculations of the gravitational SF exists since quite a while ago
\cite{QW,MST,DW,ModeSum}, most actual computations so far have been restricted
to toy models based on scalar or electromagnetic fields, with very little
progress on the gravitational problem of relevance to LISA. What held progress
up, mainly, is a certain gap within the standard glossary of techniques
comprising black hole perturbation theory. Here we intend to
supplement the necessary piece of perturbative technology needed to facilitate
calculations of the gravitational SF. Although our motivation bears primarily
on the SF problem, we expect the perturbation approach to be developed here to be
useful in a wider class of problems in mathematical and numerical relativity.

What has been hindering calculations of the gravitational SF can
be described, in most simple terms, as follows. The term ``gravitational SF''
refers to the effective local force exerted on a point mass particle through
interaction with its own gravitational field. (The notion of a ``point particle'',
in our general-relativistic context, builds on the clear separation of length scales
in the extreme-mass-ratio inspiral problem---see, e.g., the discussion in
\cite{PoissonRev}.) Technically, the SF appears as a correction term,
quadratic in the particle's mass, in the geodesic equation of motion.
To obtain the SF one first needs to address the subtle issue of
``regularization'': telling the part of the particle's field whose back-reaction
affects the motion, from a remaining, singular piece that can be safely
``removed'' without affecting the motion. A formal procedure for identifying
the correct ``regularizable'' piece (for mass particles in any vacuum curved
spacetime) has been established in works by Mino {\em et al.} \cite{MST},
Detweiler and whiting \cite{DW}, and others.

The above regularization procedure involves the introduction of a suitable
local frame for the particle, and also the imposition of an appropriate
{\em gauge} condition for the particle's field, the latter being treated as a
perturbation over the fixed background of the massive black hole.
Recall here that the metric perturbation (MP) is subject to a gauge freedom,
associated with infinitesimal coordinate transformations.
The full MP induced by the particle is gauge dependent, and so is the form of its
singular, ``regularizable'' piece. This local piece is most conveniently
constructed within the so-called ``Lorenz gauge'': an analogue of
the familiar Lorenz gauge of electromagnetism, wherein the covariant
divergence of the (trace-reversed) MP is taken to vanish [see Eq.\ (\ref{II-20})
below]. The Lorenz gauge condition conforms with the isotropic form of the
singularity very close to the particle (as viewed from a suitable local frame),
and the Lorenz-gauge MP correctly reflects the ``inverse-distance'' behavior
of this singularity. This local isotropy of the Lorenz gauge is an essential
feature that should not be taken for granted: Expressed in a poorly chosen gauge,
the physical point singularity may take a complicated form, that could
render analysis intractable. Indeed, it has been shown that with the
standard gauge choices commonly made in studies of black hole
perturbations (see below) the physical point singularity can take a
distorted form, and may even cease to be isolated \cite{BOgauge}.

Once the ``correct'' singular piece of the MP is identified, the construction
of the SF proceeds by removing this piece from the full (retarded) field of the particle,
and then calculating the force exerted on the particle by the remaining finite
field. (The full field is obtained, in principle, by solving the perturbation
equations with an energy-momentum source term representing the point particle.)
A practical way to subtract the singular piece from the full field is offered
by the {\em mode sum scheme} \cite{ModeSum}, in which one first decomposes the
full MP into
multipole harmonics (defined with respect to the background black hole geometry)
and then carries out the subtraction mode by mode---thereby avoiding the need
to deal with singular fields. The mode-by-mode subtraction scheme, like the
original construction in \cite{QW,MST}, is formulated {\em in the Lorenz
gauge}, and relies on having at hand the full Lorenz-gauge MP near the
particle.

Unfortunately, the standard toolkit of black hole perturbation theory
does not include practical techniques or working tools for calculating
the MP in the Lorenz gauge. Standard formulations of black hole
perturbations employ other gauges, which are favored for their algebraic
simplicity. A most commonly favored gauge choice for analyzing perturbations of
spherically symmetric black hole spacetimes is the one introduced long
ago by Regge and Wheeler (RW) \cite{RW} (and developed further by others,
including Zerili \cite{Zerilli1,Zerilli2} and Moncrief \cite{Moncrief}).
In the RW gauge, certain projections of the MP onto a tensor-harmonic basis,
in a specific coordinate system, are taken to vanish.\footnote {There is an
equivalent formulation of the RW gauge, which does not rely on a tensor-harmonic
decomposition \cite{BOgauge,Lousto}. In this formulation one sets to zero
certain combinations of the MP components and their derivatives,
in a certain coordinate system.} Another such ``algebraic'' gauge proven useful,
is the one referred to as the {\em radiation gauge} \cite{Chrz}, where one sets
to zero the projection of the MP along a principle null direction of the
background black hole geometry. Perturbations of the {\em Kerr} geometry have
been studied almost exclusively through the powerful Teukolsky formalism
\cite{Teukolsky}, in which the perturbation is formulated in terms of the Weyl
scalars, rather than the metric. A reconstruction procedure for the MP, out of the
perturbation in the Weyl scalars, has been prescribed by Chrzanowski \cite{Chrz}
(with later supplements by Wald \cite{WaldReconst} and Ori \cite{OriReconst}).
This reconstruction is formulated within the radiation gauge, and relies
crucially on this choice of gauge.

Hence, we are in a situation where the singular, ``removable'' piece of the
MP near the particle is given in the Lorenz gauge (which best reflects
the symmetry of the point singularity), while the full field of the particle
can only be calculated in the RW or radiation gauges (which bear on the
symmetries of the black hole background). Much effort has been invested
recently in trying to resolve this problematic issue. One approach has been to try
reformulating the SF, and the subtraction scheme, entirely within the RW
gauge (in the Schwarzschild background) \cite{MinoRW,DetweilerRW}.
Another approach incorporated an ``intermediate'' gauge, in which the MP
admits the isotropic Lorenz-gauge form near the particle, but gradually
approaches the form of the RW (or radiation) gauge as one moves away from
the particle \cite{BOgauge}. Neither approaches have gone
very far, partly because of the difficulty in tackling the singular gauge
transformations involved. The complicated singular nature of these gauge
transformations reflects the fact that the RW/radiation gauges are far from
being natural gauges to describe point particle singularities. For example,
is has been demonstrated \cite{BOgauge} that the radiation-gauge MP develops a
singularity along an infinite ray emanating from the particle.
Indeed, the only actual computation of the gravitational SF carried out
so far \cite{LoustoLetter,Barack&Lousto} restricted to a special case where
the RW-gauge MP happens to coincide with the Lorenz-gauge MP (the case of
a strictly radial infall into a Schwarzschild black hole).

Of course, the above gauge problem would have never occurred, had we the
right tools for calculating the full perturbation field of the particle in
the Lorenz gauge. This would have allowed a direct approach to SF
calculations, based entirely on the Lorenz gauge and avoiding the above
gauge-related complications. In this work we begin developing such practical
tools for analysis of Lorenz-gauge perturbations. This paper deals with
MP of the Schwarzschild black hole. It formulates the
Lorenz-gauge MP equations in a form accessible to numerical time-evolution
treatment, and presents calculations of the Lorenz-gauge field in the
example of a mass particle moving in a circular geodesic in the strong
field of the black hole. The paper also discusses, in brief, how we
might go about carrying out Lorenz-gauge calculations in the {\em Kerr}
spacetime. This important task is currently subject to intensive study.

It may be worthwhile to summarize here, in bullet points, the main
strengths of our Lorenz-gauge approach to black hole perturbations.
The following points also give cues to the variety of problems where
such an approach can prove useful.
\begin{itemize}
\item
A regularization scheme for the SF in Kerr spacetime
has only been prescribed in the Lorenz gauge. It is not clear yet how
one goes about formulating and calculating the gravitational SF in
other gauges. Having at hand the Lorenz-gauge MP allows one, immediately,
to compute the gravitational SF, without having to resort to cumbersome gauge
adjustments.
\item
In our approach one solves directly for the MP components, without
resorting to complicated reconstruction procedures. This feature becomes
important in problems where knowledge of the Weyl scalars (or RW/Zerilli's
variables, or Moncrief's variables)
is insufficient, and direct access to the MP itself is necessary.
\item
Related to the last point is the fact that, in our approach, the MP reconstruction
is algebraic (see Sec.\ \ref{Subsec:Reconst} below), and does not involve
differentiation of the field variables. This comes to be a great advantage in
numerical applications, where differentiation often results in loss of numerical accuracy.
In particular, in SF calculations it is necessary to resolve the MP
near the particle with a great precision. The higher the order of the derivatives
involved in constructing the SF are, the tougher become the resolution requirements.
If the SF is to be constructed from the Weyl scalars (or from Moncrief's variables,
as in \cite{Barack&Lousto}), this construction
would necessitate taking three successive derivatives---two to reconstruct the
MP, and a further one to obtain the force exerted by the MP. This has proven
to be very demanding computationally \cite{Barack&Lousto}. In our approach,
one need only take a single derivative of the numerical field variables.
\item
Yet another advantage of working with the Lorenz-gauge MP components as
field variables, is that these behave more regularly near point particles
than do Teukolsky's or Moncrief's variables. This has a simple manifestation
when considering the multipole decomposition of the MP: The individual multipole
modes of the Lorenz-gauge MP are {\em continuous} at the particle; only their first
derivatives are discontinues there. The multipole modes of Teukolsky's or Moncrief's
variables, on the other hand, are themselves discontinuous at the particle, and
so are, in general, the modes of the MP in the RW gauge. Obviously, this better
regularity of the Lorenz-gauge MP comes to be a great advantage when it
comes to numerical implementation.
\item
The Lorenz-gauge perturbation equations take a fully hyperbolic form.
(Compare with the situation in the RW or radiation gauges, where the set
of perturbation equations split into a subset of hyperbolic field equations,
and a subset of elliptic equations that constrain the evolution.)
This makes the Lorenz-gauge formulation especially convenient for
numerical applications which are based on time-domain evolution. The supplementary
gauge conditions indeed take the form of elliptic ``constraint'' equations,
but these can be made to hold automatically. We shall discuss
this point in Sec.\ \ref{SecII}.
\item
The Lorenz gauge condition does not interfere with the local isotropic
symmetry of point particle singularities. This makes it well suited---in
gravitational perturbation theory as in electromagnetism---for study of
the local field near such particles. Other gauge choices (the RW and radiation
gauges are examples) may artificially distort this symmetry, requiring a
more cautious treatment.
\item
The use of (generalized) ``harmonic coordinates'', closely related to the
choice of Lorenz gauge in a perturbative context, has lately been a popular
trend in Numerical Relativity (e.g., \cite{hyper}). It has been partially
responsible for the significant progress made recently in fully non-linear
simulations of binary black hole mergers \cite{Pretorius}. There is still
much to understand about the underlying mathematics that makes this formulation
so successful. The simpler realm of perturbation theory can provide a good
test bed for these ideas, and help gaining insight into some of the
mathematical features of hyperbolic formulations.
\end{itemize}

The obvious down side of the Lorenz-gauge formulation
is that one can no longer benefit from the kind of algebraic simplicity
that has made the RW/radiation gauges so popular. There is no known way
to fully decouple between the various tensorial components of the
Lorenz-gauge MP, and one has to treat the perturbation equations as
a coupled set. This, seemingly, has discouraged the development
of a Lorenz-gauge MP formulation in the past. Part of our aim here
is to demonstrate that such formulation is tractable, and accessible to
numerical implementation, {\em despite} the lack of algebraic simplicity.

This paper is structured as follows.
In Sec.\ \ref{SecII} we formulate MP theory in the Lorenz gauge.
Specializing to the case of a point mass source in Schwarzschild geometry,
we decompose the field equations (and the gauge conditions) into tensor
harmonic multipoles, and obtain a set of 10 hyperbolic wave equations
for each multipole mode (each $l,m$). This set couples between the various
tensorial components of the MP; however, the MP variables are chosen in
such a way that the equations are decoupled at their principle parts.
Using the supplementary gauge conditions we modify the original wave
equations to incorporate a ``constraint damping'' scheme: A mechanism
designed to guarantee that violations of the gauge conditions (due,
e.g., to numerical errors) are damped automatically during numerical evolution.
In Sec.\ \ref{SecIII} we implement the above formulation for the case of a
source particle in a circular geodesic orbit
around the black hole. We give analytic solutions for the monopole mode of
the MP, and for all of its axially-symmetric, odd-parity modes.
To solve for the rest of the modes, we present a numerical code, based
on characteristic time-evolution in 1+1 dimensions.
Section \ref{SecIV} presents a series of validation tests for our code.
In particular, we use our Lorenz-gauge solutions to extract the fluxes of
energy (in each of the multipole modes) radiated to infinity in gravitational
waves, and compare them with the values given in the literature.
Finally, in Sec.\ \ref{SecV}, we discuss the applicability of our approach
to more general orbits in Schwarzschild, and to orbits around Kerr black holes.

\section{Formulation} \label{SecII}

\subsection{Linearized Einstein equations in the Lorenz gauge}

Let $g_{\mu\nu}$ be the metric in a given ``background'' spacetime, which
we assume to be Ricci-flat [i.e., $R_{\alpha\beta}(g_{\mu\nu})=0$].
Let then $h_{\mu\nu}$ represent a small gravitational perturbation away
from $g_{\mu\nu}$, produced by a given energy-momentum distribution
$T_{\alpha\beta}$. Linearization of the Einstein equations,
$G_{\alpha\beta}(g_{\mu\nu}+h_{\mu\nu})=8\pi T_{\alpha\beta}$,
in the perturbation $h_{\mu\nu}$ about the background $g_{\mu\nu}$ yields the
general form \cite{Zerilli2}
\begin{equation}\label{II-5}
\Box h_{\alpha\beta}
-g_{\alpha\beta}\Box h
+h_{;\alpha\beta}
+2R^{\mu}{}_{\alpha}{}^{\nu}{}_{\beta}h_{\mu\nu}
-h_{\alpha\mu}{}^{\!;\mu}{}_{\!;\beta}
-h_{\beta\mu}{}^{\!;\mu}{}_{\!;\alpha}
+g_{\alpha\beta}h_{\mu\nu}{}^{\!;\mu\nu}
=-16\pi T_{\alpha\beta},
\end{equation}
where a semicolon denotes covariant differentiation in the background metric
$g_{\mu\nu}$, $\Box\equiv {}_{;\lambda}{}^{\!;\lambda}$ is the covariant
D'Alambertian operator, $h\equiv g^{\mu\nu}h_{\mu\nu}$ is the trace of $h_{\mu\nu}$,
and indices are raised and lowered using the background metric $g_{\mu\nu}$.
Here, and throughout this paper, we follow the conventions of Ref.\ \cite{MTW};
hence, the metric signature is $({-}{+}{+}{+})$, the connection coefficients are
$\Gamma^{\lambda}_{\mu\nu}=\frac{1}{2}g^{\lambda\sigma}(g_{\sigma\mu,\nu}
+g_{\sigma\nu,\mu}-g_{\mu\nu,\sigma}$), the Riemann tensor is
$R^{\alpha}{}_{\!\lambda\mu\nu}=\Gamma^{\alpha}_{\lambda\nu,\mu}
-\Gamma^{\alpha}_{\lambda\mu,\nu}+\Gamma^{\alpha}_{\sigma\mu}\Gamma^{\sigma}_{\lambda\nu}
-\Gamma^{\alpha}_{\sigma\nu}\Gamma^{\sigma}_{\lambda\mu}$,
the Ricci tensor and scalar are $R_{\alpha\beta}=R^{\mu}{}_{\!\alpha\mu\beta}$
and $R=R_{\alpha}{}^{\!\alpha}$, and the Einstein equations are
$G_{\alpha\beta}=R_{\alpha\beta}-\frac{1}{2}g_{\alpha\beta}R=8\pi T_{\alpha\beta}$.
We shall use standard geometrized units, with $c=G=1$.

It is convenient to re-express the MP equations (\ref{II-5})
in terms of the new variables
\begin{equation}\label{II-10}
\bar h_{\alpha\beta}=h_{\alpha\beta}-\frac{1}{2}g_{\alpha\beta}h.
\end{equation}
[Note $\bar h(\equiv \bar h_{\mu}{}^{\!\mu})=-h$; hence $\bar h_{\alpha\beta}$ is
referred to as the ``trace-reversed'' MP.] Imposing the
Lorenz gauge condition,
\begin{equation}\label{II-20}
\bar h_{\alpha\beta}{}^{\!;\beta}=0,
\end{equation}
the MP equations reduce to the compact form
\begin{equation}\label{II-30}
\Box\bar h_{\alpha\beta}(x)+2R^{\mu}{}_{\alpha}{}^{\nu}{}_{\beta}\bar h_{\mu\nu}
=-16\pi T_{\alpha\beta}.
\end{equation}
This is a ``linear, diagonal second order hyperbolic'' system, which
admits a well posed initial-value formulation on a spacelike Cauchy hypersurface
(see, e.g., Theorem 10.1.2 of \cite{Wald}). Furthermore, if the gauge conditions
(\ref{II-20}) are satisfied on the initial Cauchy surface, then they are guaranteed
to hold everywhere [assuming that Eqs.\ (\ref{II-30}) are satisfied everywhere,
and that ${T_{\alpha\beta}}^{;\beta}=0$].\footnote{The conditions (\ref{II-20})
do not fully specify the gauge: There is a residual gauge freedom within the
family of Lorenz gauges, $h_{\alpha\beta}\to h_{\alpha\beta}+\xi_{\alpha;\beta}
+ \xi_{\beta;\alpha}$,  with any $\xi^{\mu}$ satisfying $\Box\xi^{\mu}=0$.
It is easy to verify that the form of both (\ref{II-20}) and (\ref{II-30})
is invariant under such gauge transformations.}

The main motivation for this work comes from problems
where the source of perturbation can be regarded as a ``point particle'',
with a definite trajectory defined on the background metric. (See \cite{PoissonRev}
for a discussion of how the notions of a ``point particle'' and a ``trajectory''
can be made sense of in a general-relativistic context.)
In this case, the source term in the MP equations takes the form
~~~~~~~~~~~~~~~~~~~~~~~~~~~~~~~~~~~~~~~~~~~~~~~~~~~~~~~~~~~~~~~~~~~~~~~
\begin{equation}\label{II-40}
T_{\alpha\beta}(x^{\mu})=\mu \int_{-\infty}^{\infty}
(-g)^{-1/2}\,\delta^4[x^{\mu}-x_{\rm p}^{\mu}(\tau)]u_{\alpha}u_{\beta}\,d\tau
\quad\text{(particle case)},
\end{equation}
where $\mu$ is the mass of the particle, $g$ is the determinant of $g_{\mu\nu}$,
$x_{\rm p}^{\mu}(\tau)$ denotes the particle's worldline (parametrized by proper
time $\tau$), and $u^{\alpha}\equiv dx^{\alpha}_{\rm p}/d\tau$ is a tangent 4-velocity
defined along the worldline.

\subsection{Multipole decomposition}

In the rest of this work we restrict our discussion to perturbations of
the Schwarzschild black hole spacetime. The line element in the background
geometry is then given by
~~~~~~~~~~~~~~~~~~~~~~~~~~~~~~~~~~~~~~~~~~~~~~~~~~~~~~~~~~~~~~~~~~~~~~~
\begin{equation}\label{II-50}
ds^2=-fdt^2+f^{-1}dr^2+r^2(d\theta^2+\sin^2\theta\,d\varphi^2),
\end{equation}
where $f\equiv 1-2M/r$, $M$ is the mass of the black hole, the event horizon is
at $f(r)=0$, and $t,r,\theta,\varphi$ are the standard Schwarzschild coordinates,
which we adopt throughout this paper. We shall proceed by decomposing the MP
into multipole harmonics. This will be achieved by projecting
$\bar h_{\alpha\beta}$ onto a basis of 2nd-rank tensor harmonics, defined, in the
background Schwarzschild geometry, on 2-spheres $t,r={\rm const}$ \cite{Thorne}.
The spherical symmetry of the background geometry will guarantee that the individual
multipole harmonics (``$l,m$ modes'') are eigenvectors of the wave operator
in Eq.\ (\ref{II-30}), and hence evolve independently. However, the ten
tensorial components of (each $l,m$ mode of) the perturbation will generally
remain coupled.

We will adopt here the basis of tensor harmonics defined in Appendix \ref{AppA},
which represents a slightly modified version of the one introduced by Zerilli
\cite{Zerilli1}.
We denote our tensor harmonics by $Y^{(i)lm}_{\alpha\beta}(\theta,\varphi;r)$,
where $l$  and $m$ are the multipole and azimuthal numbers, respectively,
$i=1,\ldots,10$ labels the ten elements of the tensorial basis,\footnote{
At $l=0$ and $l=1$ there are actually fewer than ten independent basis elements:
There are only {\it four} independent elements at $l=0$, and {\em eight} independent
elements at each of the three $l=1$ modes---see Sec.\ \ref{Subsec:structure} below
for a more detailed discussion.}
and $\alpha,\beta$ are tensorial indices.\footnote
{To avoid confusion we note that, despite the similar notation, the
$Y^{(i)lm}_{\alpha\beta}$'s adopted here differ from the basis used by one of the
authors in \cite{Barack}.}
The harmonics $Y^{(i)lm}_{\alpha\beta}$ depend only on the Schwarzschild coordinates
$\theta$ and $\varphi$, except for simple multiplicative powers of $r$ and
$f(r)$ that are included in their definition (see Appendix \ref{AppA})---the
former in the interest of dimensional balance, and the latter for regulating the
behavior at the event horizon.
With this definition, the noncoordinate-basis components  $Y^{(i)lm}_{\hat\alpha\hat\beta}
\equiv (g^{\alpha\alpha} g^{\beta\beta})^{1/2}Y^{(i)lm}_{\alpha\beta}$ (no summation
over repeated indices) are all dimensionless, and become $r$-independent at the
limit $M/r\to 0$. Also, these harmonics are regular at the event horizon [$f(r)=0$], in
the sense that they attain finite (generally nonzero) values there when expressed in
regular, ``horizon penetrating'' coordinates (like the Kruskal coordinates).

The $Y^{(i)lm}_{\alpha\beta}$'s constitute an orthonormal set, in the sense that
\begin{equation} \label{II-60}
\int d\Omega\, \eta^{\alpha\mu}\eta^{\beta\nu}[Y_{\mu\nu}^{(i)lm}]^*
Y_{\alpha\beta}^{(j)l'm'} =\delta_{ij}\delta_{ll'}\delta_{mm'}
\end{equation}
(for any $i,j=1,\ldots, 10$),
where $\eta^{\alpha\mu}\equiv{\rm diag}\left(1,f,r^{-2},r^{-2}\sin^{-2}\theta\right)$,
an asterisk denotes complex conjugation, and the integration is carried
over a $2$-sphere of constant $r$ and $t$.
The seven harmonics $i=1,\ldots,7$ constitute a basis for all covariant 2nd-rank
symmetric tensors of {\em even} parity, while the remaining three harmonics $i=8,9,10$
span all such tensors which are of {\em odd} parity.
Any covariant 2nd-rank symmetric tensor $t_{\alpha\beta}$ can hence be expanded as
$t_{\alpha\beta}=\sum_{l,m}\sum_{i=1}^{10}
t^{(i)lm}(r,t) Y^{(i)lm}_{\alpha\beta}$,
where the time-radial coefficients are given by
$t^{(i)lm}(r,t)=\int d\Omega\,t_{\alpha\beta}\eta^{\alpha\mu}\eta^{\beta\nu}
[Y_{\mu\nu}^{(i)lm}]^* $.

We expand the trace-reversed metric perturbation in the above tensor harmonics as
\begin{equation}\label{II-70}
\bar h_{\alpha\beta}=\frac{\mu}{r}\sum_{l,m}\sum_{i=1}^{10}
a^{(i)l} \bar h^{(i)lm}(r,t)
\,Y^{(i)lm}_{\alpha\beta}(\theta,\varphi;r).
\end{equation}
The coefficients $a^{(i)l}$ are introduced for the purpose of simplifying
the form of Eqs.\ (\ref{II-150}) below, and are given by
\begin{equation}\label{II-80}
a^{(i)l}=\frac{1}{\sqrt{2}}\times\left\{\begin{array}{ll}
1, & i=1,2,3,6,        \\
{[}l(l+1)]^{-1/2},  & i=4,5,8,9, \\
{[}\lambda l(l+1)]^{-1/2},  & i=7,10,
\end{array}
\right.
\end{equation}
where
\begin{equation}\label{II-90}
\lambda\equiv (l+2)(l-1).
\end{equation}
The dimensionless scalar fields $\bar h^{(i)lm}(r,t)$ will serve as
integration variables for the numerical time-evolution of the mode-decomposed
perturbation equations. For that reason, the above construction is careful in making sure
that these fields are well behaved (i.e., attain finite, generally nonzero values)
both at spacial infinity and along the event horizon. The factor $1/r$ in front of the
expansion (\ref{II-70}) was inserted to guarantee $\bar h^{(i)lm}\propto{\rm const}$
as $r\to\infty$. The $\bar h^{(i)lm}$'s are guaranteed to be regular at the
horizon since both the physical perturbation $\bar h_{\alpha\beta}$ and the
harmonics $Y^{(i)lm}_{\alpha\beta}$ are regular there.

In a similar manner, we expand the energy-momentum tensor as
\begin{equation}\label{II-100}
T_{\alpha\beta}=\sum_{l,m}\sum_{i=1}^{10}
T^{(i)lm}(r,t) Y^{(i)lm}_{\alpha\beta}(\theta,\varphi;r),
\end{equation}
with the time-radial coefficients given by
\begin{equation}\label{II-110}
T^{(i)lm}(r,t)=\int d\Omega\,\eta^{\alpha\mu}\eta^{\beta\nu}
[Y_{\mu\nu}^{(i)lm}]^* T_{\alpha\beta}.
\end{equation}
For the point particle source with energy-momentum given in Eq.\ (\ref{II-40}),
this becomes
\begin{equation}\label{II-120}
T^{(i)lm}(r;x_{\rm p})=\frac{\mu}{u^t r_{\rm p}^2}
u_{\alpha}u_{\beta}\eta^{\alpha\mu}(x_{\rm p})
\eta^{\beta\nu}(x_{\rm p})Y_{\mu\nu}^{(i)lm*}(\Omega_{\rm p}) \delta(r-r_{\rm p})
\quad\text{(particle case)},
\end{equation}
where, recall, $x_{\rm p}(\tau)$ denotes the particle's trajectory, and $u^{\alpha}$
its 4-velocity.

We now wish to obtain equations for the various fields $\bar h^{(i)lm}(r,t)$
(numbering ten, for each given $l,m$), that are fully separated with respect
to $l,m$, and are uncoupled with respect to $i$ at their principle part.
This is achieved by first substituting both expansions (\ref{II-70}) and
(\ref{II-100}) into the field equations (\ref{II-30}), and then constructing
certain combinations of the resulting equations. For example, to obtain an
equation for $\bar h^{(1)lm}$, one has to add the $tt$ component of
Eq.\ (\ref{II-30}) to the $rr$ component of that equation multiplied
by $f^2$. For some of the other $\bar h^{(i)lm}$'s
one has to combine certain derivatives of the field equations. The full list of
necessary combinations is given in Appendix \ref{AppB}. This procedure
yields a set of 10 equations, where the principle part of the $i$'th equation
involves solely the $i$'th MP function, $\bar h^{(i)lm}$. One
proceeds by showing that, in each of the 10 equations, the angular dependence
on both sides is simply $\propto Y^{lm}$, and then using the orthogonality
of the spherical harmonics to separate the equations into multipole modes.

To write the resulting separated equations in a convenient form, we introduce
the standard 2-dimensional scalar-field wave operator (including centrifugal
potential),
\begin{equation}\label{II-140}
\square_{\rm sc}^{2d}\equiv \partial_{uv}
+\frac{f}{4}\left[\frac{f'}{r}+\frac{l(l+1)}{r^2}\right],
\end{equation}
where $f'\equiv df/dr=2M/r^2$ and $v,u$ are the Eddington-Finkelstein
null coordinates, defined by $v=t+r_*$ and $u=t-r_*$, with $dr_*/dr=f^{-1}$.
(Throughout this paper it is to be understood that partial derivatives with
respect to $v$ or $u$ are taken with fixed $u$ or $v$, respectively, while
partial derivatives with respect to $t$ or $r$ are taken with fixed $r$ or
$t$, respectively.)
The separated field equations then take the form
\begin{equation}\label{II-130}
\square_{\rm sc}^{2d} \bar h^{(i)lm}+
{\tilde{\cal M}}^{(i)l}_{\;(j)}\bar h^{(j)lm}=4\pi\mu^{-1} [rf/a^{(i)}]T^{(i)lm}
\equiv S^{(i)lm}\quad (i=1,\ldots,10).
\end{equation}
The quantities ${\tilde{\cal M}}^{(i)l}_{\;(j)}$ are differential operators,
of the first order at most, that couple between the various $\bar h^{(j)}$'s
(with same $l,m)$. The explicit form of the ${\tilde{\cal M}}^{(i)l}_{\;(j)}$'s
can be found in Appendix \ref{AppC}. As expected, one finds that the seven
equations for the even-parity modes $\bar h^{(1,\ldots, 7)}$ decouple from
the remaining 3 equations for the odd-parity modes $\bar h^{(8,9,10)}$:
We have ${\tilde{\cal M}}^{l(i)}_{\;(j)}=0$ for any $i=1,\ldots,7$ with
$j=8,9,10$, and for any $i=8,9,10$ with $j=1,\ldots,7$. Further reduction
of the system will be achieved in the next step, where we reemploy the gauge
conditions.

\subsection{Gauge conditions and constraint damping}

The field equations (\ref{II-30}) are supplemented by the gauge conditions
$Z_{\alpha}\equiv{\bar h_{\alpha\beta}}^{\ \ \,;\beta}=0$
of Eq.\ (\ref{II-20}).  In a mode-decomposed form, these four conditions
read
\begin{mathletters}\label{II-160} \label{gauge}
\begin{equation}\label{gauge1}
H_{1}^{lm}(r,t)\equiv
-\bar h^{(1)}_{,t}-\bar h^{(3)}_{,t}+f\left(
\bar h^{(2)}_{,r}+\bar h^{(2)}/r- \bar h^{(4)}/r\right)=0,
\end{equation}
\begin{eqnarray}\label{gauge2}
H_{2}^{lm}(r,t)\equiv
\bar h^{(2)}_{,t}
-f \left(\bar h^{(1)}_{,r}-\bar h^{(3)}_{,r}\right)
+(1-4M/r)\bar h^{(3)}/r
-(f/r)\left(\bar h^{(1)}-\bar h^{(5)}-2f \bar h^{(6)}\right)=0,
\end{eqnarray}
\begin{eqnarray}\label{gauge3}
H_3^{lm}(r,t)\equiv
\bar h^{(4)}_{,t}-f\left( \bar h^{(5)}_{,r}+2\bar h^{(5)}/r+l(l+1)\,
\bar h^{(6)}/r-\bar h^{(7)}/r\right)=0,
\end{eqnarray}
\begin{eqnarray}\label{gauge4}
H_4^{lm}(r,t)\equiv
\bar h^{(8)}_{,t}-f\left(\bar h^{(9)}_{,r}
+2\bar h^{(9)}/r-\bar h^{(10)}/r \right)=0.
\end{eqnarray}
\end{mathletters}
[To obtain these equations, insert the expansion (\ref{II-70}) into the
equations $Z_t=0$, $Z_r=0$, $(\sin\theta\, Z_{\theta})_{,\theta}
+(Z_{\varphi}/\sin\theta)_{,\varphi}=0$, and
$Z_{\theta,\varphi}-Z_{\varphi,\theta}=0$, respectively; then show that the angular
part in all cases is proportional to $Y^{lm}$ and use the orthogonality
of the spherical harmonics to separate the equations.]
These conditions, relating the various $\bar h^{(i)}$'s and their
first derivatives, are to supplement the separated field equations
(\ref{II-130}).

One now faces one of the standard problems of Numerical Relativity:
the fact that the combined set of 10 field equations and 4 gauge conditions
(or ``constraints'') is over-determined. Given initial data, the solutions
are determined uniquely, in principle, by evolving the 10 field equations;
how do we then make sure that the gauge conditions are satisfied as
well during the evolution? In our case of Lorenz-gauge perturbations,
and in the continuum limit, theory has it that if the gauge conditions are
satisfied on the initial Cauchy surface, then the field equations will
preserve them throughout the evolution. To see why this is true, simply
take the divergence of Eq.\ (\ref{II-30}), which, with the help of
the contracted Bianchi identities and assuming Ricci flatness of the
background geometry and $T_{\alpha\beta}^{\;\;\; ;\beta}=0$, yields a simple
homogeneous wave equation for the divergence of $\bar h_{\alpha\beta}$:
$\Box Z_{\alpha}=0$. Thus, in the continuum limit, if $\bar h_{\alpha\beta}$
satisfies a well-posed initial value problem and $Z_{\alpha}=0$ on the
initial surface, then $Z_{\alpha}=0$ is guaranteed throughout the evolution.
However, in actual numerical time-evolution implementations, gauge condition
violations due to finite differentiating and/or roundoff errors can grow
out of control even provided appropriate initial data. Moreover, in most
cases it is practically impossible to devise initial data that satisfy the
gauge conditions precisely, and at the same time are consistent with the
field equations.

Gundlach {\it et al.} \cite{Gundlach} recently proposed a general scheme
for dealing with the above problem (in the wider context of fully
non-linear Numerical Relativity), which employs the idea of {\em constraint
damping}. Adapted to our problem of Lorenz-gauge perturbations, the idea
would be to add to the linearized Einstein equations, Eqs.\ (\ref{II-30}),
a term of the form $-\kappa(t_{\alpha}Z_{\beta}+t_{\beta}Z_{\alpha})$,
where $\kappa$ is a positive constant and $t_{\alpha}$ is a future-directed
timelike vector field. Obviously, the new system of 10+4 equations is equivalent
(in the continuum limit) to the original system. Also, the added term
does not alter the principle part of the field equations and hence does
not interfere with their neat hyperbolic form. However, the evolution
equation for $Z_{\alpha}$ now becomes
$\Box Z_{\alpha}-\kappa(t_{\alpha}Z_{\beta}+t_{\beta}Z_{\alpha})^{;\beta}=0$,
which includes a damping term. One then expects that, under a range of
circumstances \cite{Gundlach}, $Z_{\alpha}$ would ``automatically'' damp to
zero during the evolution (with a timescale set by $\kappa$, if $t_{\alpha}$
is taken to be of unit length).

Here we will be inspired by the above scheme, but will allow ourselves
an amount of freedom in executing it: We will seek to add terms
$\propto Z_{\alpha}$ to our field equations, that would assure
efficient damping of the constraints, and at the same time would
lead to simplification in the final form of the decoupled field
equations (putting in mind simple numerical implementation).
The ultimate ``justification'' for the specific form selected for
the added terms would come from numerical experiments with circular
orbits, described in the following Sections. 

We take $t_{\alpha}=-\delta_{\alpha}^t$, and add to the field equations
(\ref{II-30}) a term $\kappa(\delta^t_{\alpha}\tilde Z_{\beta}+
\delta^t_{\beta}\tilde Z_{\alpha})$, where $\kappa=\kappa(r)=f'=2M/r^2$
and $\tilde Z_{\alpha}=(Z_r,2Z_r,Z_{\theta},Z_{\varphi})$ (Schwarzschild
components). This choice simplifies the form of the 10 field equations,
and leads to partial decoupling as we describe below. Our $\kappa$ is not
constant, but is expected to suit its purpose as it varies slowly, on a
scale of the background curvature, which is much larger than the typical
scale for constraint violations in evolution of perturbations from a point
particle. (As our numerical experiments demonstrate, the latter scale tends
to relate to the radius of curvature associated with the particle---see,
for example, Figs.\ \ref{fig:gauge} in the next section.) The seemingly odd form of
$\tilde Z_{\alpha}$ (note it has $Z_r$ for its $t$ component) has been
selected based on numerical experiments with circular orbits. It turns out
(see Sec.\ \ref{SecIV-B}) to yield efficient damping of all 4 constraints
$Z_{\alpha}=0$, but we shall not attempt here to explain that
on theoretical ground.

At the level of the mode-decomposed equations (\ref{FE}), the addition
of the above term to the field equations amounts to adding $(f'/2)H_2$
at $i=1,2$, adding $(f'/4) H_3$ at $i=4,5$, and adding $(f'/4) H_4$ at
$i=8,9$. This brings the separated field equations to their final form:
\begin{equation}\label{FE}
\square_{\rm sc}^{2d} \bar h^{(i)lm}+
{\cal M}^{(i)l}_{\;(j)}\bar h^{(j)lm}=S^{(i)lm}\quad (i=1,\ldots,10),
\end{equation}
where the terms ${\cal M}^{(i)}_{\;(j)}\bar h^{(j)}$ are given
explicitly (omitting the indices $l,m$ for brevity) by
\begin{mathletters}\label{II-150}
\begin{equation} \label{M1}
{\cal M}^{(1)}_{\;(j)}\bar h^{(j)}=
\frac{1}{2}ff'\bar h^{(3)}_{,r}
+\frac{f}{2r^2}(1-4M/r)\left(\bar h^{(1)}-\bar h^{(5)}\right)
-\frac{1}{2r^2}\left[1-6M/r+12(M/r)^2\right]\bar h^{(3)}
+\frac{f^2}{2r^2}(6M/r-1)\bar h^{(6)},
\end{equation}
\begin{equation} \label{M2}
{\cal M}^{(2)}_{\;(j)}\bar h^{(j)}=
\frac{1}{2}ff'\bar h^{(3)}_{,r}
+f'\left(\bar h^{(2)}_{,v}-\bar h^{(1)}_{,v}\right)
+\frac{f^2}{2r^2}\left(\bar h^{(2)}-\bar h^{(4)}\right)
+\frac{1}{2}(f'/r)\left[(1-4M/r)\bar h^{(3)}-f\left(\bar h^{(1)}-\bar h^{(5)}
-2f\bar h^{(6)}\right)\right],
\end{equation}
\begin{eqnarray} \label{M3}
{\cal M}^{(3)}_{\;(j)}\bar h^{(j)}=
\frac{1}{2}ff'\bar h^{(3)}_{,r}
+\frac{1}{2r^2}\left[1-8M/r+10(M/r)^2\right]\bar h^{(3)}
-\frac{f^2}{2r^2}\left[\bar h^{(1)}-\bar h^{(5)}-(1-4M/r)\bar h^{(6)}
\right],
\end{eqnarray}
\begin{equation} \label{M4}
{\cal M}^{(4)}_{\;(j)}\bar h^{(j)}=
\frac{1}{2}f'\left(\bar h^{(4)}_{,v}-\bar h^{(5)}_{,v}\right)
-\frac{1}{2}\,l(l+1)\,(f/r^2)\bar h^{(2)}
-\frac{1}{4}f'f/r\left[3\bar h^{(4)}+2\bar h^{(5)}-\bar h^{(7)}+l(l+1)\bar h^{(6)}\right],
\end{equation}
\begin{eqnarray} \label{M5}
{\cal M}^{(5)}_{\;(j)}\bar h^{(j)}=
\frac{f}{r^2}\left[
(1-4.5M/r)\bar h^{(5)}-\frac{1}{2}l(l+1)\left(\bar h^{(1)}-\bar h^{(3)}\right)+
\frac{1}{2}(1-3M/r)\left(l(l+1)\bar h^{(6)}-\bar h^{(7)}\right)
\right],
\end{eqnarray}
\begin{equation} \label{M6}
{\cal M}^{(6)}_{\;(j)}\bar h^{(j)}=
-\frac{f}{2r^2}\left[\bar h^{(1)}-\bar h^{(5)}
-(1-4M/r)\left(f^{-1}\bar h^{(3)}+\bar h^{(6)}\right)\right],
\end{equation}
\begin{equation} \label{M7}
{\cal M}^{(7)}_{\;(j)}\bar h^{(j)}=
-\frac{f}{2r^2}\left(\bar h^{(7)}
+\lambda\,\bar h^{(5)}\right),
\end{equation}
\begin{equation} \label{M8}
{\cal M}^{(8)}_{\;(j)}\bar h^{(j)}=
\frac{1}{2}f'\left(\bar h^{(8)}_{,v}-\bar h^{(9)}_{,v}\right)
-\frac{1}{4}f'f/r\left(3\bar h^{(8)}+2\bar h^{(9)}-\bar h^{(10)}
\right),
\end{equation}
\begin{equation} \label{M9}
{\cal M}^{(9)}_{\;(j)}\bar h^{(j)}=
\frac{f}{r^2}\left(1-4.5M/r\right)\bar h^{(9)}
-\frac{f}{2r^2}\left(1-3M/r\right)\,\bar h^{(10)},
\end{equation}
\begin{equation} \label{M10}
{\cal M}^{(10)}_{\;(j)}\bar h^{(j)}=
-\frac{f}{2r^2}\left(\bar h^{(10)}+\lambda\,\bar h^{(9)}\right).
\end{equation}
\end{mathletters}
Recall in these equations $f=1-2M/r$, $f'=2M/r^2$, $\lambda=(l+2)(l-1)$,
$\partial_r$ is taken with fixed $t$, and $\partial_v$ is taken with fixed $u$.

\subsection{Hierarchical structure of the separated field equations}
\label{Subsec:structure}

Following the above manipulations, the five equations for the even parity modes
$i=1,3,5,6,7$ no longer couple to the remaining two equations for $i=2,4$.
Similarly, in the odd-parity subset, the two equations for $i=9,10$ no longer
couple to the third equations, for $i=8$. One can then solve the
set of field equations (\ref{FE}) in a hierarchical manner, starting with the
5 even-parity equations for $\bar h^{(1,3,5,6,7)}$ and the 2 odd-parity equations
for $\bar h^{(9,10)}$, and then using the solutions as source terms in the
equations for $\bar h^{(2,4)}$ (even parity) and $\bar h^{(8)}$ (odd parity).
This simplification is achieved regardless of the form of the source term
$T_{\alpha\beta}$. Note that the functions $\bar h^{(2,4,8)}$ are those
constructing the MP components $\bar h_{tr}$, $\bar h_{t\theta}$ and
$\bar h_{t\varphi}$ [see Eqs.\ (\ref{1-83}) below], which are associated with
the shift vector on the surface $t=\rm const$.

The multipole sum in Eq.\ (\ref{II-70}) [and in Eq.\ (\ref{II-100})] contains
the two modes $l=0,1$. At these modes there are fewer than ten independent
tensor-harmonic basis elements. At $l=0$, the ``vectorial'' elements
$Y^{(4,5,8,9)}$ and ``tensorial'' elements $Y^{(7,10)}$ vanish identically,
and the monopole MP is then composed, in general, of only the four ``scalar''
elements $Y^{(1,2,3,6)}$ (which are all even-parity). The system of field
equations (\ref{FE}) thus reduces, at $l=0$, to a hierarchical set of 3
coupled equations for $\bar h^{(1,3,6)}$, plus a single equation for $\bar h^{(2)}$.
As to the dipole, $l=1$ mode: Here, the two ``tensorial'' elements $Y^{(7,10)}$
vanish identically, and one is left with a hierarchical set of $4+2$ equations
for the even parity modes, and a second hierarchical set of $1+1$ equations
for the odd parity modes.

Table \ref{TableI} summarizes the hierarchical structure of our separated
field equations, for the different values of $l$.
\begin{table}[htb]
\centerline{$\begin{array}{c|cccc|cccc}\hline\hline
\mbox{}  & \multicolumn{4}{c|}{\text{Even parity}} & \multicolumn{4}{c}{\text{Odd parity}} \\
\hline\hline
l=0     & 1,3,6       & \to & 2   & [m=0]            & \mbox{}& \mbox{} & \text{---} & \mbox{}\\
l=1     & 1,3,5,6   & \to & 2,4 & [m=\pm 1]        & 9       & \to        & 8       & [m=0]\\
l\geq 2 & 1,3,5,6,7 & \to & 2,4 & [l+m \text{ even}]& 9,10   & \to        & 8       & [l+m \text{ odd}]\\
\hline\hline
\end{array}$}
\caption{\protect\footnotesize
The hierarchical structure of the decoupled field equations (\ref{FE})
[numbers in this table refer to ``$i$'' values of the tensorial-harmonic modes
$\bar h^{(i)}$]: The full set of 10 equations first decouple into two subsets of 7
equations (even parity modes, $i=1,\ldots,7$), and 3 equations
(odd parity modes, $i=8,9,10$). Both even and odd sectors then further reduce
into smaller subsets of equations (in a hierarchical sense):
For modes with $l\geq 2$, the even-parity sector reduces to a subset of
5 equations, for $i=1,3,5,6,7$, whose solutions are then used as source
terms in solving for $i=2,4$. Similarly, the odd-parity sector decouples
into two (hierarchical) subsets of 2 and 1 equations.
In the monopole and dipole cases the system is simpler:
At $l=0$ the MP is purely even-parity, and one solves two (hierarchical)
subsets with 3 and 1 equations. At $l=1$ one deals with $4+2$ equations
in the even-parity sector, and $1+1$ equations in the odd-parity sector.
When the source of the perturbation is an orbiting point particle,
we can choose to work in a Schwarzschild coordinate system in which the orbit
is confined to the equatorial plane ($\theta=\pi/2$). In this case, modes
with even values of $l+m$ will be of pure {\em even} parity, while modes
with odd values of $l+m$ will be of pure {\em odd} parity.
The relevant $m$ modes that contribute to each of the entries of the table,
in the case of a source particle in an equatorial orbit, are indicated in
square brackets.
}
\label{TableI}
\end{table}

\subsection{Reconstruction of the metric perturbation} \label{Subsec:Reconst}

Finally, it is useful to have at hand explicit formulas for reconstructing
the various components of the original metric perturbation $h_{\alpha\beta}$,
given the functions $\bar h^{(i)lm}(r,t)$.
Using $h_{\alpha\beta}=\bar h_{\alpha\beta}-\frac{1}{2}g_{\alpha\beta}\bar h$,
together with Eq.\ (\ref{II-70}) and Appendix \ref{AppA}, we find
\begin{equation}\label{1-50}
h_{\alpha\beta}=\frac{\mu}{2r}\sum_{l=0}^{\infty}
\sum_{m=-l}^{l} h^{lm}_{\alpha\beta},
\end{equation}
with
\begin{eqnarray}\label{1-83}
h^{lm}_{tt}&=& \left(\bar h^{(1)}+f\bar h^{(6)}\right)Y^{lm}, \nonumber\\
h^{lm}_{tr}&=& f^{-1}\bar h^{(2)}Y^{lm}, \nonumber\\
h^{lm}_{rr}&=& f^{-2}\left(\bar h^{(1)}-f\bar h^{(6)}\right)Y^{lm}, \nonumber\\
h^{lm}_{t\theta}&=& r\left(\bar h^{(4)}Y^{lm}_{\rm V1}
                   +\bar h^{(8)}Y^{lm}_{\rm V2}\right),\nonumber\\
h^{lm}_{t\varphi}&=& r\sin\theta\left(\bar h^{(4)}Y^{lm}_{\rm V2}
                   -\bar h^{(8)}Y^{lm}_{\rm V1}\right),\nonumber\\
h^{lm}_{r\theta}&=& rf^{-1}\left(\bar h^{(5)}Y^{lm}_{\rm V1}
                   +\bar h^{(9)}Y^{lm}_{\rm V2}\right), \nonumber\\
h^{lm}_{r\varphi}&=& irf^{-1}\sin\theta\left(\bar h^{(5)}Y^{lm}_{\rm V2}
                   -\bar h^{(9)}Y^{lm}_{\rm V1}\right), \nonumber\\
h^{lm}_{\theta\theta}&=& r^2\left(f^{-1}\bar h^{(3)}Y^{lm}
 +\bar h^{(7)}Y^{lm}_{\rm T1}+\bar h^{(10)}Y^{lm}_{\rm T2}\right), \nonumber\\
h^{lm}_{\theta\varphi}&=& ir^2\sin\theta\left(\bar h^{(7)}Y^{lm}_{\rm T2}
                   -\bar h^{(10)}Y^{lm}_{\rm T1}\right), \nonumber\\
h^{lm}_{\varphi\varphi}&=&
             r^2\sin^2\theta\,\left(f^{-1}\bar h^{(3)}Y^{lm}
             -\bar h^{(7)}Y^{lm}_{\rm T1}-\bar h^{(10)}Y^{lm}_{\rm T2}\right),
\end{eqnarray}
where we have omitted the indices $lm$ from $\bar h^{(i)lm}$ for brevity,
and where $Y^{lm}_{\rm V1}$, $Y^{lm}_{\rm V2}$, $Y^{lm}_{\rm T1}$, and
$Y^{lm}_{\rm T2}$ are angular functions constructed from the spherical
harmonics through
\begin{eqnarray} \label{eqIII30}
Y^{lm}_{\rm V1} &\equiv & \frac{1}{l(l+1)}\, Y^{lm}_{,\theta},  \nonumber\\
Y^{lm}_{\rm V2} &\equiv & \frac{1}{l(l+1)}\,\sin^{-1}\theta\,Y^{lm}_{,\varphi}, \nonumber\\
Y^{lm}_{\rm T1} &\equiv & \frac{1}{\lambda l(l+1)} \left[
\sin\theta \left(\sin^{-1}\theta\, Y^{lm}_{,\theta}\right)_{,\theta}
-\sin^{-2}\theta\, Y^{lm}_{,\varphi\varphi}\right], \nonumber\\
Y^{lm}_{\rm T2} &\equiv & \frac{2}{\lambda l(l+1)}\,
\left(\sin^{-1}\theta\, Y^{lm}_{,\varphi}\right)_{,\theta}.
\end{eqnarray}

Note Eq.\ (\ref{1-50}) gives the MP itself, not its trace-reversed counterpart.
In applying Eqs.\ (\ref{1-83}) to $l=0,1$, recall $Y^{l=0}_{\rm V1, V2}=
Y^{l=0}_{\rm T1, T2}\equiv 0$ and $Y^{l=1}_{\rm T1, T2}\equiv 0$.
Finally, note that the trace of the MP is simply given by
\begin{equation}\label{1-60}
h\equiv g^{\alpha\beta}h_{\alpha\beta}
=(\mu/r)\sum_{lm}(f^{-1}\bar h^{(3)}-\bar h^{(6)})Y^{lm}.
\end{equation}
This implies that the function $\bar h^{(3)}$ must vanish at the event
horizon (where $f=0$), since the trace $h$ must be regular at the horizon, and the
functions $\bar h^{(3)}$ and $\bar h^{(6)}$ are both finite there by construction.

The various fields $h^{lm}_{\alpha\beta}$ are complex quantities. However,
the sum $\sum_m h^{lm}_{\alpha\beta}$ is guaranteed to be real. From the
definition of the harmonics $Y^{(i)lm}$ in App.\ \ref{AppA}, and the
decoupled field equations (\ref{FE}), one readily verifies
the symmetry relation
\begin{equation}\label{1-65}
\left(\bar h^{(i)}Y^{(i)lm}\right)_{m\to -m}=\left(\bar h^{(i)}Y^{(i)lm}\right)^*
\end{equation}
(namely, the product $\bar h^{(i)}Y^{(i)lm}$ is invariant under simultaneous
sign-reversal of $m$ and complex conjugation),
which is valid for each $i$ and any $l$. Hence, the sum over modes $m$
in the reconstruction formula (\ref{1-50}) can be written in the form
\begin{equation}\label{1-67}
\sum_{m=-l}^{l} h^{lm}_{\alpha\beta}=
h^{l,m=0}_{\alpha\beta}+2\sum_{m=1}^{l} {\rm Re}(h^{lm}_{\alpha\beta}),
\end{equation}
which is manifestly real. In practice, this allows a more economic
implementation of the reconstruction scheme: One need only compute
the $m\geq 0$ modes to reconstruct the full MP.

\section{Implementation: Case of a particle in a circular geodesic orbit}
        \label{SecIII}

\subsection{Setup}

Consider a pointlike particle of mass $\mu$, in a circular orbit around
a Schwarzschild black hole with mass $M\gg\mu$. Neglecting SF effects,
the particle traces a geodesic $x^{\alpha}=x_{\rm p}^{\alpha}(\tau)$,
with four velocity $u^{\alpha}\equiv dx_{\rm p}^{\alpha}/d\tau$.
In what follows we work in a Schwarzschild coordinate system $t,r,\theta,\varphi$
at which the orbit is confined to the equatorial plane:
\begin{equation} \label{z}
x_{p}^{\alpha}(\tau)=\left[t(\tau),r_0={\rm const},\theta_0=\pi/2,\varphi(\tau)\right].
\end{equation}
This geodesic is completely parametrized by the radius $r_0$, or,
alternatively, by an ``angular velocity''
\begin{equation} \label{omega}
\omega\equiv d\varphi/dt=\sqrt{M/r_0^3}.
\end{equation}
The circular geodesic can also be parameterized by the (conserved)
specific energy, given by
\begin{equation} \label{E}
{\cal E}\equiv -u_t=f_0(1-3M/r_0)^{-1/2},
\end{equation}
where $f_0\equiv 1-2M/r_0$. The four velocity of the particle is
then given (in Schwarzschild coordinates) by
\begin{equation} \label{u}
u^{\alpha}=({\cal E}/f_0)[1,0,0,\omega].
\end{equation}

Our goal here is to calculate the physical (stationary) MP associated
with this orbiting particle, in the Lorenz gauge.

\subsection{Source terms}

From Eqs.\ (\ref{II-120}) and (\ref{II-130}) we obtain the source terms
for our decoupled filed equations (\ref{FE}). They read
\begin{equation} \label{Si}
S^{(i)lm}(r,t)=4\pi{\cal E}\alpha^{(i)}\delta(r-r_0)\times\left\{
\begin{array}{ll}
Y^{lm*}(\pi/2,\omega t),            & i=1\text{---}7  \text{ (even)}, \\
Y^{lm*}_{,\theta}(\pi/2,\omega t),  & i=8\text{---}10 \text{ (odd)},
\end{array} \right.
\end{equation}
where the coefficients $\alpha^{(i)}$ are given by
\begin{eqnarray}\label{1-80}
\alpha^{(1)} &=& \alpha^{(3)}=f_0^2/r_0,      \nonumber\\
\alpha^{(2)} &=& \alpha^{(5)}=\alpha^{(9)}=0, \nonumber\\
\alpha^{(4)} &=& 2if_0m\omega,                 \nonumber\\
\alpha^{(6)} &=& r_0\omega^2,                 \nonumber\\
\alpha^{(7)} &=& r_0\omega^2[l(l+1)-2m^2],    \nonumber\\
\alpha^{(8)} &=& 2f_0\omega,                  \nonumber\\
\alpha^{(10)}&=& 2imr_0\omega^2.
\end{eqnarray}

Note the special case $m=0$:
For these axially-symmetric modes the field equations for both
$\bar h^{(9)}$ and $\bar h^{(10)}$ are sourceless, and (since these two equations
do not couple to any of the other $\bar h^{(i)}$'s) one finds
\begin{equation} \label{m0odd}
\bar h^{(9)}_{m=0}=\bar h^{(10)}_{m=0}=0.
\end{equation}
The entire axially-symmetric odd-parity perturbation
is then described by a single function $\bar h^{(8)}$, satisfying a closed-form
equation. Since, in the circular orbit case, $m=0$ modes are static, this
equation is in fact an ordinary differential equation (ODE), and is readily
solvable analytically. Below we construct the analytic
solutions for these axially-symmetric odd-parity modes.

In the even-parity sector, both functions $\bar h^{(2)}$ and $\bar h^{(4)}$ have
vanishing source terms at $m=0$. Inspecting the field equations in their form
(\ref{II-130}), with Eqs.\ (\ref{tildeM2}) and (\ref{tildeM4}), we observe
that $\bar h^{(2)}$ and $\bar h^{(4)}$ couple not only to each other,
but also to $\bar h^{(1)}$ and $\bar h^{(5)}$. However, since this coupling
occurs through $t$ derivative terms, and since in our circular-orbit case
$m=0$ modes are static, these coupling terms in fact vanish, and we find
\begin{equation} \label{m0even}
\bar h^{(2)}_{m=0}=\bar h^{(4)}_{m=0}=0.
\end{equation}
Hence, the axially-symmetric even-parity part of the perturbation is
described by the 5 functions $\bar h^{(1,3,5,6,7)}$, which satisfy a coupled
set of ODEs.

\subsection{Analytic solutions for the axially-symmetric, odd-parity modes}

As we explained above, in the circular-orbit case, axially-symmetric ($m=0$)
odd-parity modes of the MP are constructed, at each $l$, from the single
function $\bar h^{(8)}_{m=0}$, which is $t$-independent in this case.
Denoting $\bar h^{(8)}_{m=0}\equiv \phi_l(r)$, the field equation
for $\bar h^{(8)}$ [Eqs.\ (\ref{FE}) with (\ref{M8})] takes the form
\begin{equation} \label{ODE}
\phi''_l+V_l(r)\phi_l=-4f^{-2}S^{(8)}_{m=0}=\beta_l\times\delta(r-r_0),
\end{equation}
where a prime denotes $d/dr$,
\begin{equation}
V_l(r)=-f^{-1}(r)\left[\frac{l(l+1)}{r^2}-\frac{4M}{r^3}\right],
\end{equation}
and the coefficient $\beta_l$ is given by
\begin{eqnarray}
\beta_l&=&-32\pi f_0^{-1}{\cal E}\omega Y^{l,m=0}_{,\theta}(\theta=\pi/2) \nonumber\\
&=&
\left\{\begin{array}{ll}
16 f_0^{-1}{\cal E}\omega(-1)^{(l-1)/2}[\pi(2l+1)]^{1/2}l!!/(l-1)!!,
& \text{$l$ odd}, \\
0, & \text{$l$ even}.
\end{array}\right.
\end{eqnarray}
Hence, we have $\phi_l(r)\equiv 0$ for all modes with even values
of $l$, and we need only consider modes with odd $l$ values.

Two independent homogeneous solutions to Eq.\ (\ref{ODE}) are given,
for $l\geq 2$ (we will discuss the mode $l=1$ separately below), by
\begin{eqnarray}\label{homosol}
\phi_{l\geq 2}^{\rm EH}(r)&=&\frac{x}{1+x}\sum_{n=0}^{l+1}a_n^l x^n, \nonumber\\
\phi_{l\geq 2}^{\infty}(r)&=&\phi_l^{\rm EH}\ln f+\frac{1}{1+x}\sum_{n=0}^{l+1}b_n^l x^n,
\end{eqnarray}
where
\begin{equation}
x\equiv r/(2M)-1,
\end{equation}
and the coefficients read
\begin{equation}
a_n^l=\frac{l(l+1)(l+n-1)!}{(l-n+1)!(n+1)!n!}, \quad \quad
b_n^l=\sum_{k=0}^{l-n+1}(-1)^k\frac{a_{n+k}^l}{k+1}.
\end{equation}
These solutions have the following asymptotic behavior at the horizon
($r\to 2M$, $x\to 0$, $f\to 0$) and at infinity ($r,x\to\infty$):
\begin{equation}
\phi_{l\geq 2}^{\rm EH}\propto\left\{
\begin{array}{ll}
f,       & r\to 2M, \\
r^{l+1}, & r\to\infty,
\end{array}\right.
\end{equation}
\begin{equation}
\phi_{l\geq 2}^{\infty}\propto\left\{
\begin{array}{ll}
f\ln f,       & r\to 2M, \\
r^{-l},       & r\to\infty.
\end{array}\right.
\end{equation}
The solution $\phi_l^{\rm EH}$ is regular (analytic) at the horizon
but diverges at $r\to\infty$, whereas the solution $\phi_l^{\infty}$
is regular at $r\to\infty$ but irregular at the horizon (it vanishes
there, but it is non-differentiable). Matching these solutions at the
particle's location ($r=r_0$), we hence construct a {\em unique}
regular, continuous solution for the inhomogeneous Eq.\ (\ref{ODE}):
\begin{equation} \label{phi}
\phi_{l\geq 2}=-2M\lambda\beta_l\times\left\{\begin{array}{ll}
\phi_l^{\rm EH}(r)\phi_l^{\infty}(r_0), & r\leq r_0, \\
\phi_l^{\infty}(r)\phi_l^{\rm EH}(r_0), & r\geq r_0,
\end{array}\right.
\end{equation}
where, recall, $\lambda=(l+2)(l-1)$.
In obtaining this solution we have used the fact that the Wronskian
$W\equiv\phi^{\rm EH}[\phi^{\infty}]'-[\phi^{\rm EH}]'\phi^{\infty}$
must be constant [since the ODE (\ref{ODE}) contains no $\phi_l'$ term].
Evaluating it at $x=0$ one then easily obtains
$W=-b_0^l/(2M)=-(2M\lambda)^{-1}$.

At $l=1$, the function $\phi_{l}^{\infty}$ of Eq.\ (\ref{homosol})
fails to be a solution of the homogeneous part of Eq.\ (\ref{ODE})
(although $\phi_{l=1}^{\rm EH}$ still is a solution).
Instead, the general homogeneous solution takes the simple form
\begin{equation} \label{phil1general}
\phi_{l=1}=ar^2+b/r,
\end{equation}
where $a$ and $b$ are constants. [Note that for $l=1$ the effective
potential in Eq.\ (\ref{ODE}) reduces to simply $V(r)=-2/r^2$.]
The coefficients $a$ and $b$ are determined uniquely by requiring
regularity at the horizon\footnote{There is a subtlety here: The
homogeneous solution $\propto 1/r$ is regular everywhere for any
finite $M$ and $r_0$. However, we do wish to require that the solution
remains regular even at the limit $M\to 0$ (taken with fixed $r_0$, which
is mathematically equivalent to the limit $r_0\to\infty$, taken with fixed
$M$). This excludes the solution $\propto 1/r$ at $r<r_0$, as
it grows unboundedly at the horizon in that limit.}
and at infinity, and imposing continuity at
$r=r_0$, along with a `jump' condition for the derivative there:
[$\phi'_{l=1}]_{r_0}=\beta_{l=1}$. This yields
\begin{equation} \label{phil1}
\phi_{l=1}=-\frac{1}{3}r_0\beta_{l=1}\times\left\{
\begin{array}{ll}
(r/r_0)^2,      & r\leq r_0, \\
(r_0/r),        & r\geq r_0,
\end{array} \right.
\end{equation}
with $\beta_{l=1}=16\sqrt{3\pi} f_0^{-1}{\cal E}\omega$.

We can now write down explicitly the solution for the axially-symmetric,
odd-parity part of the MP $h_{\alpha\beta}$ itself. From the reconstruction
equations (\ref{1-50}) [with Eqs.\ (\ref{1-83})], recalling $Y_{\rm V2}^{lm}=0$
at $m=0$, we find that the only non-vanishing axially-symmetric, odd-parity
components are $h_{t\varphi}=h_{\varphi t}$, given by
\begin{eqnarray} \label{htphi}
h_{t\varphi}[m=0,\text{odd}]&=&
-\frac{\mu}{2}\sum_{\text{odd } l}\sin\theta\, \bar h^{(8)}_{m=0}Y_{\rm V1}^{l,m=0}
\nonumber\\
&=&-\sum_{\text{odd } l}\frac{\mu}{2l(l+1)}\,\phi_l(r)
\sin\theta\,Y_{,\theta}^{l,m=0}(\theta).
\end{eqnarray}

The lowest multipole contribution to the sum in Eq.\ (\ref{htphi})
is a ``conservative'' piece coming from the dipole mode, $l=1$.
It reads
\begin{eqnarray} \label{htphi1}
h_{t\varphi}[m=0,l=1]&=&
-\frac{1}{4}\mu\,\phi_{l=1}\sin\theta\,Y_{,\theta}^{l=1,m=0} \nonumber\\
&=&-2\mu f_0^{-1}{\cal E}\omega r_0 \sin^2\theta\times\left\{
\begin{array}{ll}
(r/r_0)^2,      & r\leq r_0, \\
(r_0/r),        & r\geq r_0.
\end{array} \right.
\end{eqnarray}
This agrees with the odd-parity dipole solution first obtained by
Zerilli \cite{Zerilli2}, which, as pointed out recently by Detweiler
and Poisson \cite{DP}, is a Lorenz-gauge solution. [In comparing
our solution (\ref{htphi1}) with Eq.\ (4.1) of \cite{DP}, note
$f_0^{-1}{\cal E}\omega r_0^2=[Mr_0/(1-3M/r_0)]^{1/2}$ is the particle's
specific angular momentum, denoted $\tilde L$ in \cite{DP}.]
This piece of the MP describes the shift in the angular momentum
content of the perturbation across the surface $r=r_0$.

\subsection{Analytic solution for the monopole ($l=0$) mode}

The lowest multipole contribution to the MP comes from the monopole,
$l=m=0$ mode. This ``conservative'' piece of the MP (which is purely
even parity) describes the shift in the mass parameter of the perturbation
across the surface $r=r_0$. As was the case with the $m=0$, odd parity
modes considered above, at $l=0$, too, the field equations (\ref{FE})
simplify enough that one can obtain the solution {\em analytically},
with moderate effort. This solution was derived recently
by Detweiler and Poisson \cite{DP}.\footnote{
In fact, Detweiler and Poisson did not obtain their solution by directly
tackling the Lorenz-gauge equations. Rather, they started with the $l=0$
solution derived by Zerilli \cite{Zerilli2} in a non-Lorenz gauge, and then,
essentially, solved the gauge transformation equations that take Zerilli's
solution to the Lorenz gauge.}
For the sake of completeness, and since Detweiler and Poisson do not
actually write down their solution explicitly (as they are interested
mainly in the SF exerted by the monopole, not the monopole
MP itself), we bring this solution here.

At $r\leq r_0$, the non-vanishing components of the Lorenz-gauge
monopole perturbation are given by
\begin{mathletters}\label{monopolein}
\begin{equation}
h_{tt}^{l=0}(r\leq r_0)=-\frac{AfM}{r^3}P(r),
\end{equation}
\begin{equation}
h_{rr}^{l=0}(r\leq r_0)=\frac{A}{r^3f}Q(r),
\end{equation}
\begin{equation}
h_{\theta\theta}^{l=0}(r\leq r_0)=
(\sin\theta)^{-2} h_{\varphi\varphi}^{l=0}(r\leq r_0)=AfP(r),
\end{equation}
\end{mathletters}
where
\begin{equation}
A=\frac{2\mu {\cal E}}{3Mr_0f_0}\left[M-(r_0-3M)\ln f_0\right],
\end{equation}
\begin{equation}
P(r)= r^2+2Mr+4M^2, \quad\quad
Q(r)= r^3-Mr^2-2M^2r+12M^3.
\end{equation}

At $r\geq r_0$, the solutions read
\begin{mathletters}\label{monopoleout}
\begin{eqnarray}
h_{tt}^{l=0}(r\geq r_0)&=&
\frac{2\mu{\cal E}}{3r^4r_0f_0}\left\{3r^3(r_0-r)+M^2(r_0^2-12Mr_0+8M^2)+
\right. \nonumber\\
&& \left.
(r_0-3M)\left[-rM(r+4M)+rP(r)f\ln f+8M^3\ln(r_0/r)\right]
\right\},
\end{eqnarray}
\begin{eqnarray}
h_{rr}^{l=0}(r\geq r_0)&=&
-\frac{2\mu{\cal E}}{3r^4r_0f_0f^2}\left\{
-r^3r_0-2Mr(r_0^2-6Mr_0-10M^2)+3M^2(r_0^2-12Mr_0+8M^2)+
\right. \nonumber\\ && \left.
(r_0-3M)\left[
5Mr^2+(r/M)Q(r)f\ln f
-8M^2(2r-3M)\ln(r_0/r)\right]\right\},
\end{eqnarray}
\begin{eqnarray}
h_{\theta\theta}^{l=0}(r\geq r_0)&=&
(\sin\theta)^{-2} h_{\varphi\varphi}^{l=0}(r\geq r_0)=
-\frac{2\mu{\cal E}}{9rr_0f_0}\left\{
3r_0^2M-80M^2r_0+156M^3+
\right. \nonumber\\ && \left.
(r_0-3M)\left[-3r^2-12Mr
+3(r/M)P(r)f\ln f+44M^2+24M^2\ln(r_0/r)\right]
\right\}.
\end{eqnarray}
\end{mathletters}
It can be readily verified that these solutions match continuously at
$r=r_0$. It can also be checked that they satisfy both the field equations
(\ref{FE}) and the gauge conditions (\ref{gauge}).
For this, note that at $l=0$, in the circular orbit case, the only
non-vanishing $\bar h^{(i)}$'s are $\bar h^{(1,3,6)}$, and
use the relations
$\bar h^{(1)}_{l=0}=2\sqrt{\pi}\mu^{-1}\,r(h_{tt}+f^2h_{rr})$,
$\bar h^{(6)}_{l=0}=2\sqrt{\pi}\mu^{-1}\,rf^{-1}(h_{tt}-f^2h_{rr})$, and
$\bar h^{(3)}_{l=0}=4\sqrt{\pi}\mu^{-1}(f/r)h_{\theta\theta}$.

The above monopole solution is regular both at the event horizon
and at infinity, taking the asymptotic forms
\begin{equation} \label{monopoleEH}
\left. \begin{array}{ll}
h_{tt}^{l=0}=    -\frac{2}{3} Af +O(f^2) \nonumber\\
f^2 h_{rr}^{l=0}= \frac{2}{3} Af +O(f^2) \nonumber\\
h_{\theta\theta}^{l=0} = 12M^2Af +O(f^3)
\end{array} \right\}
\quad \text{as $r\to2M$ ($f\to 0$)},
\end{equation}
\begin{equation} \label{monopoleInfty}
\left. \begin{array}{ll}
h_{tt}^{l=0}=    -\frac{2\mu{\cal E}}{r_0}f_0^{-1}(1-r_0/r) \nonumber\\
h_{rr}^{l=0}= \frac{2\mu{\cal E}}{r}   \nonumber\\
r^{-2}h_{\theta\theta}^{l=0} = \frac{2\mu{\cal E}}{r}f_0^{-1}(1-3M/r_0)
\end{array} \right\}
+O(1/r^2),\quad \text{as $r\to\infty$}.
\end{equation}
Detweiler and Poisson show \cite{DP} that this is a {\em unique} Lorenz-gauge
solution which is regular both at the event horizon and at infinity: Any
gauge transformation of this solution within the class of Lorenz gauges
would lead to irregular behavior at one (or both) of these asymptotic
domains.\footnote{Note, however, the peculiar feature of the above solution,
$h_{tt}\to{\rm const}[=-2\mu{\cal E}(r_0f_0)^{-1}]$ at $r\to\infty$, which
means that the perturbed metric, expressed in Schwarzschild coordinates,
does not tend to the Minkowski metric at infinity. (Recall, however, that
this peculiarity merely relates to the choice of gauge; the underlying
perturbed geometry is, of course, asymptotically flat.)
}

Physically, the monopole perturbation of Eqs.\ (\ref{monopolein}) and (\ref{monopoleout})
describes a shift in the Schwarzschild mass across $r=r_0$. This is most
clearly evident from Zerilli's form of the monopole solution \cite{Zerilli1}
(which differs from the above Lorenz-gauge solution only by a gauge transformation
\cite{DP}), where it is easily seen that the geometry described by
$g_{\alpha}+h_{\alpha\beta}^{l=0}$ is that of a Schwarzschild black hole with
mass $M$ at $r<r_0$, and that of yet another Schwarzschild black hole, with mass
$M+\mu{\cal E}$, at $r>r_0$.

\begin{figure}[htb]
\input{epsf}
\centerline{\epsfysize 5cm \epsfbox{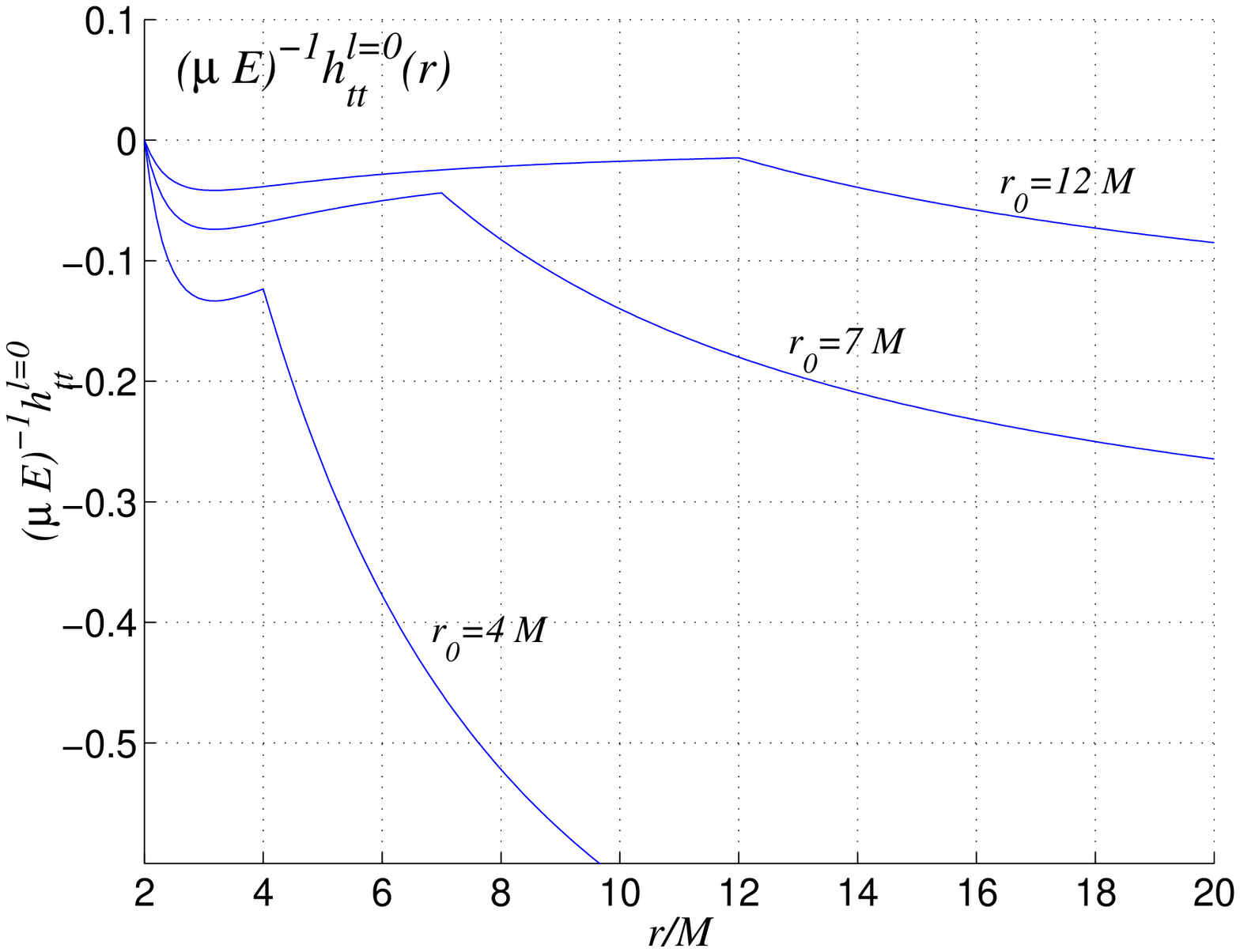}
\epsfysize 5cm \epsfbox{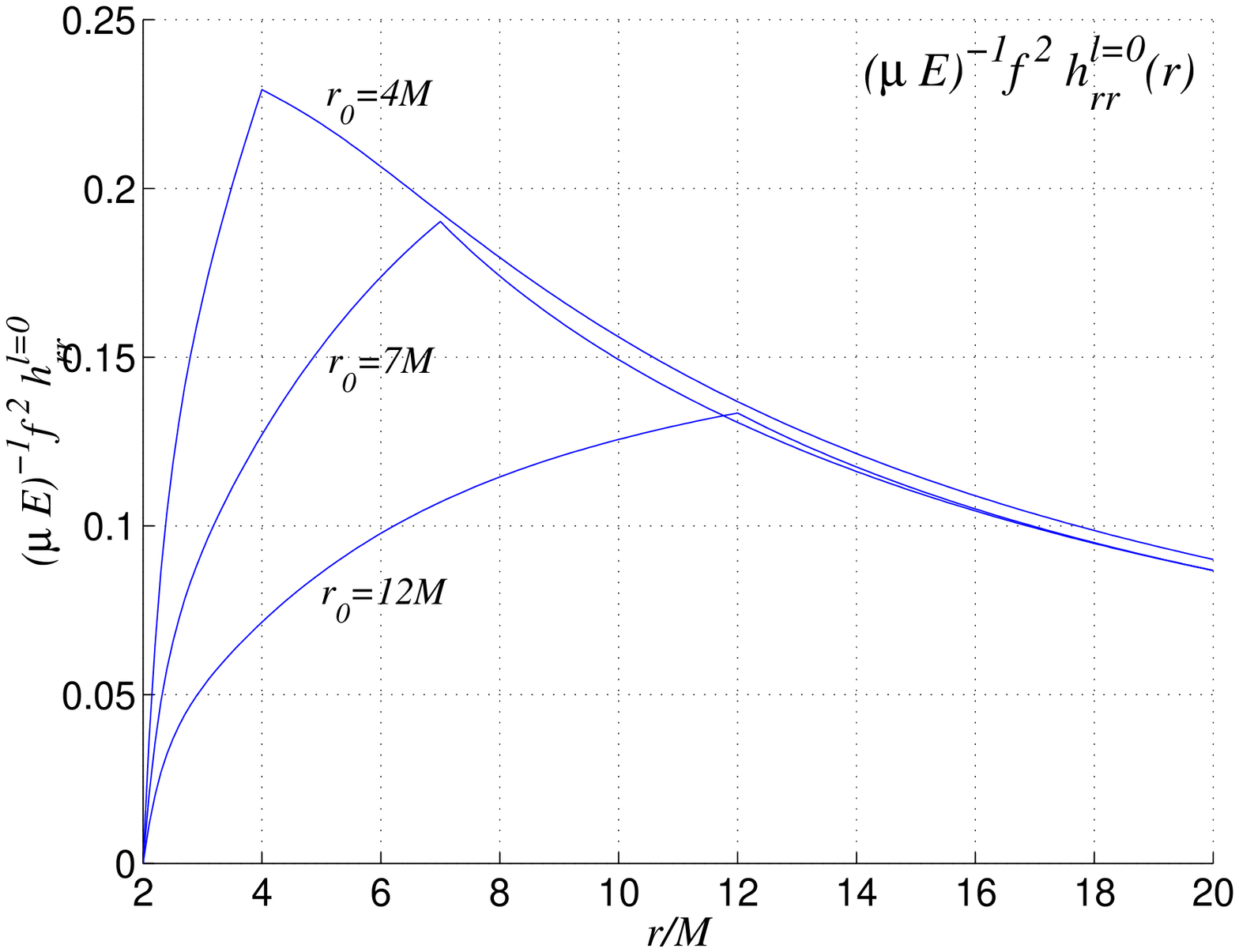}}
\centerline{\epsfysize 5cm \epsfbox{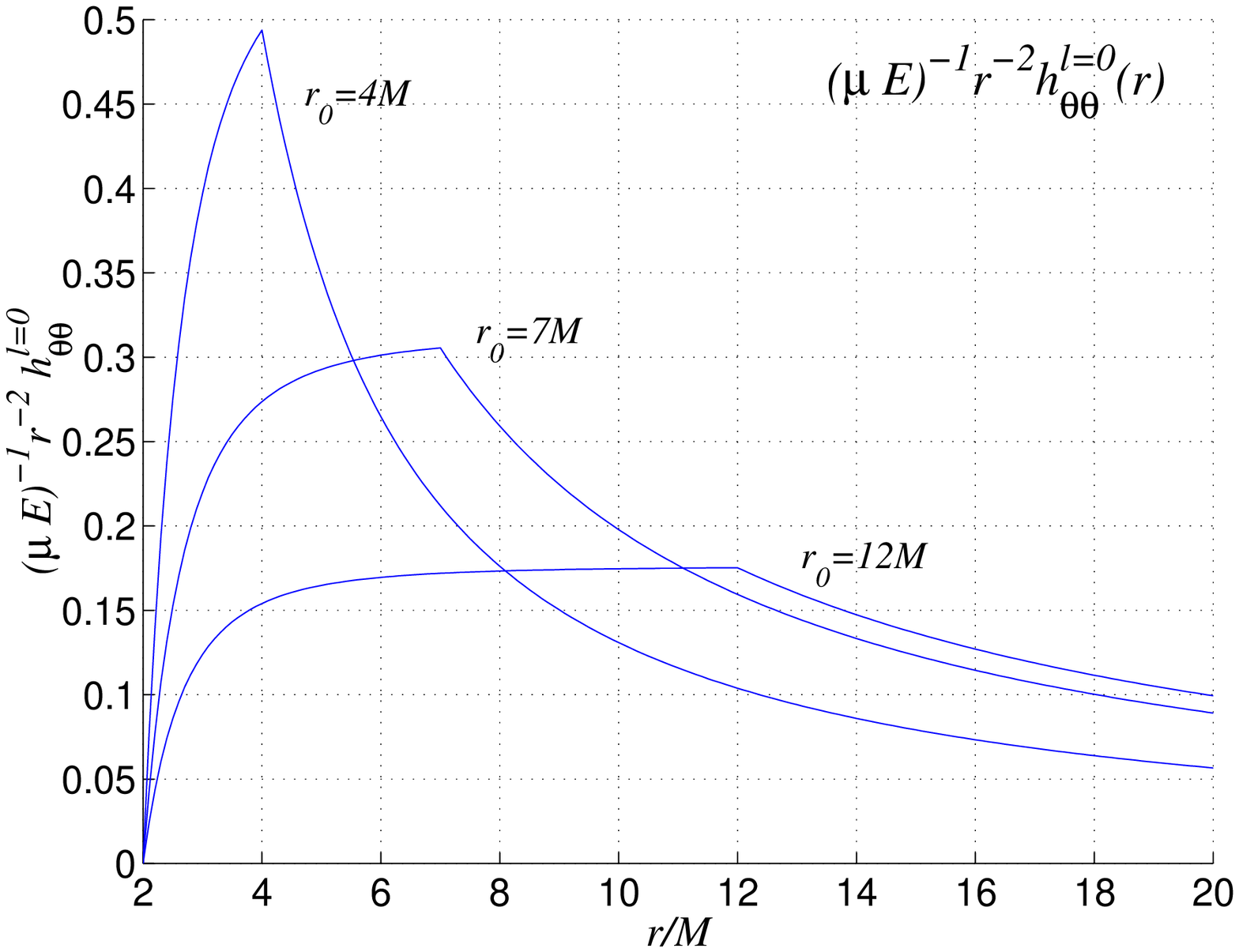}}
\caption{\protect\footnotesize
The Lorenz-gauge monopole solution, Eqs.\ (\ref{monopolein}) and
(\ref{monopoleout}), plotted here for three values of the orbital radius,
$r_0=4M,7M,12M$. By construction (Detweiler and Poisson, Ref.\ \protect\cite{DP})
this solution is continuous across the orbit, and well behaved both at
the event horizon and at infinity. It is a unique Lorenz-gauge monopole
solution with these properties. (The quantity $E$ is the specific energy
parameter, denoted elsewhere in this paper by $\cal E$.)
}
\label{fig-monopole}
\end{figure}

The above Lorenz-gauge monopole solution is plotted in Fig.\ \ref{fig-monopole},
for a sample of $r_0$ values.

\subsection{Numerical solutions for the rest of the modes}

For the modes considered so far (those with $m=0$ and $l=0,1,3,5,\ldots$)
the field equations simplify enough to allow a fully analytic treatment
(as least within the simplicity of the circular orbit case).
To solve for the rest of the modes we will resort to numerical methods,
employing the full machinery of the formalism developed in Sec.\ \ref{SecII}.
In the rest of this section we briefly describe our numerical method,
and plot a sample of numerical solutions. In the next section we shall
present various validation tests for our numerical code.

We need to integrate, numerically, the set of coupled field equations
(\ref{FE}), with the source terms given in Eq.\ (\ref{Si}). Recall that
for each given $l,m$, we are facing two sets of 7 (even parity) and 3
(odd parity) equations, which couple in the manner described in Table
\ref{TableI} (but recall that no coupling occurs at the principal parts
of the equations). We choose to integrate these equations in the {\em time
domain}, i.e., without introducing a Fourier decomposition. A frequency-domain
analysis (of the sort employed many times in the past---see, e.g., \cite{Poisson})
is likely to be more efficient, numerically, for studying circular orbits.
Our strong motivation in developing a time-domain evolution code stems from
the fact that such a code is readily extensible to any type of orbit, with
arbitrary eccentricity. A time-domain code can handle radial plunge trajectories
as efficiently as it handles circular orbits.
Frequency-domains codes, on the other hand, quickly loose their efficiency
with growing eccentricity, as the number of Fourier modes one must sum
over grow rapidly with eccentricity. Orbits with eccentricities greater
than $\sim 0.7$ are essentially intractable with a frequency-domain
treatment \cite{Glam&Ken}. Another reason to opt for a time evolution
approach, especially having in mind SF applications, is the following:
A frequency domain analysis of the field equations requires careful
implementation of boundary conditions, which becomes increasingly difficult
with increasing $l$ values. The technical reason for this is explained,
e.g., by Hughes in \cite{Hughes}. With a time domain evolution one can
avoid this difficulty, either by expanding the spacial boundaries of
the numerical domain such as to dismiss any boundary effects, or by
using characteristic evolution (as described below) which avoids
boundaries altogether. Hence, a time domain approach should allow
better accessibility to the higher $l$ modes.

A suitable numerical method, for time evolution of the field equations
with a particle source represented by a delta function, was first presented by
Lousto and Price \cite{Lousto&Price}. It has since been implemented in
a variety of a cases: A radially-falling scalar charge \cite{Barack&Burko},
a radially-falling mass particle \cite{Barack&Lousto,Martel&Poisson},
a mass particle in a circular orbit \cite{LoustoUP}, and, lately, a mass
particle in eccentric and parabolic orbits \cite{Martel}.
In the scalar field case \cite{Barack&Burko} the source term in the
decoupled field equations [the scalar-field equivalent of our Eqs.\
(\ref{FE})] is simply proportional to a delta function
$\delta[r-r_{\rm p}(\tau)]$.
However, in the gravitational field case, all above works
\cite{Barack&Lousto,Martel&Poisson,LoustoUP,Martel} employed the
Regge-Wheeler--Zerilli--Moncrief's formulation, where the source
involves also terms proportional to $d\delta[r-r_{\rm p}(\tau)]/dr$.
As a consequence, solutions to the Regge-Wheeler--Zerilli--Moncrief
equations turn out {\em discontinuous} across the orbit, which, of course,
complicates considerably the implementation of the above numerical method.
One great advantage of working with the Lorenz-gauge MP, is that the
source term in the field equations involves
only a delta function, no derivatives thereof---as in the simple
scalar case. The solutions are then continuous everywhere, and the
implementation of the numerical integration scheme of Ref.\ \cite{Lousto&Price}
becomes must simpler.

\begin{figure}[htb]
\input{epsf}
\centerline{\epsfysize 7cm \epsfbox{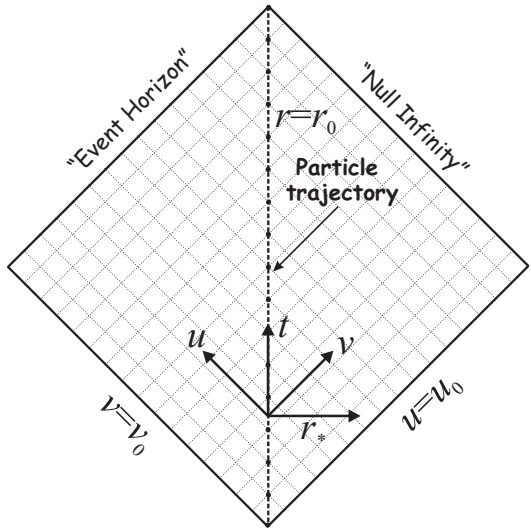}}
\caption{\protect\footnotesize
Numerical grid for characteristic evolution of the field equations.
See the text for description.
}
\label{fig-grid}
\end{figure}

Our numerical code is based on characteristic evolution and uses double-null
coordinates $v\equiv t+r_*$ and $u\equiv t-r_*$ (like in \cite{Barack&Burko},
but unlike, e.g., in \cite{Martel}). The numerical domain is a two dimensional
fixed-step grid, as illustrated in Fig.\ \ref{fig-grid}. The evolution starts
with characteristic data on $v=v_0$ and $u=u_0$, where we take
$v_0=r_*(r_0)$ and $u_0=-r_*(r_0)$. That is, we take $t=0$ at the initial
instance of the evolution [represented by the vortex $(v_0,u_0)$],
and take the two initial null surfaces to be the ingoing and outgoing light
rays emanating from the particle at that instance. The location of the
event horizon on this grid is approximated by a large value of $u$ (with
constant $v$), while large value of $v$ (with constant $u$) approximates
the location of null infinity. The trajectory of the particle is represented
by the vertical line $v-u=2r_*(r_0)$, connecting the lower- and upper-most
vertices of the grid. Since our numerical evolution variables are continuous
at the trajectory, there are no complication involved in taking the trajectory
to cut through grid points, as we do here; in fact, we found this setup most
convenient.

The numerical evolution is based on the finite difference scheme developed in
\cite{Lousto&Price}---as simplified to the case where the source term contains
no derivatives of a delta function, and generalized to deal with a set of
simultaneous equations. Since the scheme has been described in detail
elsewhere, we shall not elaborate on it here, but rather refer the reader to
\cite{Lousto&Price} or \cite{Martel&Poisson}. The finite difference
algorithm is second order convergent. The value of the solution at a
grid point with coordinates $(v_i,u_i)$ is approximated, within an error
quadratic in the step size, based on the values at the three grid points
$(v_{i-1},u_i)$, $(v_{i},u_{i-1})$, and $(v_{i-1},u_{i-1})$, which have
been solved for in previous steps. For terms containing derivatives
$\partial_v$ we also use information from the grid points $(v_{i},u_{i-2})$
and $(v_{i-1},u_{i-2})$ (which is necessary to maintain second order
convergence). The numerical evolution proceeds along successive lines
$v=\rm const$, from $v_0$ to large values of $v$ (``null infinity''),
where along each such line the integration proceeds from $u=u_0$ to
large values of $u$ (the ``event horizon'').

Obviously, we do not know how to prescribe exact initial data for our
problem. (In a our stationary scenario, knowledge of the exact solution
at one particular moment would be equivalent to having at hand the
solution at all times!) This is a standard problem of Numerical
Relativity. However, unlike the situation in fully non-linear simulations
of (say) equal-mass black hole mergers, here we have the luxury of being
able to stably run our evolution for as long as it takes for any
effect of ``corrupted'' initial data to dissipate off the numerical domain,
and for the solution to settle down to its correct, ``physical'' value.
To explain the basic idea, consider first the case of a scalar field, where
the evolution is not constrained. Suppose that we prescribe
some initial data on the initial surface (e.g., in our characteristic initial
data formulation, determine the magnitude of the scalar field along $v=v_0$
and $u=u_0$), and then numerically integrate the inhomogeneous scalar field
equation, with a source particle. Suppose that the solution we thus obtain
represents, in the continuum limit, an exact solution of the inhomogeneous
field equation. This solution would then differ from the true, ``physical''
solution by a homogeneous solution of the field equation. We may hence view
the numerical solution as a superposition of the ``physical'' solution and
a spurious homogeneous perturbation. However, homogeneous (i.e., vacuum)
perturbations of black hole spacetimes always decay at late time (the
``no hair'' theorem). One need only wait long enough until the spurious waves
radiate away and the ``true'' solution is revealed. On theoretical ground,
we expect the spurious waves to die off in time with a power-law tail, where
the power index depends on the multipole number of the mode in question.
In the circular orbit case, experiment shows that one practically has to wait
for about one orbital period before the spurious waves clear
out (see \cite{Barack&Burko,Barack&Lousto,Martel}, and also Figs.\
\ref{fig:solutions:fixedr}--\ref{fig:solutions:highl} below).

The above picture becomes slightly more involved in our case, where the
evolving fields are components of the MP, which are subject to certain
constraints in the form of gauge conditions. The initial data are now
no longer freely specifiable, but are required to satisfy the gauge
conditions---or otherwise our solution would not be guaranteed to satisfy
the gauge conditions even at late time. Alternatively, one may incorporate
in the numerical evolution procedure a machinery for damping out constraint
violations, like the one described above.

Here we follow the second strategy: We use a version of the field equations
that has constraint damping built into it [i.e., above Eqs.\ (\ref{FE})], and
free ourselves from the need to devise exact initial data. This option has two
obvious advantages: Firstly, the required exact initial data would depend
on the source in question, and will have to be developed case by case.
On the other hand, once an effective constraint-damping scheme is established
for circular orbits, it should be equally effective in other cases (say,
eccentric orbits). Secondly, and more importantly, a successful
constraint-damping scheme should take care of both initial constraint
violations, {\em and} constraint violations from numerical errors.
Specifying constraint-obeying initial data, on the other hand, would not
guarantee, by itself, that constraint violations due to numerical
(e.g., finite differentiation) errors remain small.

As initial data for our numerical evolution we take, most simply,
\begin{equation} \label{IC}
\bar h^{(i)lm}(v=v_0)=\bar h^{(i)lm}(u=u_0)=0,
\end{equation}
for any $l,m$ and each of $i=1,\ldots,10$.

\subsection{Sample numerical results}

\begin{figure}[htb]
\input{epsf}
\centerline{\epsfysize 7cm \epsfbox{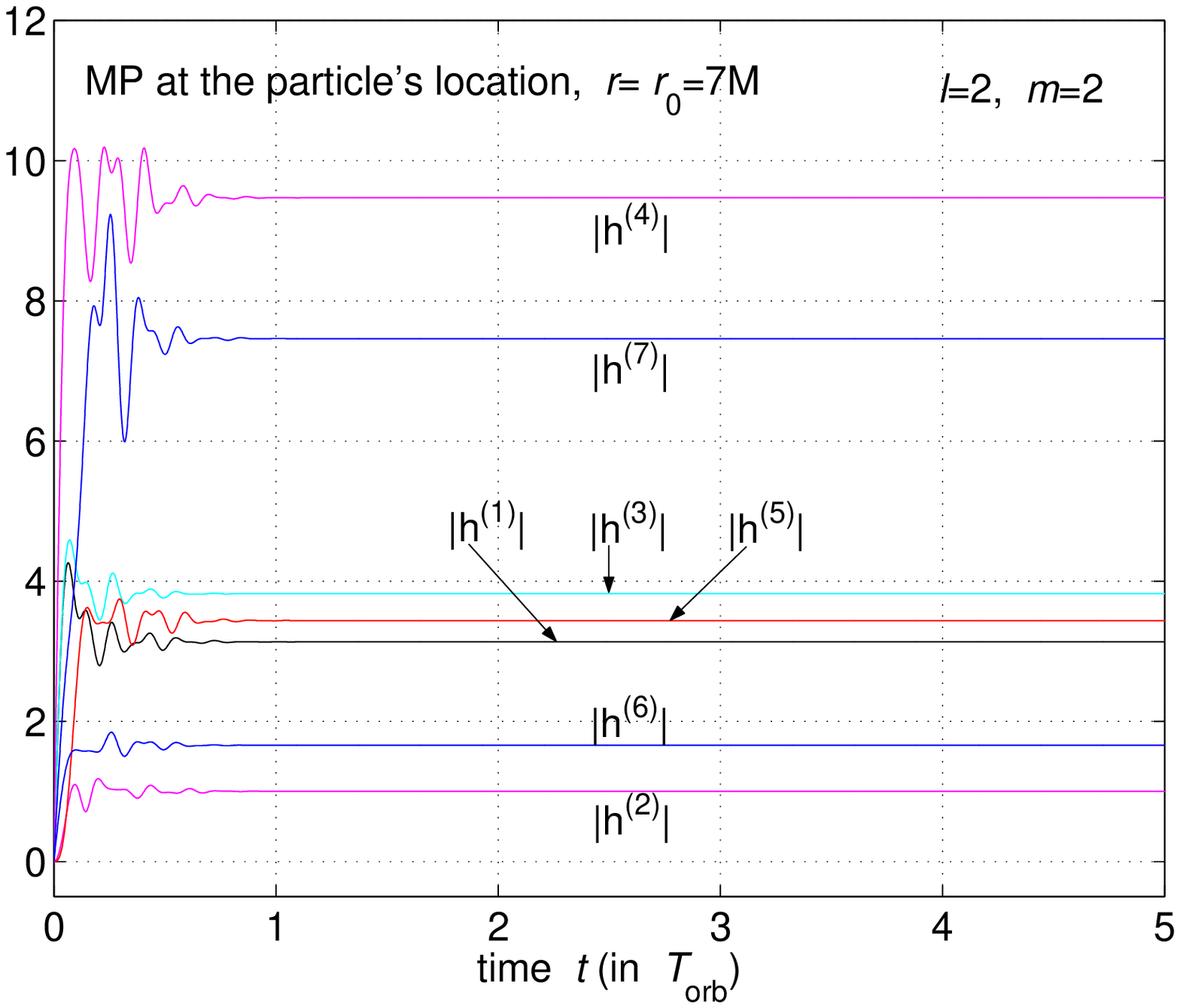}
\epsfysize 7cm \epsfbox{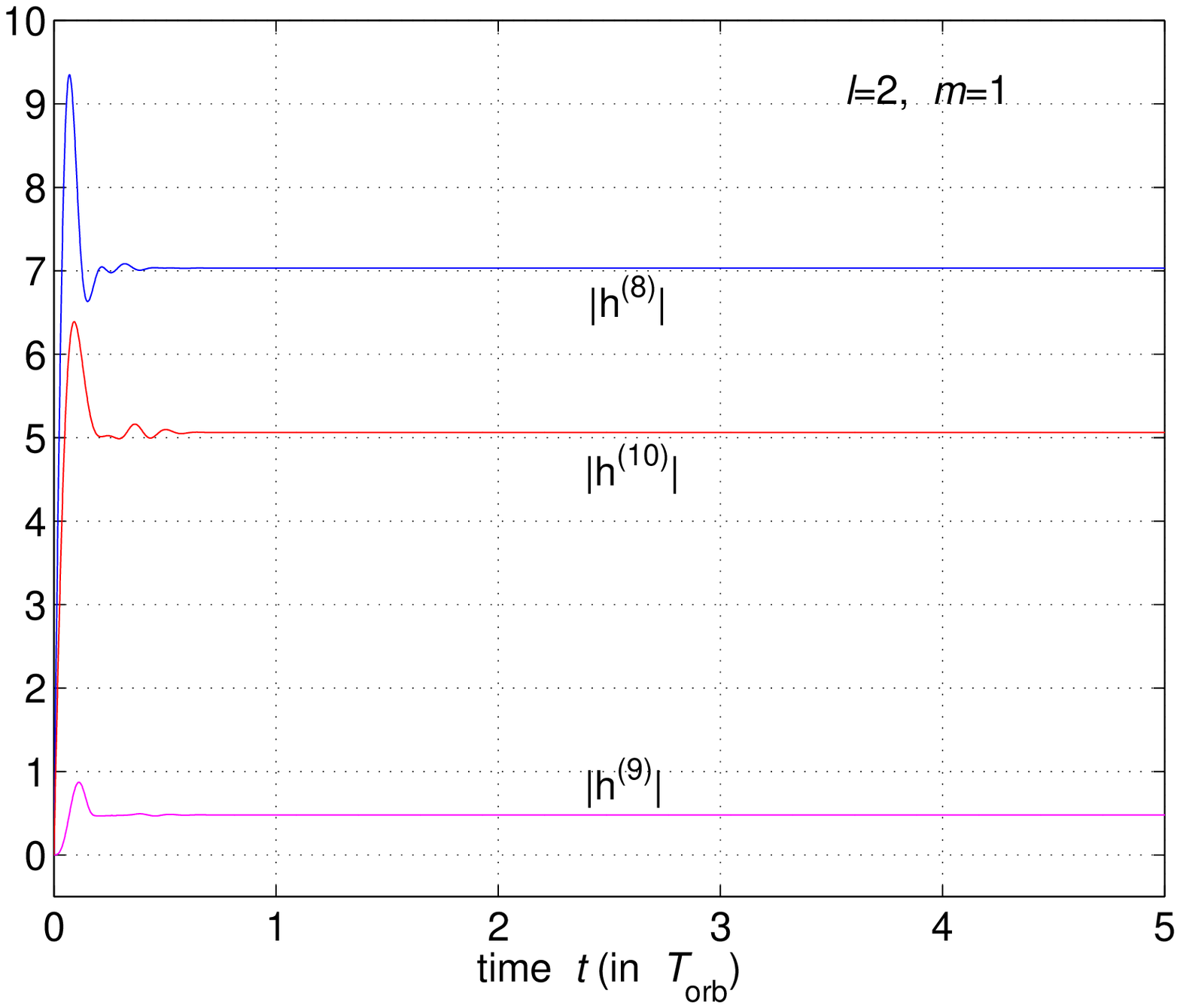}}
\caption{\protect\footnotesize
Numerical solutions for the (dimensionless) Lorenz-gauge MP functions
$\bar h^{(i)l=2,m=1,2}$, evaluated at the particle's location, for a particle
in a circular orbit at $r=r_0=7M$. In the Lorenz gauge (unlike in the
Regge--Wheeler gauge, for example) the mode-decomposed MP is continuous
at the particle, and has a definite value there.
The early stage of the
time evolution is dominated by transient spurious waves associated
with the imperfection of the initial data. This part of the evolution
(which, of course, is to be discarded in interpreting the physics content
of the numerical results) lasts around one orbital period of evolution
time, after which the inherent physical behavior is unveiled.
Our code evolves separately the real and imaginary parts of the complex
functions $\bar h^{(i)lm}$, which are both needed to construct the full MP
[through formula (\ref{1-50})]; for compactness, we show here only
the moduli $\left|\bar h^{(i)lm}\right|
= \left\{[{\rm Re}(\bar h^{(i)lm})]^2+[{\rm Im}(\bar h^{(i)lm})]^2\right\}^{1/2}$.
The mode $l=m=2$ is of even parity, and is constructed from the seven
even-parity functions $\bar h^{(1,\ldots,7)}$; the odd-parity mode $l=2$, $m=1$
is constructed from the three remaining functions, $\bar h^{(8,9,10)}$.
}
\label{fig:solutions:fixedr}
\end{figure}

\begin{figure}[htb]
\input{epsf}
\centerline{\epsfysize 7cm \epsfbox{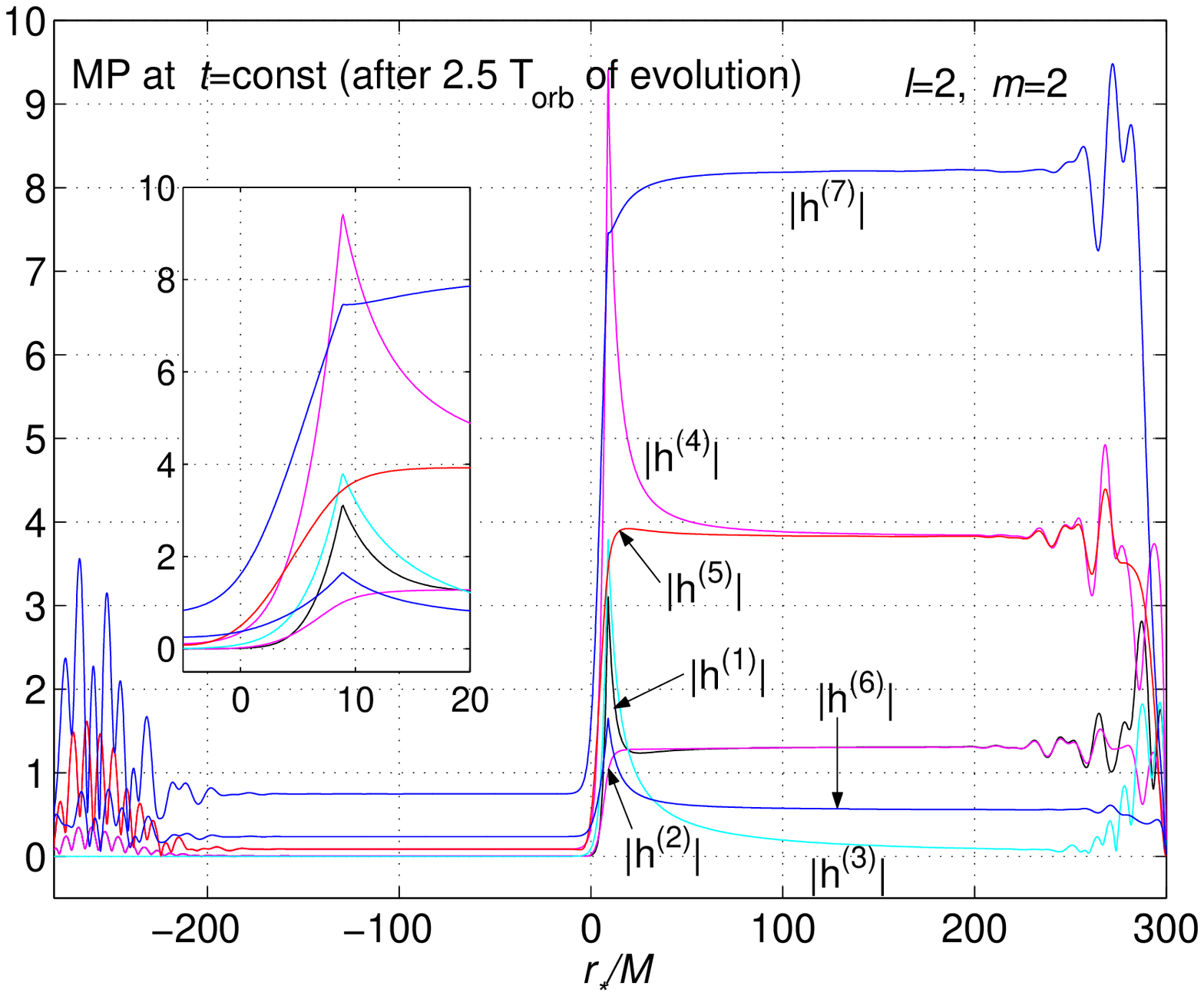}
\epsfysize 7cm \epsfbox{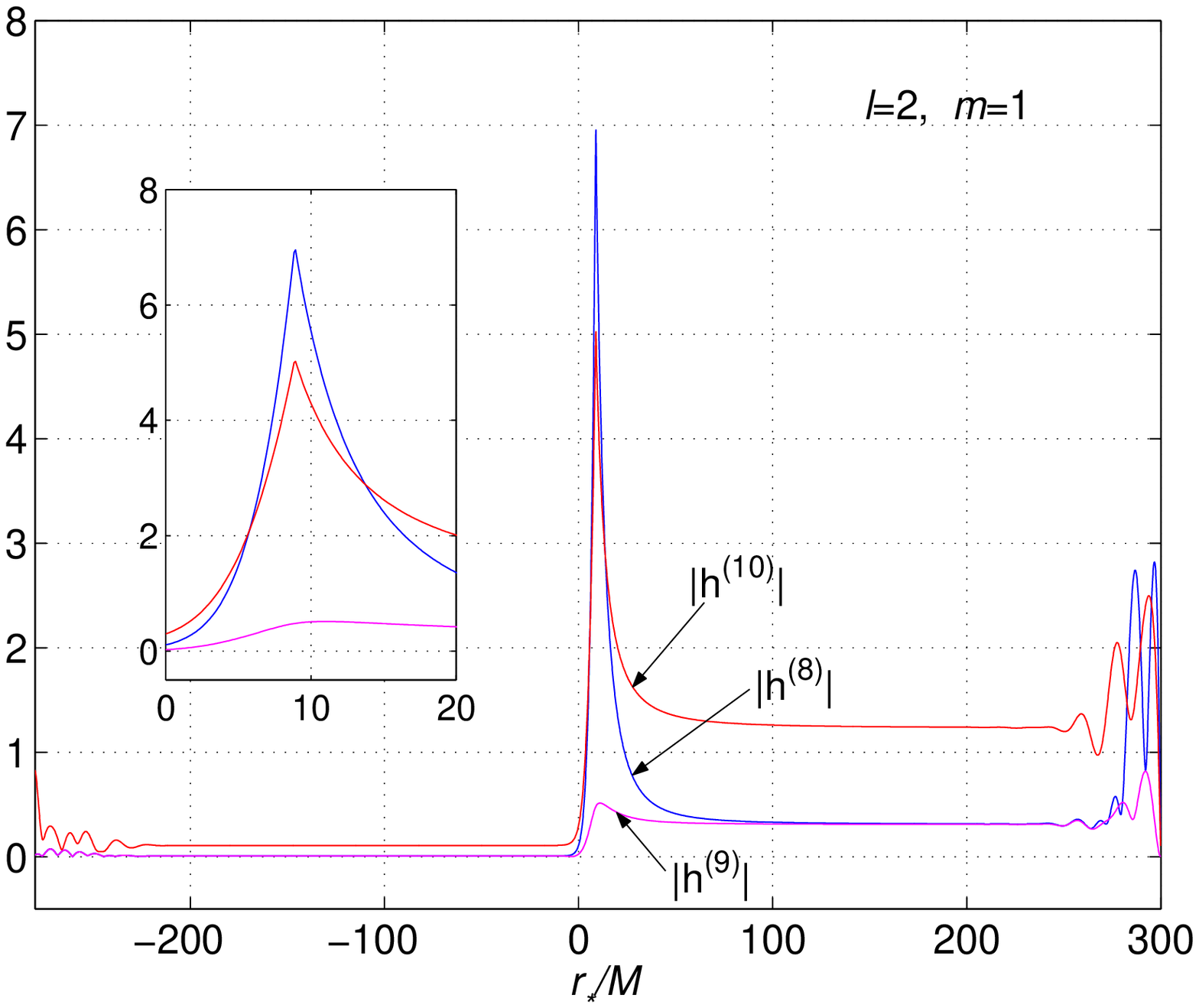}}
\caption{\protect\footnotesize
A different slice cut through the solutions of Fig.\ \ref{fig:solutions:fixedr},
this time showing the behavior of the fields $\bar h^{(i)l=2,m=1,2}$ on a
$t$=const slice, 2.5 $T_{\rm orb}$ into the evolution.
The wavy feature on the left- and right-hand sides (small and large
values of $r_*$, respectively) are associated with the spurious initial waves
propagating inward (toward the black hole) and outward (to infinity),
and are to be discarded. While all functions $\bar h^{(i)lm}$ are continuous
across the particle (located at $r=7M$, $r_*\simeq 8.83M$), those functions
whose corresponding sources $S^{(i)lm}$ are non-zero have discontinues
$r$ derivatives there. The fields $\bar h^{(2,5,9)lm}$, whose sources
$S^{(2,5,9)lm}$ vanish, have continuous derivatives across the particle.
The insets expand the peak area of the plots, for better clarity.
The code resolves the gradients of the MP fields at the particle with good
accuracy (cf.\ Figs.\ \ref{fig:res} in the next section). These gradients
are needed for SF calculations.
Note the significant damping in the MP amplitude at $r_*\protect\lesssim 0$.
This, presumably, is the effect of the well known potential barrier
surrounding the black hole. The high-frequency spurious waves, on the other
hand, penetrate this barrier with ease.
}
\label{fig:solutions:fixedt}
\end{figure}

\begin{figure}[htb]
\input{epsf}
\centerline{\epsfysize 7cm \epsfbox{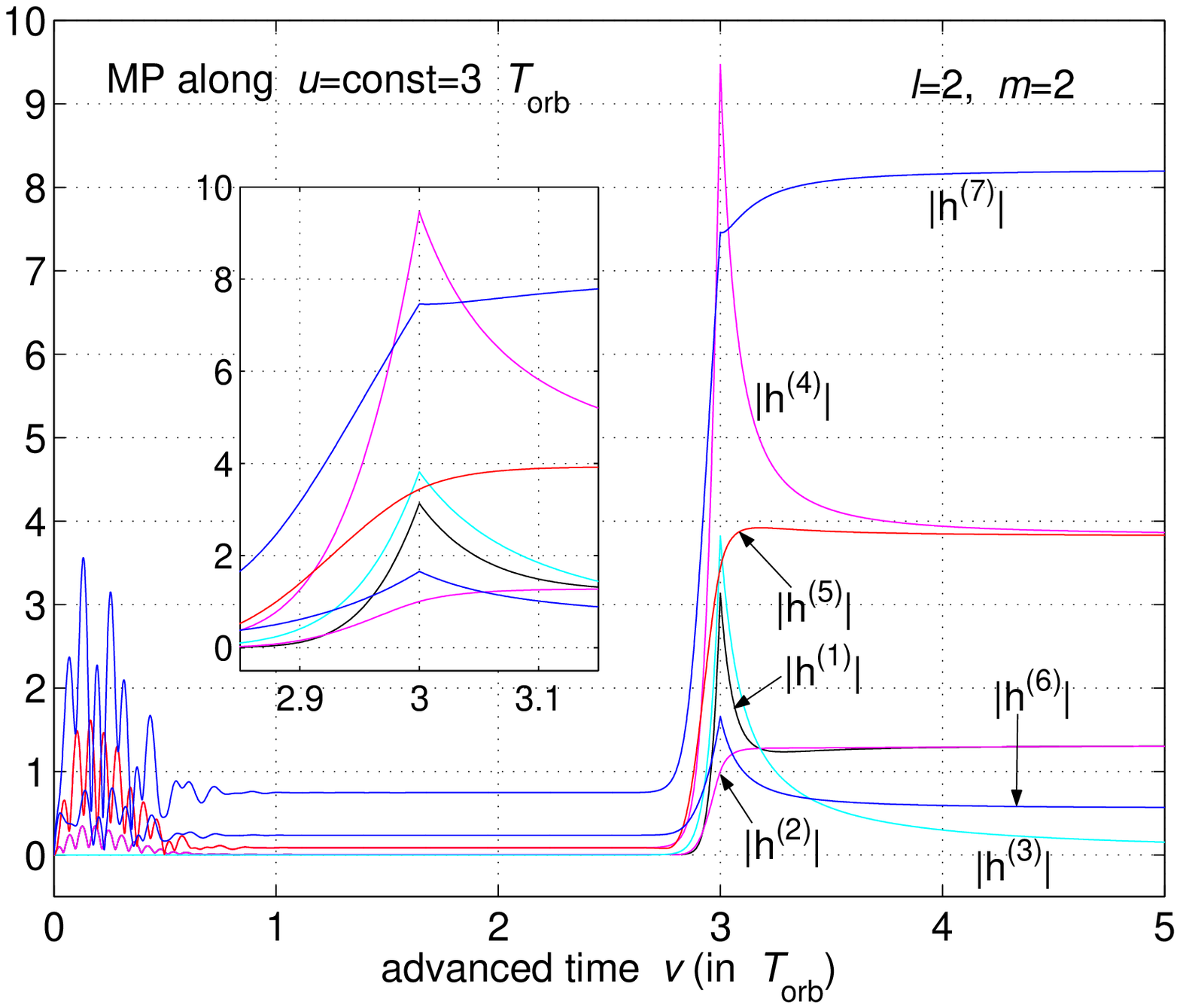}
\epsfysize 7cm \epsfbox{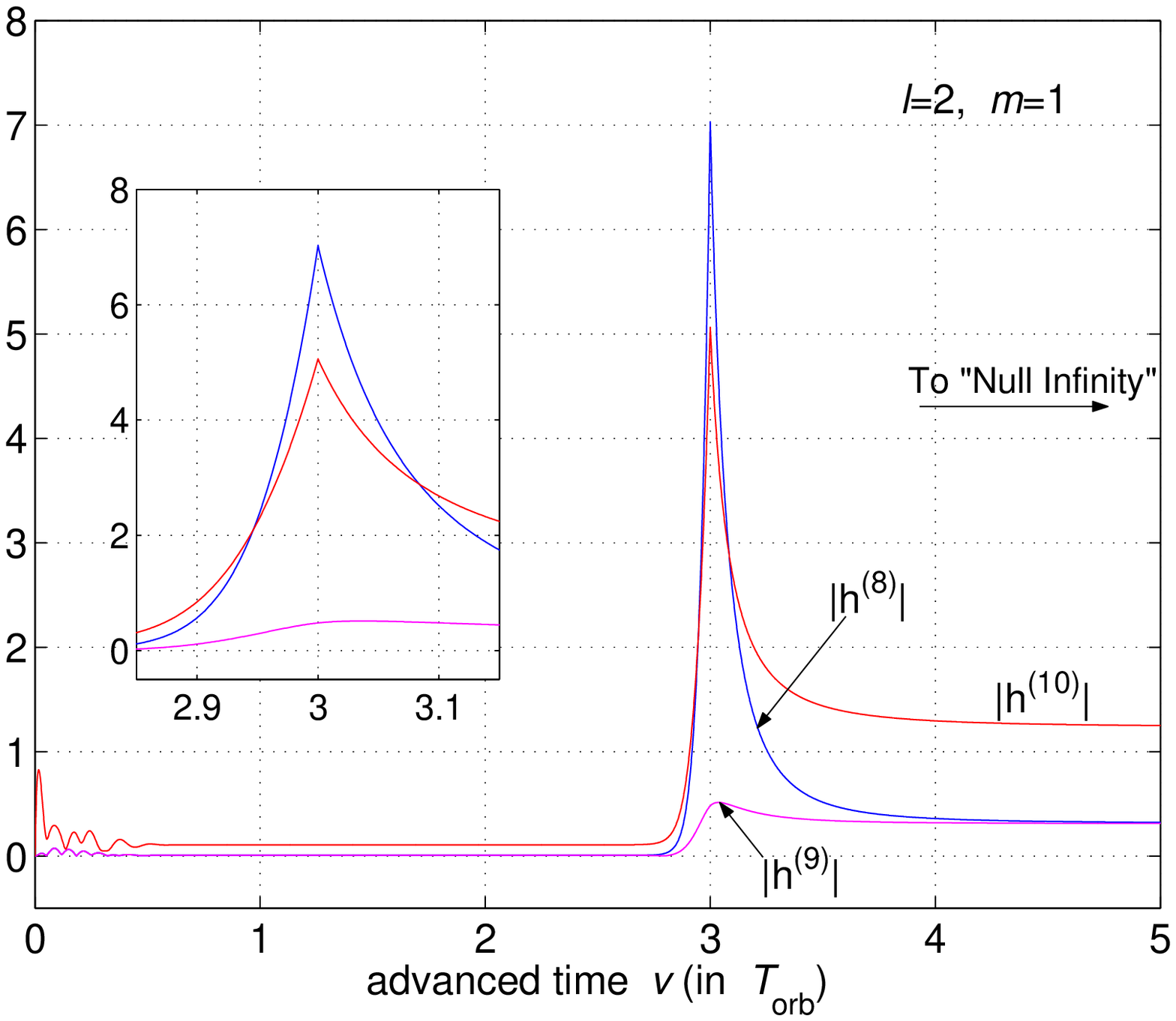}}
\caption{\protect\footnotesize
Another slice cut through the solutions of Figs.\ \ref{fig:solutions:fixedr}
and \ref{fig:solutions:fixedt}, this time showing the behavior along
an outgoing ray, $u={\rm const}(=3\;T_{\rm orb})$. Once again, the insets
expand the peak areas. All fields $\bar h^{(i)lm}$ approach constant values
at late advanced times (``null infinity''). The fluxes of energy and angular
momentum carried by gravitational waves to infinity are straightforwardly
extracted from these values (if fact, only $|\bar h^{(7)lm}|$ and $|\bar h^{(10)lm}|$
are needed for this purpose), as we discuss in Sec.\ \ref{Subsec:flux}.
Note how at large $v$ we seem to have
$|\bar h^{(3)}|\sim 0$,
$|\bar h^{(2)}|\sim |\bar h^{(1)}|$, $|\bar h^{(4)}|\sim |\bar h^{(5)}|$, and
$|\bar h^{(8)}|\sim |\bar h^{(9)}|$. In Sec.\ \ref{Subsec:flux} we explain
how these empirical asymptotic relations are predicted theoretically.
}
\label{fig:solutions:fixedu}
\end{figure}

\begin{figure}[htb]
\input{epsf}
\centerline{\epsfysize 7cm \epsfbox{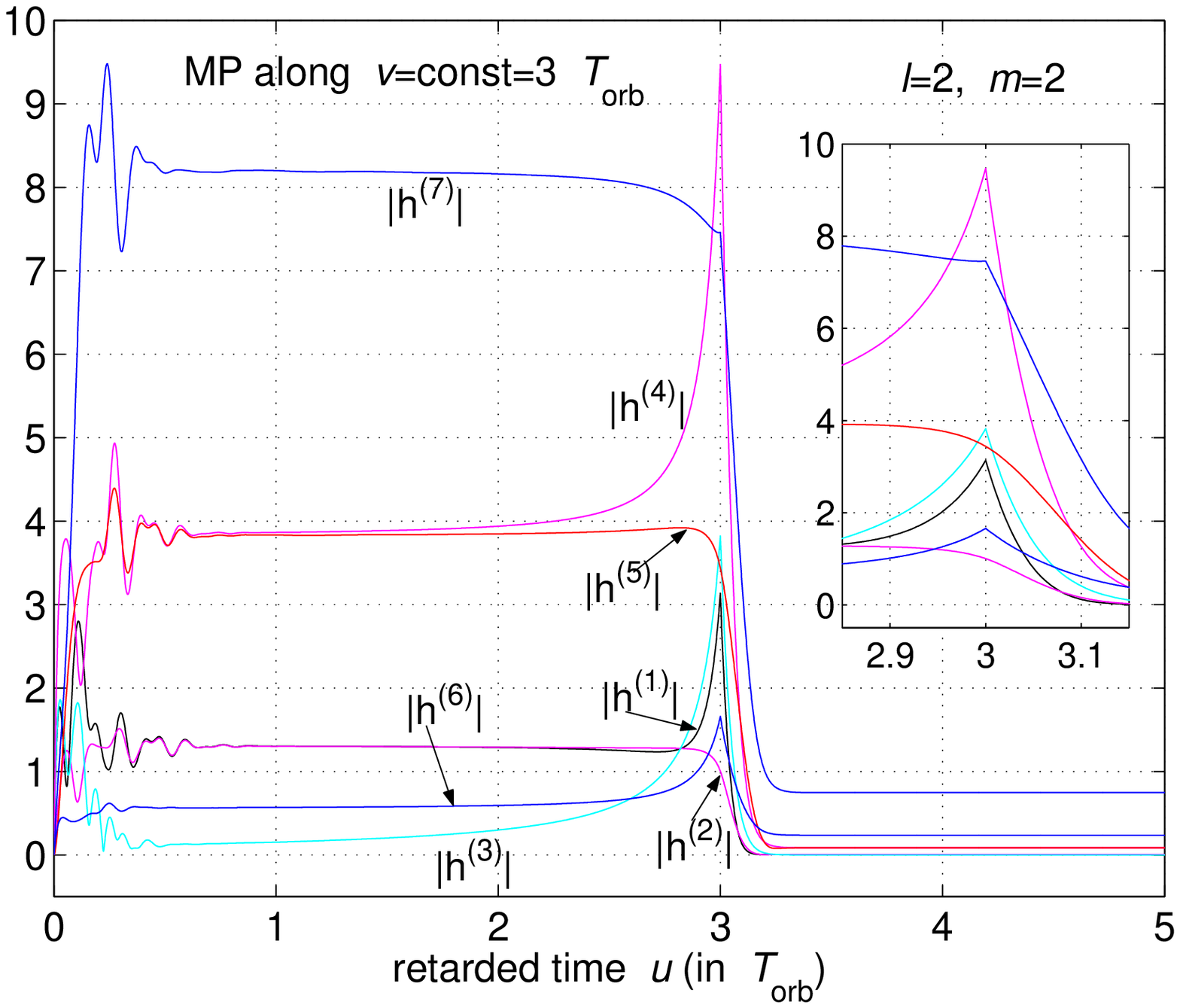}
\epsfysize 7cm \epsfbox{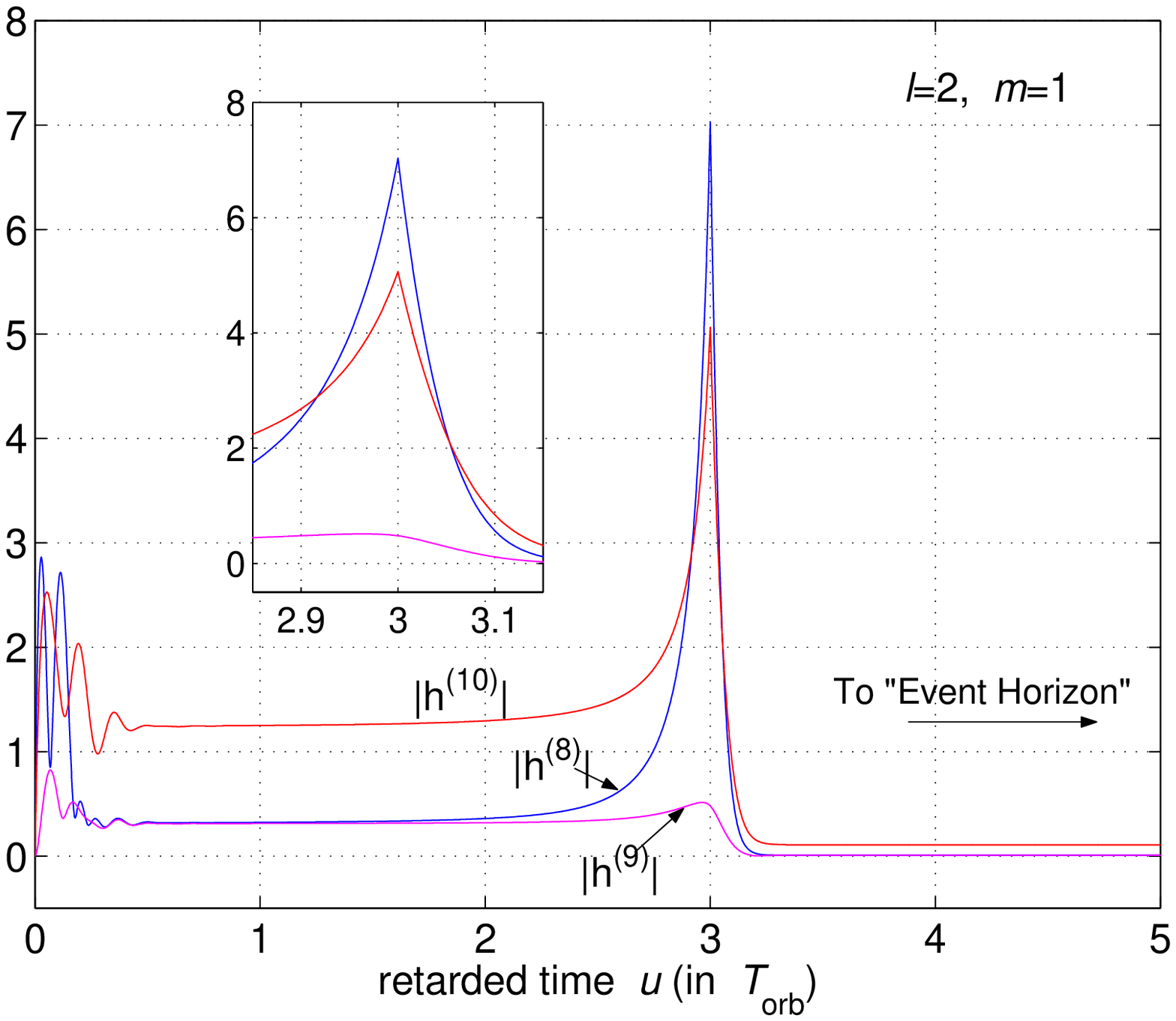}}
\caption{\protect\footnotesize
One more slice cut through the solutions of Figs.\ \ref{fig:solutions:fixedr},
\ref{fig:solutions:fixedt} and \ref{fig:solutions:fixedu}, showing the behavior
along an incoming ray, $v={\rm const}(=3\;T_{\rm orb})$. With increasing
retarded time, approaching the ``event horizon'', all fields $|\bar h^{(i)lm}|$
settle at constant values. Although hard to tell from these plots, these
values are generally nonzero (with the exception of $\bar h^{(3)lm}$, which
tends to zero at the horizon---see the text for explanation).
}
\label{fig:solutions:fixedv}
\end{figure}

\begin{figure}[htb]
\input{epsf}
\centerline{\epsfysize 7cm \epsfbox{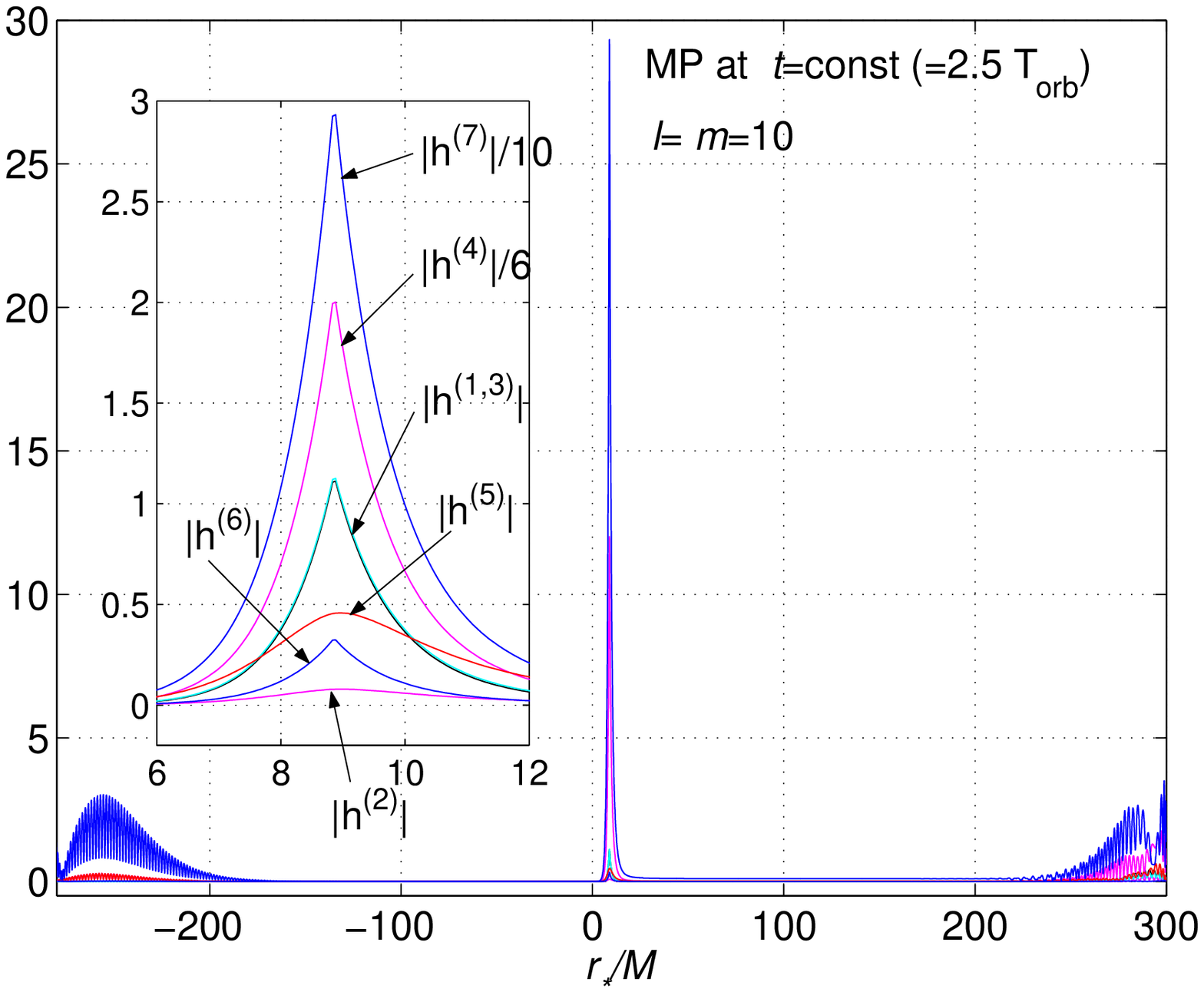}
\epsfysize 7cm \epsfbox{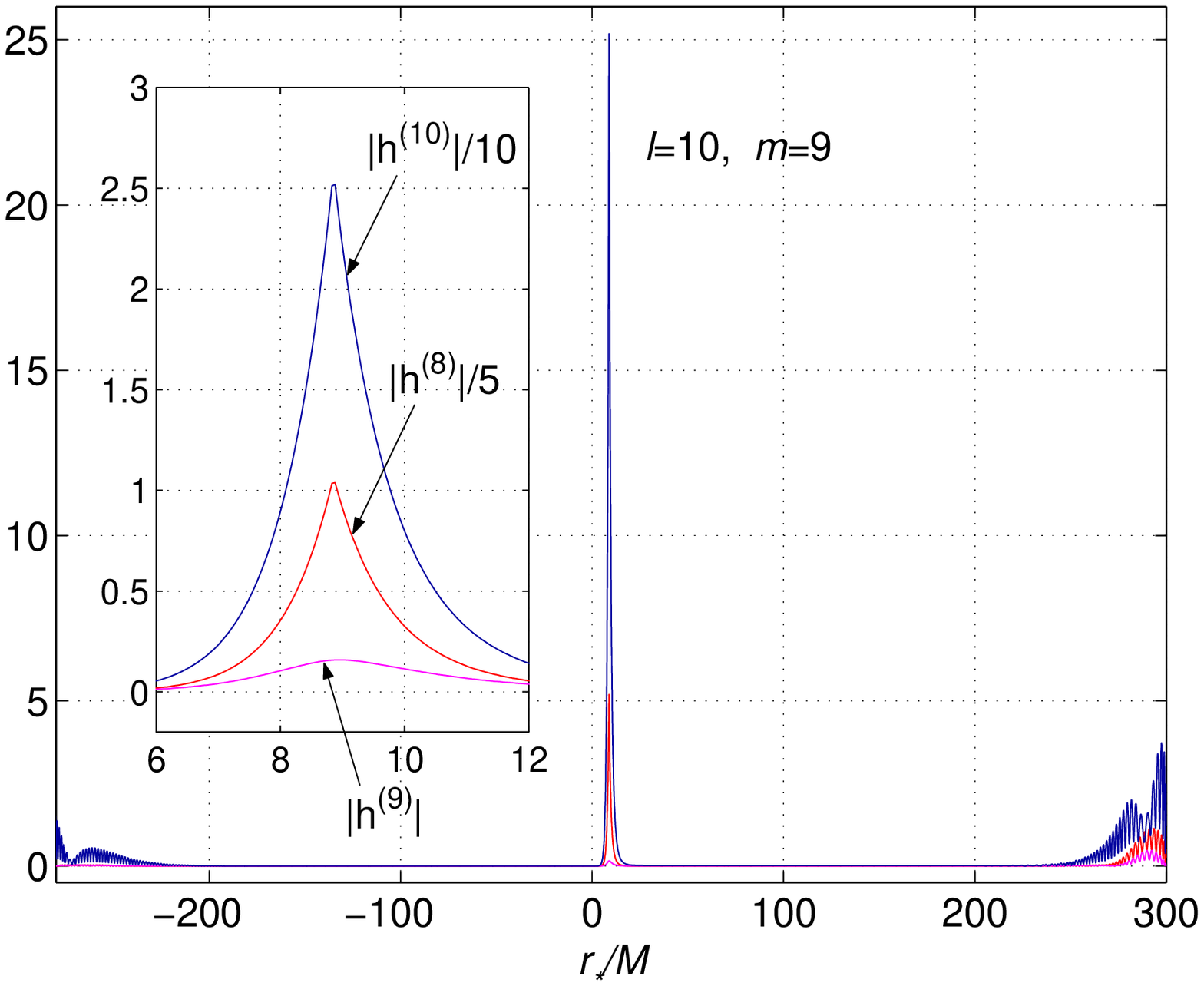}}
\caption{\protect\footnotesize
The MP functions  $\bar h^{(i)lm}$ for $l=10$ and $m=9,10$, on a $t$=const slice
(compare with Fig.\ \ref {fig:solutions:fixedt}). Spacial gradients of the
MP get larger with growing $l$, but at $l=10$ the particle is still perfectly
resolvable (our code performs well for multipole numbers up to $\sim 20$).
The seemingly noisy features on both ends of the plots are, in fact,
numerically robust: They are transient spurious waves of the type
also apparent in Figs.\ \ref{fig:solutions:fixedr}--\ref{fig:solutions:fixedv},
but now occurring at much higher frequencies.
}
\label{fig:solutions:highl}
\end{figure}

Figures \ref{fig:solutions:fixedr}--\ref{fig:solutions:highl} show
typical numerical solutions for the various functions $\bar h^{(i)lm}$.
For these plots we have taken the particle to move in a
strong-field circular geodesic orbit at $r_0=7M$ (corresponding
to a period of $T_{\rm orb}\sim 116M$). The fields are evolved for
an amount of time equivalent to 5 $T_{\rm orb}$, which we find
more than sufficient to allow the transient spurious waves die off.
The linear resolution for these runs is set at a few grid points
per $M$ in both $v$ and $u$, which translates to a few hundreds points
per wave cycle at $m=1$. With this resolution, the magnitude of the
various $\bar h^{(i)lm}$'s is resolved at a fractional accuracy better
than $\sim 10^{-4}$ even near the particle's location, where the fields'
gradients are largest (cf.\ Fig.\ \ref{fig:res} in the next section).
With these specifications, a single $l,m$ mode takes of order a second
to run on a standard laptop.

The graphs in Figs.\ \ref{fig:solutions:fixedr}--\ref{fig:solutions:fixedv}
show the various $\bar h^{(i)lm}$'s for $l=2$ and $m=1,2$. Plotted are the
absolute values, $\left|\bar h^{(i)lm}\right|
= \left\{[{\rm Re}(\bar h^{(i)lm})]^2+[{\rm Im}(\bar h^{(i)lm})]^2\right\}^{1/2}$.
The different Figures present different slice-cuts
through the 2-dimensional numerical grid: $r={\rm const}(=7M)$,
$t=\rm const$, $u=\rm const$ (i.e., an outgoing ray approaching null
infinity), and $v=\rm const$ (an incoming ray approaching the event horizon).
Note that in Fig.\ \ref{fig:solutions:fixedr} we show the fields $\bar h^{(i)lm}$
evaluated {\em at the particle's location}; recall these fields are continuous
and admit definite values there. The last plot, in Fig.\ \ref{fig:solutions:highl},
shows the numerical solutions for $l=10$ and $m=9,10$. High-$l$ calculations
are more demanding computationally, as the spacial scale of variation of
the fields decreases with increasing $l$.
In its present from, our code can handle multipoles up to $l\sim 20$;
this should be sufficient for high precision calculations of the SF,
through the mode-sum scheme.

Here are some features to notice when examining
Figs.\ \ref{fig:solutions:fixedr}--\ref{fig:solutions:highl}:
(i) The early stage of the fields' evolution is dominated by spurious
waves associated with the imperfection of the initial data. These
are transients, and die off almost entirely within one orbital period
of evolution time (the rate of decay of the spurious waves seems roughly
independent of $l,m$). Typically, one can safely read off the values of
the fields after $\sim 2\; T_{\rm orb}$ of evolution time. In performing
precision calculations (e.g., of the SF) for periodic orbits,
it is easy (and advisable) to monitor the level of ``contamination''
from transient waves, by comparing the values of the fields at, say,
$t=T_{\rm orb},2\; T_{\rm orb},3\; T_{\rm orb},\ldots$.
(ii) As emphasized above, the fields $\bar h^{(i)lm}$ are all continuous
through the particle's location. This is manifested in Figs.\
\ref{fig:solutions:fixedt}, \ref{fig:solutions:fixedu},
\ref{fig:solutions:fixedv}, and \ref{fig:solutions:highl}.
Those fields $\bar h^{(i)lm}$ whose evolution equations have non-vanishing
source terms (in our circular-orbit case, $i=1,3,4,6,7,8$) have
discontinuous spacial derivatives across the particle, as expected on
theoretical grounds. Those functions that are sourceless ($i=2,5,9,10$)
have continuous derivatives there. The values of the fields and
their gradients at the particle provide sufficient information for
calculating the gravitational SF. Our code resolves these
values with great accuracy.
(iii) The various $\bar h^{(i)lm}$'s approach finite values at null infinity
and toward the event horizon, as they are designed to do by construction.
These finite values are generally non-zero, except that $\bar h^{(3)lm}$
vanishes toward both ends. The reason for the vanishing of $\bar h^{(3)lm}$
at the horizon has been explained above [see the discussion surrounding
Eq.\ (\ref{1-60})]. The reason for the vanishing of this function at
$r\to\infty$ will be discussed in Sec.\ \ref{Subsec:flux} below.

Finally, we remind that, given the $\bar h^{(i)lm}$'s, the MP itself
is constructed algebraically through formula (\ref{1-50}).


\section{Code validation} \label{SecIV}

In this Section we present a series of tests we have performed to check the
validity of our numerical evolution code. These include (i) a test of
numerical convergence, (ii) confirmation that the numerical solutions
satisfy the Lorenz gauge conditions, and (iii) comparison of the flux of
energy radiated in gravitational waves to infinity, as extracted from our
solutions, with the values obtained using other methods. The second and third
of these tests check both our new formulation of the MP equations, and the
validity of its numerical implementation.

\subsection{Numerical convergence}

\begin{figure}[htb]
\input{epsf}
\centerline{\epsfysize 7cm \epsfbox{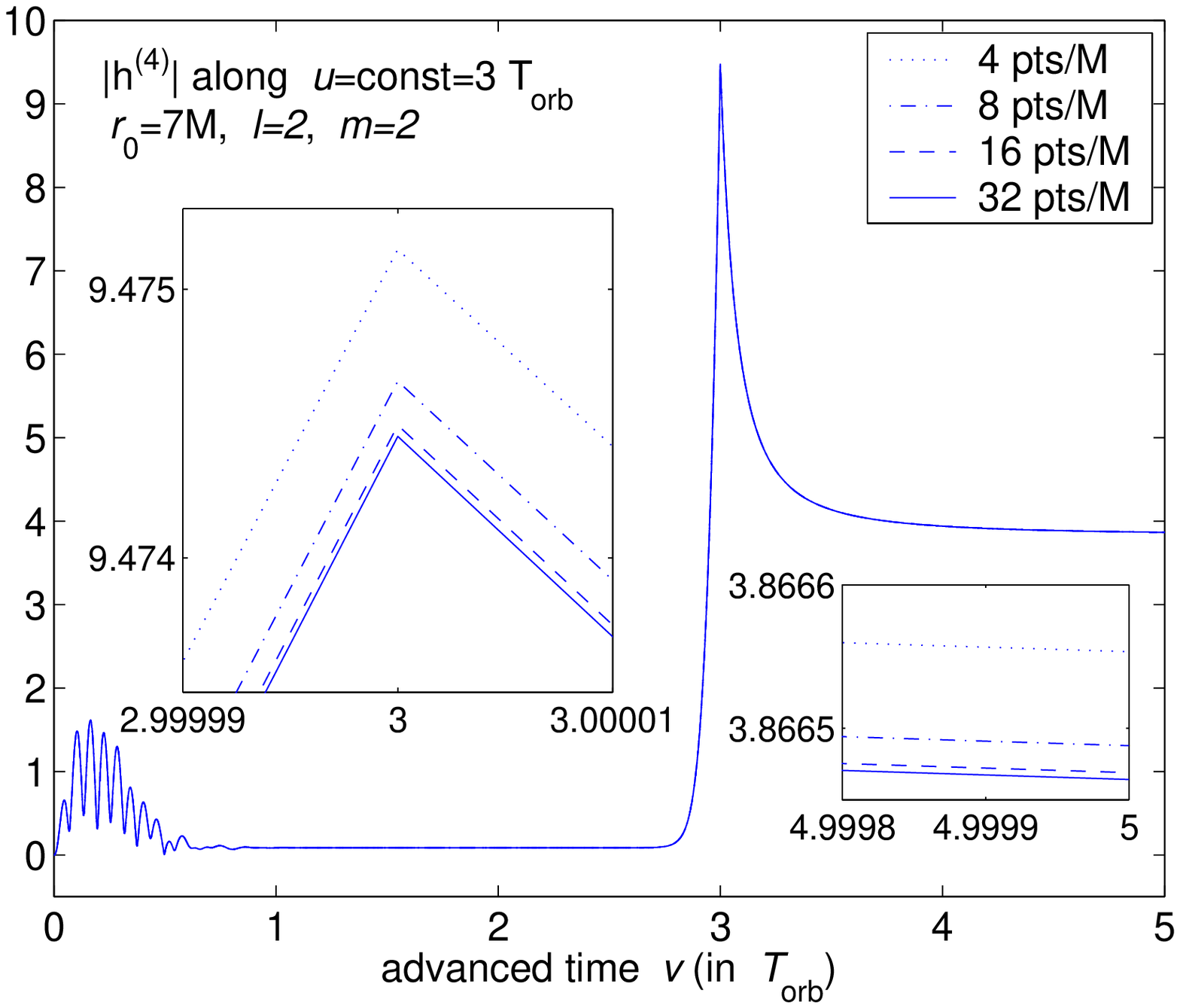}
\epsfysize 7cm \epsfbox{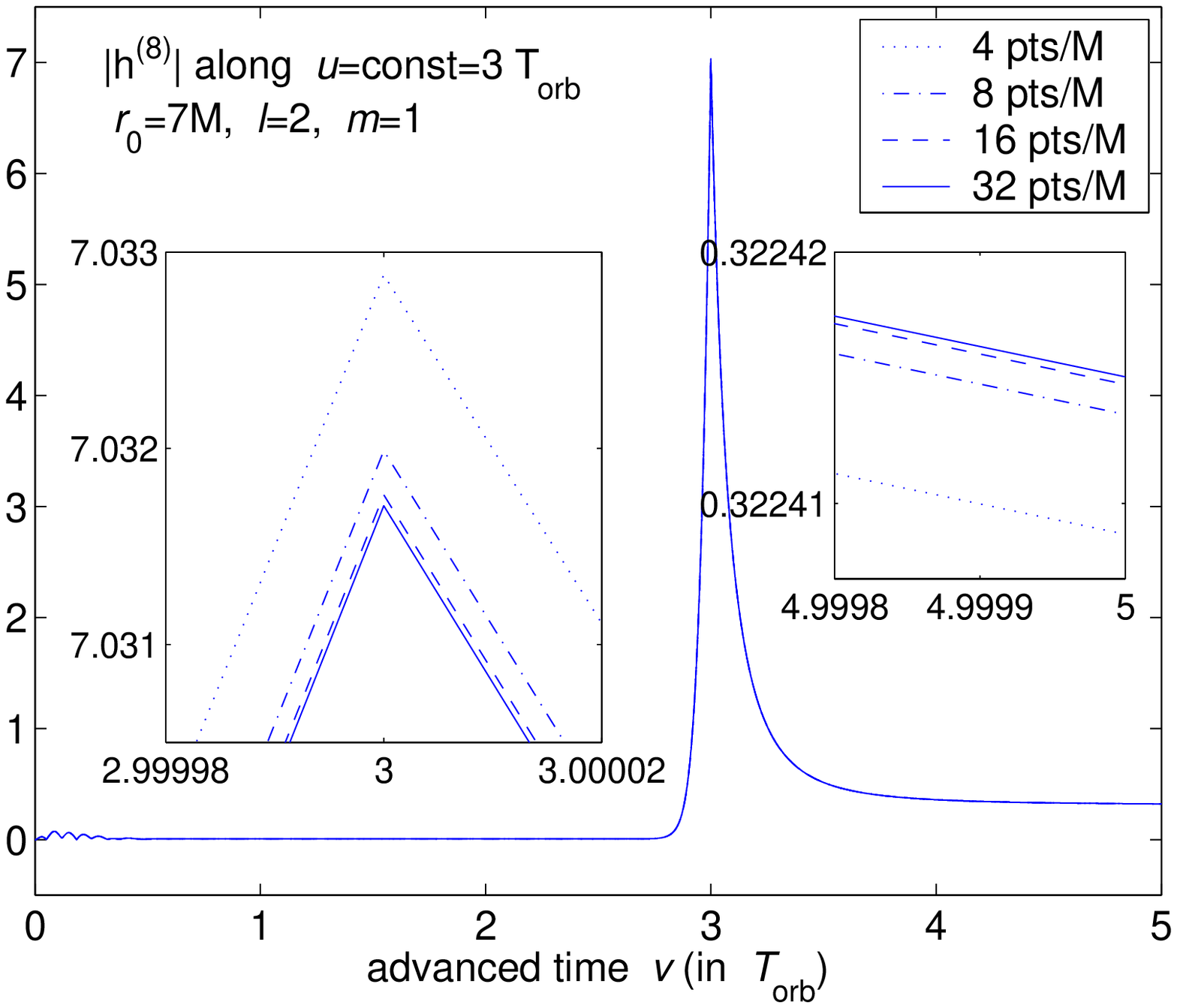}}
\caption{\protect\footnotesize
Illustration of numerical convergence: Shown here are numerical solutions
for (the absolute values of) the functions $\bar h^{(4)l=2,m=2}$ (left panel) and
$\bar h^{(8)l=2,m=1}$ (right panel) along an outgoing ray
$u={\rm const}=3\; T_{\rm orb}$---the solutions also shown in Fig.\ \ref{fig:solutions:fixedu}.
In each of the figures we have superposed 4 numerical solutions, obtained
with different numerical resolutions: 4, 8, 16, and 32 grid points per $M$ (this
corresponds to the linear resolution; the numbers of grid points per unit grid
area $M^2$ are the squares of these values). The insets show, greatly magnified,
two details from each plot: one near the peak (the particle's location), and the
other at the far right end (``null infinity''). It clearly appears that the
solutions are numerically convergent, and that the convergence rate is faster
than linear. Note also that, with the range of resolutions used here, the
magnitude of the fields is already resolvable to within fractional errors of
$\sim 10^{-5}$. This is the case even near the particle, where the fields'
gradients are largest.
}
\label{fig:res}
\end{figure}

\begin{figure}[htb]
\input{epsf}
\centerline{\epsfysize 7cm \epsfbox{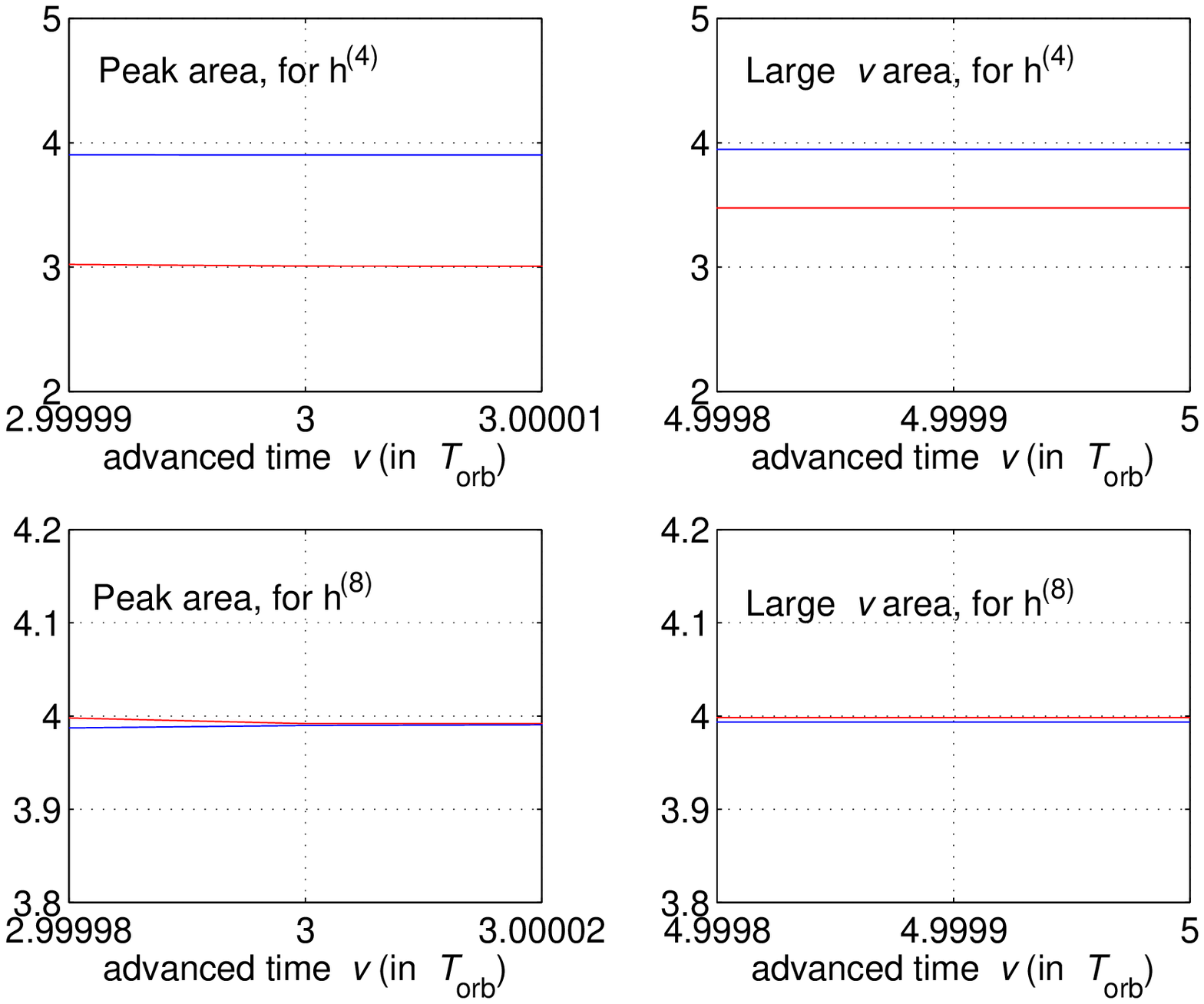}}
\caption{\protect\footnotesize
Demonstration of second-order numerical convergence. These four plots
provide a  quantitative test of the solution's numerical convergence rate.
The left and right panels of the upper (lower) row correspond, respectively,
to the left and right insets in the left (right) panels of Fig.\ \ref{fig:res}.
The red (pale) line in each of the plots shows the ratio
$\left(\left|\bar h^{(i)}[4 {\rm pts}/M]-\bar h^{(i)}[8 {\rm pts}/M]\right|\right)
/\left(\left|\bar h^{(i)}[8 {\rm pts}/M]-\bar h^{(i)}[16 {\rm pts}/M]\right|\right)$,
where $i=4,8$ and $\bar h^{(i)}[4 {\rm pts}/M]$, for example,
represents the numerical solution obtained with resolution of 4 grid points per
$M$. The blue (dark) lines similarly show the ratio
$\left(\left|\bar h^{(i)}[8 {\rm pts}/M]-\bar h^{(i)}[16 {\rm pts}/M]\right|\right)
/\left(\left|\bar h^{(i)}[16 {\rm pts}/M]-\bar h^{(i)}[32 {\rm pts}/M]\right|\right)$.
Both ratios approach the value of 4 with growing resolution, which indicates
second-order convergence.
}
\label{fig:conv}
\end{figure}

Our code is second-order convergent in the fields $\bar h^{(i)}$.
This is demonstrated in Figs.\ \ref{fig:res} and \ref{fig:conv} for
the fields $\bar h^{(4)}$ and $\bar h^{(8)}$. (We have chosen these two functions
for our demonstration here since the evolution of these specific fields
couples to that of all other fields $\bar h^{(i)}$.)

\subsection{Preservation of the Lorenz gauge conditions}\label{SecIV-B}

As discussed above, we do not impose the Lorenz gauge conditions actively
in the numerical evolution scheme. It is therefore important to verify
that (or, rather, monitor how well) our numerical solutions indeed satisfy
these four conditions, given by $H_1=H_2=H_3=H_4=0$ [see Eqs.\ ({\ref{gauge})].

To check this, our code contains a subroutine that constructs the
functions $H$ (out of the $\bar h^{(i)}$'s and their first derivatives) along
specified slice cuts through the two-dimensional numerical grid.
If the gauge conditions are fully damped along the specified slice cut,
we should expect all of these four quantities to converge to zero with
decreasing numerical step size (and, since the $H$'s are calculated to
second-order accuracy, we expect this convergence to be quadratic).
On the other hand, if there exists a finite-size constraint violation,
one or more of the $H$'s should converge (quadratically, again) to a
non-zero value.

Fig.\ \ref{fig:gauge} shows the behavior of the four functions $H$
along an outgoing ray at late retarded time ($u=3\; T_{\rm orb}$),
in the example of the modes $l=m=2$ and $l=2$, $m=1$. The functions
$H_1$, $H_2$, and $H_3$ are constructed from the even-parity MP modes,
and are not trivially zero for $l=m=2$. The function $H_4$ is constructed
from the odd-parity MP modes, and is not trivially zero for $l=2$, $m=1$.
We normalize the $H$'s (which have units of 1/distance) by $(m\omega)^{-1}$,
the typical length scale on which the MP varies (away from the particle).
The plots demonstrate that, at least in the part of the evolution
later than $t\sim T_{\rm orb}$, all four gauge conditions are satisfied
with great accuracy, to within numerical error. Similar results are
obtained when examining the late time behavior along incoming rays
(nearing the horizon) or along slices of fixed $r$.

We emphasize that the non-zero values of the functions $H$ in Figs.\ \ref{fig:gauge}
are {\em not} associated with constraint violations that have not yet
been fully damped---these would have tended to vanish rapidly with increasing
(advanced) time. Rather, the finite-size values of the $H$'s are merely
discretization errors (coming both from the calculation of the MP fields
$\bar h^{(i)}$, and from the construction of the $H$'s), which tend to zero with
increasing numerical resolution. Similarly, the large numerical values
of the $H$'s near the particle should not be interpreted as indicative of
a higher level of constraint violation there, since at the particle, too,
all functions $H$ converge to zero with increasing resolution. (Obviously,
larger numerical errors arise in the $H$'s near the particle, since
the field gradients are larger there.)
\begin{figure}[htb]
\input{epsf}
\centerline{
\epsfysize 6.5cm \epsfbox{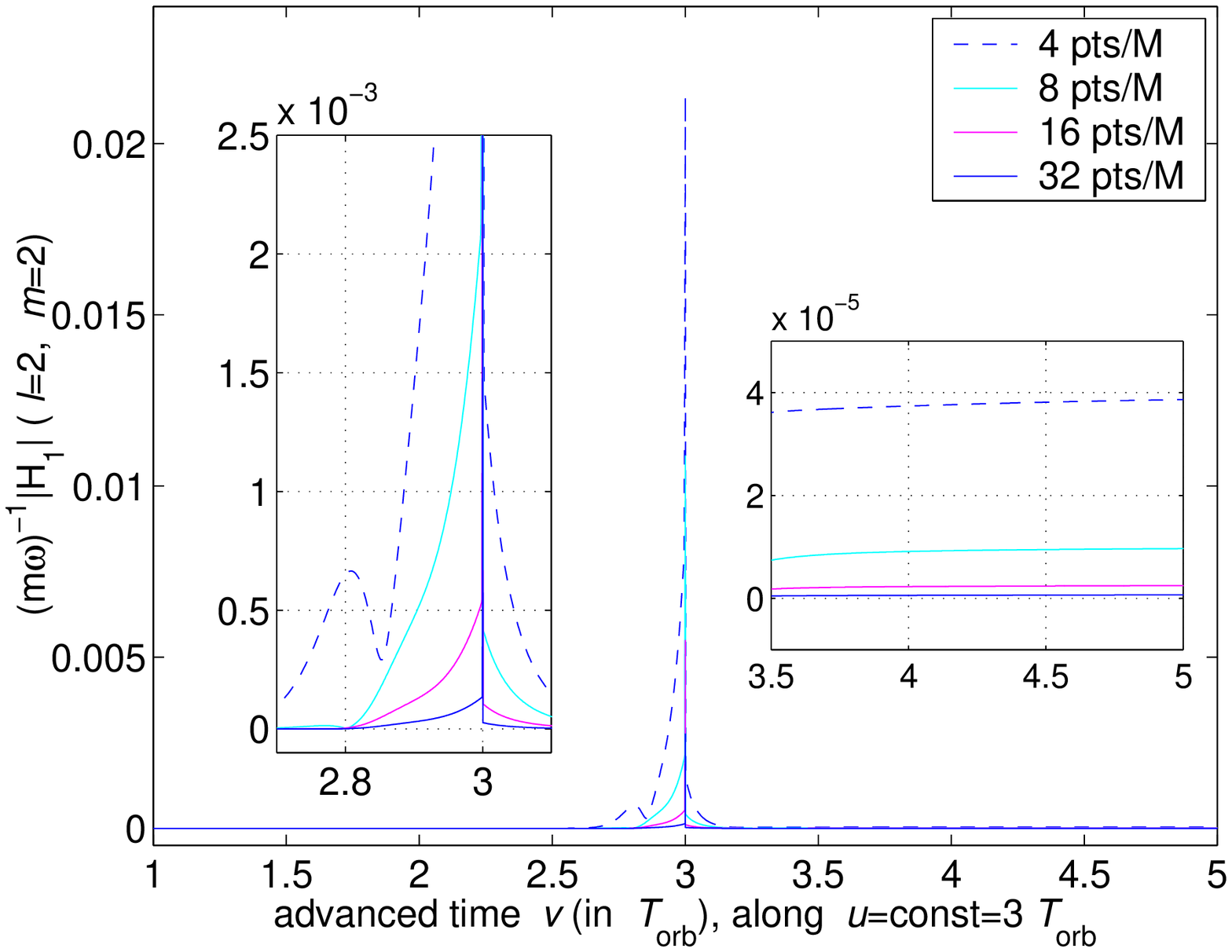}
\epsfysize 6.5cm \epsfbox{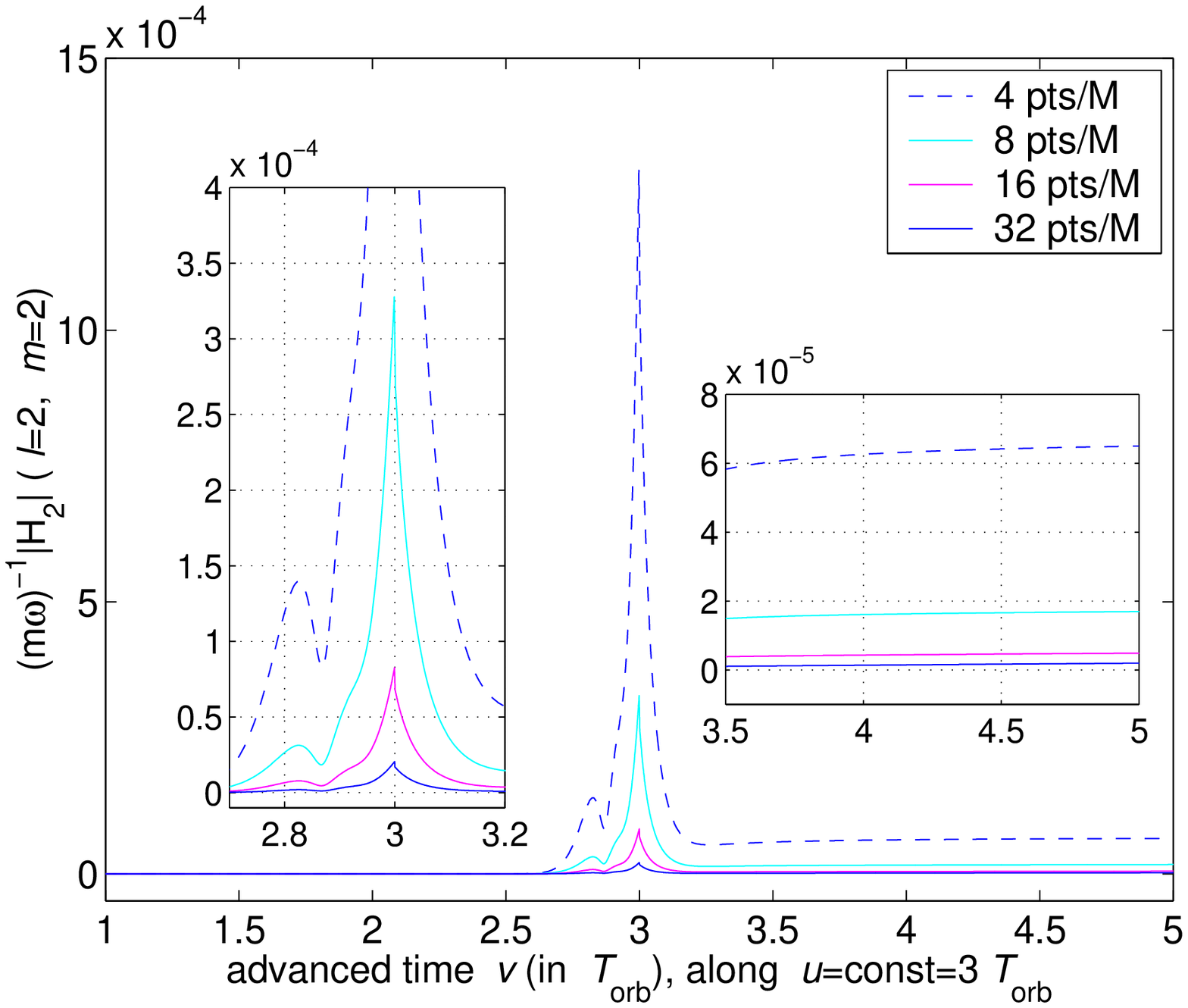}}
\centerline{
\epsfysize 6.5cm \epsfbox{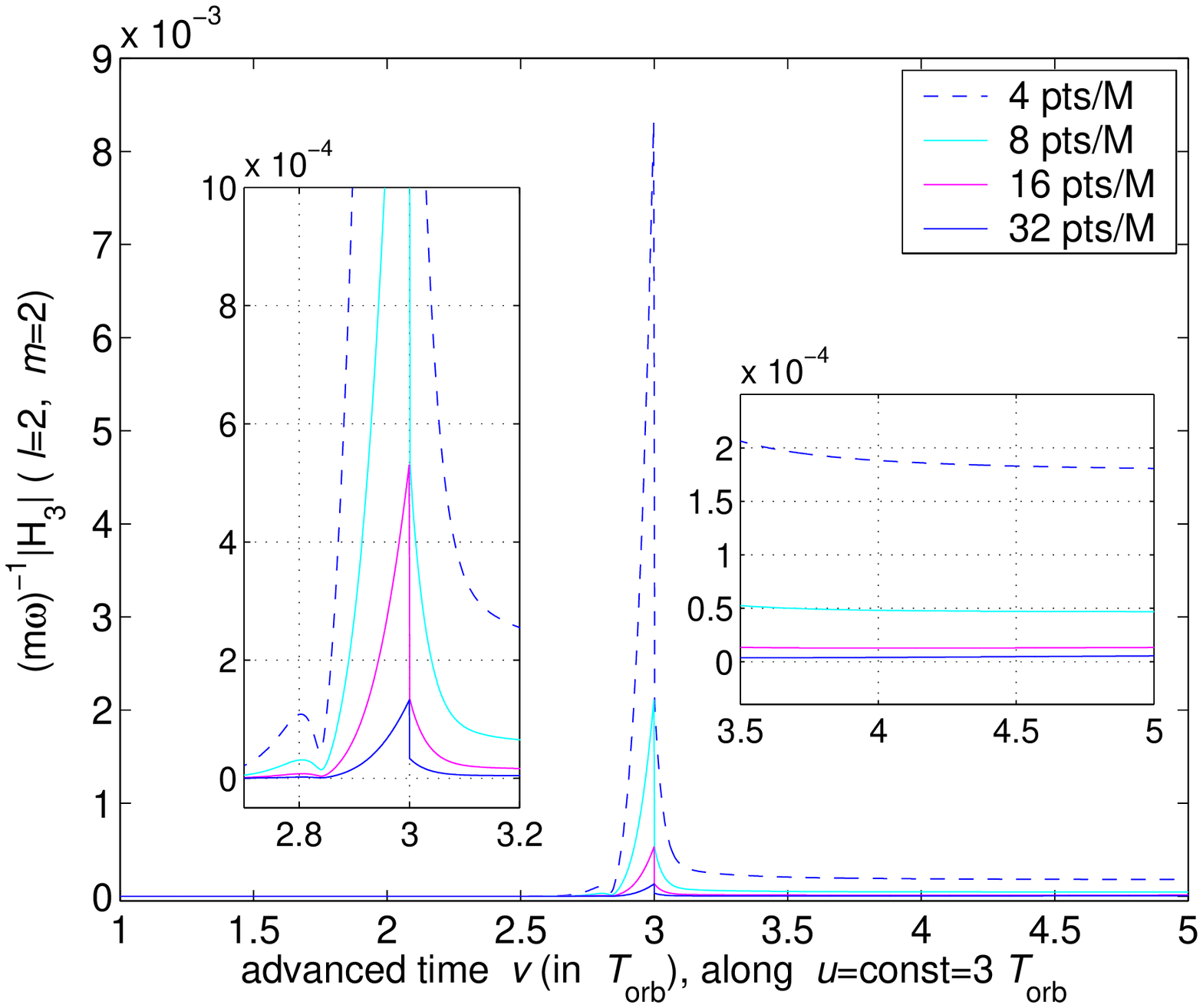}
\epsfysize 6.5cm \epsfbox{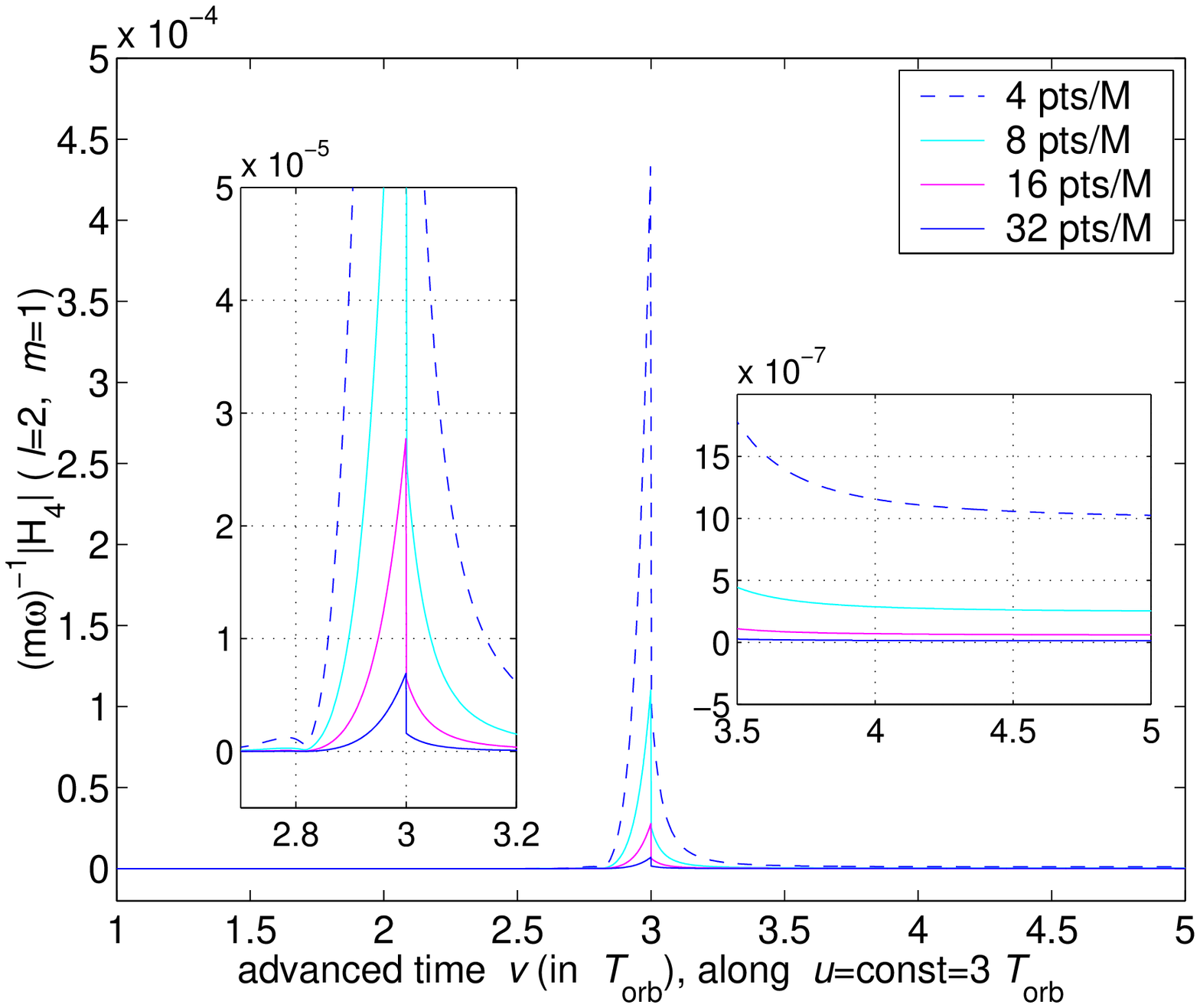}}
\caption{\protect\footnotesize
Verifying that our numerical solutions satisfy the Lorenz gauge conditions.
Plotted here are the (absolute values of the) four functions $H_1$, $H_2$, $H_3$,
and $H_4$ of Eqs.\ (\ref{gauge}), normalized by $(m\omega)^{-1}$, along an outgoing
ray $u=3\; T_{\rm orb}$.
As in previous figures, the particle is moving on a circular geodesic orbit
with $r_0=7M$. We consider here the modes $l=m=2$ (for $H_1$, $H_2$, and $H_3$,
which involve the even-parity MP modes) and $l=2$, $m=1$ (for $H_4$, which involves
the odd-parity MP modes). The Lorenz gauge condition is satisfied if (and only if)
all four functions $H$ vanish at the limit of vanishing step size.
In each of the plots we show, superposed, a sequence of four lines, obtained
with decreasing step sizes. The insets magnify specific areas in each of the plots.
All four functions $H$ appear to converge to values very close to zero, indicating
that the gauge conditions are satisfied with good accuracy. See the text for
further discussion.
}
\label{fig:gauge}
\end{figure}

\subsection{Energy flux at the wave zone} \label{Subsec:flux}

Further tests of our code could be performed by comparing with other
calculations of the Lorenz-gauge MP. The only such calculation we are
aware of was carried out by Pfenning and Poisson \cite{PP}, for
a particle in a weak-field orbit. However, the analysis in \cite{PP}
is mainly concerned with the SF acting on the particle, and
does not provide expressions for the MP itself. Another option is to
compare gauge invariant quantities derived from the MP. One such quantity
is the flux of energy radiated to infinity in gravitational waves.
For circular orbits in Schwarzschild, the energy flux (as distributed
among the $l,m$ modes) was computed previously by Poisson \cite{Poisson}
via frequency-domain numerical integration of the perturbation equations
in the standard Regge--Wheeler gauge. More recently, Martel \cite{Martel}
reproduced these fluxes (as well as the fluxes from eccentric orbits)
using a time-domain analysis of the perturbation equations, still in
the Regge-Wheeler gauge. Here we shall compare the energy fluxes
constructed from our Lorenz-gauge solutions, with those obtained
by Poisson and Martel.

We shall first need an expression for the energy flux at infinity
in terms of the Lorenz-gauge MP. Let $\dot E^{\infty}$ denote the total
energy per unit time crossing (outward) the 2-sphere $r=\rm const$ where
$r$ is very large. We assume that $r$ is large enough that the above
2-sphere resides in the ``wave-zone'', where the radius of curvature
$\cal R$ is much larger than the longest wavelength $\tilde\lambda$ of
the MP, and the perturbation takes the form of plane gravitational waves.
The overdot in $\dot E_{\infty}$, and in the expressions below, may stand
for either $\partial_t$ or $\partial_u$ (both taken with fixed $r$),
since the two are equal at the wave zone limit. Note also that at the
wave zone we can replace $h_{\alpha\beta,r}$ (fixed $t$) with $-h_{\alpha\beta,t}$
(fixed $r$), and covariant derivatives with ordinary (partial) derivatives,
the latter two differing by an amount of $O(\tilde\lambda/{\cal R})$.

The flux of energy in the gravitational waves can be obtained from Isaacson's
effective energy-momentum tensor, ${\cal T}_{\alpha\beta}$, as explained,
e.g., in Appendix B of Ref.\ \cite{Martel}. However, we must use caution
here: The standard expression for ${\cal T}_{\alpha\beta}$, as derived in
\cite{Isaacson} and used extensively in the literature (e.g., \cite{Martel}),
assumes that the MP is given in a gauge where it is traceless.
Our Lorenz gauge MP is generally not traceless,\footnote{Generally, the
supplementary gauge condition $h=0$ cannot be imposed in addition to the
Lorenz gauge condition (\ref{II-20}), unless spacetime is globally
vacuum---which is not our case here. That our solutions indeed have
$h\ne 0$ is demonstrated in Figs.\ \ref{fig:solutions:fixedr}--\ref{fig:solutions:highl},
recalling the trace is constructed from the $\bar h^{(i)}$'s through
Eq.\ (\ref{1-60}).}
and we will need to generalize the expression for the effective energy-momentum
tensor to the case $h\ne 0$. This can be done by repeating Isaacson's derivation,
starting at his Eq.\ (2.4) and going through the averaging procedure described
in his analysis---but this time keeping track of all the trace terms.
The result is
\begin{equation}\label{Isaacson}
{\cal T}_{\mu\nu}=\frac{1}{32\pi}\left\langle \bar h^{\alpha\beta}_{\;\;\;\;;\mu}
\bar h_{\alpha\beta;\nu}-\frac{1}{2}\bar h_{,\mu}\bar h_{,\nu}\right\rangle,
\end{equation}
where $\langle\cdots\rangle$ denotes an average over a region of spacetime
much larger than $\tilde\lambda$. The first term in this expression (which
is invariant under $\bar h_{\alpha\beta}\to h_{\alpha\beta}$) is the standard
Isaacson tensor. The second term is a necessary correction when the MP has
a non-zero trace. The form (\ref{Isaacson}) is now truly gauge invariant
(unlike Isaacson's original formula) and can be used to extract the
energy flux from the Lorenz-gauge MP. Given the effective energy-momentum
tensor, the energy flux at infinity is given by \cite{Martel}
\begin{equation}\label{Edot}
\dot E_{\infty}=-r^2 \int d\Omega\, {\cal T}_{tr},
\end{equation}
where the integration is carried over the above mentioned 2-sphere.
Note that the integration over $d\Omega$ automatically takes care
of averaging ${\cal T}_{\mu\nu}$ over a scale $\sim \cal R$, so
one can effectively ignore the averaging procedure in Eq.\
(\ref{Isaacson}) when evaluating the flux.

We next consider the distribution of $\dot E_{\infty}$ among the
different multipoles. To this end, we substitute for the MP  in
Eq.\ (\ref{Isaacson}) using the expansion (\ref{II-70}). In each of
the two quadratic terms we formally replace one of the MP factors
by its complex conjugate (which is allowed since the MP is a real function).
One can sort the resulting terms inside the integral $\int d\Omega$
into 3 groups, according to their angular dependence: One group of terms
comes with $ Y^{l'm'} Y^{lm*}$, the other with
$Y^{l'm'}_{,\theta}Y^{lm*}_{,\theta}+
\sin^{-2}\theta\, Y^{l'm'}_{,\varphi} Y^{lm*}_{,\varphi}$, and the
third with
$(D_2Y^{l'm'})(D_2Y^{lm*})+\sin^{-2}\theta (D_1Y^{l'm'})(D_1Y^{lm*})$.
Using the orthogonality relations (\ref{identities}), and replacing
all derivatives $\partial_r$ by $-\partial_t$, one obtains
$\dot E_{\infty}=\sum_{lm}\dot E^{lm}_{\infty}$, with the individual
multipolar contributions given by
\begin{eqnarray}\label{Edotlm}
\dot E^{lm}_{\infty}&=&\frac{\mu^2}{32\pi}\left[
\frac{1}{2}\left(|\dot{\bar h^{(1)}_{\infty}}|^2-|\dot{\bar h^{(2)}_{\infty}}|^2+2|
\dot{\bar h^{(3)}_{\infty}}|^2 \right)
 \right.\nonumber\\
&&
\left. -\frac{1}{4l(l+1)}\left(|\dot{\bar h^{(4)}_{\infty}}|^2-|\dot{\bar h^{(5)}_{\infty}}|^2+
|\dot{\bar h^{(8)}_{\infty}}|^2-|\dot{\bar h^{(9)}_{\infty}}|^2\right)  \right.\nonumber\\
&&
\left. +\frac{1}{2\lambda l(l+1)}\left(|\dot{\bar h^{(7)}_{\infty}}|^2+|\dot{\bar h^{(10)}_{\infty}}|^2
\right)\right].
\end{eqnarray}
Here, the subscript $\infty$ under the $\bar h^{(i)}$'s indicates that these
functions are to be evaluated at the wave zone.

The above expression for $\dot E^{lm}_{\infty}$ simplifies by virtue of the
gauge conditions. Consider, for example, the condition $H_1=0$ [Eq.\ (\ref{gauge1})]:
Neglecting non-derivative terms [which at the wave zone are
$O(\tilde\lambda/{\cal R})$ times smaller than the derivative terms],
and replacing $\partial_r\to-\partial_t$, we find ${\bar h^{(2)}_{\infty,t}}
=-{\bar h^{(1)}_{\infty,t}}-{\bar h^{(3)}_{\infty,t}}$.
Similarly, from $H_2=0$, $H_3=0$, and $H_4=0$ we find, respectively,
${\bar h^{(2)}_{\infty,t}}
=-{\bar h^{(1)}_{\infty,t}}+{\bar h^{(3)}_{\infty,t}}$,
${\bar h^{(4)}_{\infty,t}}=-{\bar h^{(5)}_{\infty,t}}$, and
${\bar h^{(8)}_{\infty,t}}=-{\bar h^{(9)}_{\infty,t}}$.
Altogether, we thus have
\begin{equation}\label{asymrel}
|\dot{\bar h^{(2)}_{\infty}}|^2=|\dot{\bar h^{(1)}_{\infty}}|^2, \quad\quad
|\dot{\bar h^{(3)}_{\infty}}|^2=0, \quad\quad
|\dot{\bar h^{(4)}_{\infty}}|^2=|\dot{\bar h^{(5)}_{\infty}}|^2, \quad\quad
|\dot{\bar h^{(8)}_{\infty}}|^2=|\dot{\bar h^{(9)}_{\infty}}|^2.
\end{equation}
(In our stationary scenario we have $|\dot{\bar h^{(2)}_{\infty}}|^2=
m^2\omega^2|\bar h^{(2)}_{\infty}|^2$, which means that the asymptotic
relations (\ref{asymrel}) hold between the functions $\bar h^{(i)}$ themselves,
not only between their time derivatives. That our numerical solutions satisfy
these relations at large $r$ is easily visible in Figs.\
\ref{fig:solutions:fixedt} and \ref{fig:solutions:fixedu}. This provides
yet another validity test for our code.)
Hence, Eq.\ (\ref{Edotlm}) reduces to the final form
\begin{equation}\label{flux}
\dot{E}^{lm}_{\infty}
=\frac{\mu^2}{64\pi\lambda l(l+1)}\left(
|\dot{\bar h^{(7)}_{\infty}}|^2+|\dot{\bar h^{(10)}_{\infty}}|^2
\right).
\end{equation}
Note that the flux carried by each individual mode is extracted from
a single $\bar h^{(i)}$ function: $\bar h^{(7)}$ in the case of even-parity modes
(even $l+m$), or $\bar h^{(10)}$ in the case of odd-parity modes (odd
$l+m$).

Table \ref{fluxes} lists the values of $\dot{E}_{lm}^{\infty}$ for
$l=1,\ldots,5$, as computed from our numerical Lorenz-gauge solutions
based on Eq.\ (\ref{flux}). For this table we set the particle to move
on a circular geodesic at $r_0=7.9456M$, to allow comparison with Poisson
and Martel, who provide results for this radius. The functions $\bar h^{(7)}$
and $\bar h^{(10)}$ are evaluated at $v=40\; T_{\rm orb}$ and $u=3\; T_{\rm orb}$,
corresponding to $r\sim 2100M$ and $t\sim 2500M$. It is possible to
estimate the error we make in extracting the flux at a finite $r$ by
looking at the residual values of, e.g., $|\bar h^{(3)}|$ and
$|\bar h^{(8)}|-|\bar h^{(9)}|$, which should vanish at the limit
$r\to\infty$. If we assume that the finite-$r$ error in $\bar h^{(7)}$
is the same as the largest among the residues in $|\bar h^{(3)}|$,
$|\bar h^{(1)}|-|\bar h^{(2)}|$, and
$|\bar h^{(4)}|-|\bar h^{(5)}|$, we obtain from our code
that the fractional error in the flux, for even-parity modes, is typically
of order $0.1$--$0.2\%$. In the same way, assuming that the finite-$r$ error
in $\bar h^{(10)}$ is the same as the residue in $|\bar h^{(8)}|-|\bar h^{(9)}|$,
we expect a fractional error of order $0.01$--$0.3\%$ for the various odd-parity
modes.

Our code reproduces the energy fluxes of Poisson \cite{Poisson} and
Martel \cite{Martel} with very good accuracy. Our values are no more than
$\sim 1\%$ off those of Martel, and no more than $0.07\%$ off those
of Poisson. The overall energy flux (summed over the first 5 modes) agrees
with that computed by Poisson with a mere $\sim 5\cdot10^{-5}$ relative
difference.
\begin{table}[htb]
\centerline{$\begin{array}{ll|c|c|c}\hline\hline
l & m  & \dot E^{\infty}_{lm}, \text{this paper: } &
\dot E^{\infty}_{lm}, \text{Poisson \cite{Poisson}}:     &
\dot E^{\infty}_{lm}, \text{Martel \cite{Martel}}:       \\
{} & {}
& \text{$t$-domain, from $h_{\alpha\beta}^{\rm Lorenz}$}
& \text{$f$-domain, from $h_{\alpha\beta}^{\rm RW}$}
& \text{$t$-domain, from $h_{\alpha\beta}^{\rm RW}$} \\
\hline\hline
2\quad  & 1 & 8.1654e{-}07 & 8.1633e{-}07\ \ [-0.03\%] & 8.1623e{-}07\ \ [-0.04\%]\\
{}      & 2 & 1.7061e{-}04 & 1.7063e{-}04\ \ [+0.01\%] & 1.7051e{-}04\ \ [-0.06\%]\\
3       & 1 & 2.1734e{-}09 & 2.1731e{-}09\ \ [-0.01\%] & 2.1741e{-}09\ \ [+0.03\%]\\
{}      & 2 & 2.5207e{-}07 & 2.5199e{-}07\ \ [-0.03\%] & 2.5164e{-}07\ \ [-0.17\%]\\
{}      & 3 & 2.5479e{-}05 & 2.5471e{-}05\ \ [-0.03\%] & 2.5432e{-}05\ \ [-0.18\%]\\
4       & 1 & 8.3982e{-}13 & 8.3956e{-}13\ \ [-0.03\%] & 8.3507e{-}13\ \ [-0.57\%]\\
{}      & 2 & 2.5099e{-}09 & 2.5091e{-}09\ \ [-0.03\%] & 2.4986e{-}09\ \ [-0.45\%]\\
{}      & 3 & 5.7759e{-}08 & 5.7751e{-}08\ \ [-0.01\%] & 5.7464e{-}08\ \ [-0.51\%]\\
{}      & 4 & 4.7284e{-}06 & 4.7256e{-}06\ \ [-0.06\%] & 4.7080e{-}06\ \ [-0.43\%]\\
5       & 1 & 1.2598e{-}15 & 1.2594e{-}15\ \ [-0.03\%] & 1.2544e{-}15\ \ [-0.43\%]\\
{}      & 2 & 2.7877e{-}12 & 2.7896e{-}12\ \ [+0.07\%] & 2.7587e{-}12\ \ [-1.04\%]\\
{}      & 3 & 1.0934e{-}09 & 1.0933e{-}09\ \ [-0.01\%] & 1.0830e{-}09\ \ [-0.95\%]\\
{}      & 4 & 1.2319e{-}08 & 1.2324e{-}08\ \ [+0.04\%] & 1.2193e{-}08\ \ [-1.02\%]\\
{}      & 5 & 9.4623e{-}07 & 9.4563e{-}07\ \ [-0.06\%] & 9.3835e{-}07\ \ [-0.83\%]\\
\hline
\text{total}  & {} & 2.0291{-}04 & 2.0292e{-}04\ \ [+0.005\%] & 2.0273e{-}04\ \ [-0.09\%]\\
\hline\hline
\end{array}$}
\caption{\protect\footnotesize
Comparison of energy fluxes per $l,m$-mode at infinity, as extracted from
our Lorenz-gauge MP, with the values obtained using other methods. The
values in the table give $\dot E^{\infty}_{lm}$ [in units of $(\mu/M)^2$]
for $r_0=7.9456$, with $m<0$ modes folded over onto $m>0$. The results
from Poisson's and Martel's computations are quoted from Ref.\ \protect\cite{Martel}.
Both Poisson and Martel extracted their fluxes from the MP in the Regge-Wheeler
gauge, the former using frequency domain analysis and the latter integrating
the field equations in the time domain. The values under ``this paper''
are derived using Eq.\ (\ref{flux}), with the Lorenz-gauge functions $\bar h^{(7)}$
(for modes with even $l+m$) and $\bar h^{(10)}$ (for modes with odd $l+m$) extracted
at $v=40\; T_{\rm orb}$ and $u=3\; T_{\rm orb}$ (corresponding to $r\sim 2100M$
and $t\sim 2500M$). The values in square brackets under ``Poisson'' and ``Martel''
give the relative difference, in percents, between their results and ours.
} \label{fluxes}
\end{table}

\section{Summary and discussion of future applications} \label{SecV}

We advocate here a new approach to the computation of the metric perturbation
from a small particle orbiting a black hole. The approach incorporates
the following ``principles'': (a) The particle is represented as a delta
function source term (rather than an extended distribution) in the field equations;
(b) The perturbation equations are formulated and integrated for the MP itself
(rather than generating functions thereof); (c) The MP is solved for in the
Lorenz gauge; (d) The numerical integration of the perturbation equations
is carried out in the time domain.

The main advantages offered by this approach are: (i) In applications
requiring knowledge of the MP itself, one avoids the difficult issue
of MP reconstruction; (ii) The Lorenz-gauge MP provides the most natural
description of the field near the particle, and is the most convenient to
work with in analyzing the singular structure of this field; (iii) The
Lorenz-gauge MP can be incorporated immediately into existing schemes
for calculating the gravitational SF. (iv) Time-domain numerical
integration is more efficient in dealing with eccentric orbits than
traditional frequency-domain methods, and is more easily generalizable
from one set of orbits to another.

Here we developed the above approach, and explored its feasibility,
for orbits in Schwarzschild. The core of our formulation includes the
decoupled field equations, Eqs.\ (\ref{FE}), along with the supplementary gauge
conditions, Eqs.\ (\ref{gauge}). We implemented this formulation numerically
for the case of circular orbits, using a characteristic time-evolution
code that incorporates a constraint damping scheme. We demonstrated
that such a code can efficiently resolve the singular field near the
particle, in a computationally inexpensive manner.

Since our code integrates the perturbation equations in the time domain,
it is possible to use it for analyzing {\em any} orbit in Schwarzschild:
For any given trajectory $x_{\rm p}(\tau)$ (which, if geodesic, can always
be taken to be equatorial) one merely has to specify appropriate
source functions $S^{(i)lm}[r_{\rm p}(\tau)]$ in Eq.\ (\ref{FE}), and
feed these functions into the subroutine in our code that tells the
numerical integrator where the particle is at each step of the evolution.
The numerical properties of the evolution code (stability, convergence rate,
constraint damping) are insensitive to the choice of source.  Since our
code performs well with circular orbits, it will perform well with
any other orbit.

Much more challenging is the extension to the Kerr case, which, however,
also provides the main motivation for the above approach. On a Kerr
background, the perturbation equations are no longer separable into
multipole modes in the time domain. A time-domain integrator will
therefore have to evolve the equations in 2+1 dimensions (two spacial
dimensions, and time; the azimuthal direction remains separable even
in the time domain). There exist a few working codes for integrating
Teukolsky's equation in the time domain, for vacuum perturbations in
Kerr [see, e.g., \cite{Krivan,Pazos})]. A similar code will have to be
developed for integrating the Lorenz-gauge field equations (\ref{II-30}),
with a particle source. The challenge here is not in the expected high
computational expense insomuch as it is in the
difficulty of treating the particle singularity: In 2+1 dimensions
the Lorenz gauge perturbation field is no longer continuous and finite
near the particle (as it is, conveniently, in 1+1 dimensions).
Rather, it diverges toward the particle. Although this divergence is
slow (logarithmic in proper distance from the particle), it poses a
serious problem when it comes to numerical implementation.

One possible way around this problem is to implement a procedure reminiscent
of the ``puncture'' scheme often in use in Numerical Relativity.
To sketch the basic idea, let us represent the field equation (\ref{II-30}),
symbolically, by `$\Box h=\delta$', where `$\delta$' represents a given fixed
trajectory on the black hole background, and the left-hand side is understood
to contain also the Riemann term.
We introduce a worldtube around the particle, of some fixed radius
$\rho$ that we keep as a control parameter. $\rho$ should be much smaller
than the background's radius of curvature, and much larger than the radius
of curvature associated with the particle; say, $\rho\sim \sqrt{M\mu}$.
The numerical time-evolution integrator than utilizes the following algorithm:
At any point (in the 2+1-dimensional numerical grid space) {\em outside}
the above worldtube, the integrator evolves the homogeneous equation
$\Box h=0$ as it is. However, at points {\em inside} the tube, the
integrator solves for a different, ``punctured'' field
$h_{\rm punct}\equiv h-h_{\rm sing}$, where $h_{\rm sing}$ is an analytic
approximation for the field $h$ near the singularity, chosen such that
$h_{\rm punct}$ is continuous everywhere inside the worldtube, including on
the worldline itself. Thus, inside the worldtube one solves the equation
$\Box h_{\rm punct}=\delta-\Box h_{\rm sing}$, where the source term on the right-hand
side is now distributed, but `more regular' than the original source.
The evolution across the boundary of the worldtube proceeds by matching
the external solution $h$ to $h_{\rm punct}+h_{\rm sing}$.
That this (simple) idea could work in practice has been demonstrated
recently with a toy model of a scalar field in Schwarzschild \cite{Hernandez}.

As explained in the introduction, our main drive in developing the
computational approach of this paper is the problem of calculating the
SF. With the new computational tools at hand, freeing us from
gauge complexities and issues of MP reconstruction, we can straightforwardly
implement the mode-sum scheme \cite{ModeSum}.
The scheme, we remind, requires the values of each multipole
of the Lorenz-gauge MP and its derivatives at the particle. The modes are
then ``regularized'' individually, by subtracting a certain quantity,
given analytically (the ``regularization parameters'' \cite{ModeSum}), from each.
The sum over regularized modes then gives the physical SF experienced
by the particle. Our current code can be used to calculate the SF
in this manner, for any orbit in Schwarzschild. One minor technical matter
still to be addressed is the fact that the regularization parameters
prescribed in the literature correspond (somewhat awkwardly) to a
{\em scalar}-harmonic decomposition of the force's components, whereas
the entities we calculate numerically are tensor harmonic components
of the MP, yielding the {\em vector}-harmonic components of the force.
It will be required to obtain the appropriate regularization parameters
for the vector-harmonic components, or, alternatively, expand the vector
harmonic components in scalar harmonics in order to allow them to
communicate with the standard parameters. Either of these options is
essentially straightforward to implement.

\acknowledgements

During the work on this project we have benefited from discussions with many,
including Nils Andersson, Curt Cutler, Steve Detweiler, Jonathan Gair,
Kostas Glampedakis, Yasushi Mino, and James Vickers. We are grateful to all
of them. We especially thank Eric Poisson for many valuable comments,
and Carsten Gundlach for useful advice on the issue of constraint damping.
We gratefully acknowledge the support of the NASA Center
for Gravitational Wave Astronomy at The University of Texas at Brownsville
(NAG5-13396), and the NSF for financial support from grants PHY-0140326 and
PHY-0354867. L.B.\ also acknowledges financial support from the Nuffield
Foundation, and thanks the Aspen Center for Physics for hospitality during
the Aspen 2005 Summer Workshop on LISA, where part of this research
was carried out.

\newpage

\appendix

\section{Basis of tensor harmonics} \label{AppA}

The tensor harmonics $Y^{(i)lm}_{\alpha\beta}$ adopted in
this paper are given, in Schwarzschild coordinates $t,r,\theta,\varphi$, by
\begin{mathletters} \label{eqIII20}
\begin{equation} \label{eqIII20(1)}
Y^{(1)}_{\alpha\beta}=\frac{1}{\sqrt{2}}\left(
\begin{array}{c c c c}
1 & 0 & 0 & 0 \\
0 & f^{-2} & 0 & 0 \\
0 & 0 & 0 & 0 \\
0 & 0 & 0 & 0
\end{array}\right) Y^{lm},
\quad\quad
Y^{(2)}_{\alpha\beta}=\frac{f^{-1}}{\sqrt{2}}\left(
\begin{array}{c c c c}
0 & 1 & 0 & 0 \\
1 & 0 & 0 & 0 \\
0 & 0 & 0 & 0 \\
0 & 0 & 0 & 0
\end{array}\right) Y^{lm},
\quad\quad
Y^{(3)}_{\alpha\beta}=\frac{1}{\sqrt{2}}\left(
\begin{array}{c c c c}
1 & 0 & 0 & 0 \\
0 & -f^{-2} & 0 & 0 \\
0 & 0 & 0 & 0 \\
0 & 0 & 0 & 0
\end{array}\right) Y^{lm},
\end{equation}
\begin{equation} \label{eqIII20(4)}
Y^{(4)}_{\alpha\beta}=
\frac{r}{\sqrt{2l(l+1)}}\left(
\begin{array}{c c c c}
0                 & 0 & \partial_{\theta} & \partial_{\varphi} \\
0                 & 0 &        0          &         0          \\
\partial_{\theta} & 0 &        0          &         0          \\
\partial_{\varphi}& 0 &        0          &         0
\end{array}\right) Y^{lm},
\quad\quad
Y^{(5)}_{\alpha\beta}=
\frac{rf^{-1}}{\sqrt{2l(l+1)}}\left(
\begin{array}{c c c c}
0 &        0           &        0          &          0         \\
0 &        0           & \partial_{\theta} & \partial_{\varphi} \\
0 & \partial_{\theta}  &        0          &          0         \\
0 & \partial_{\varphi} &        0          &          0
\end{array}\right) Y^{lm},
\end{equation}
\begin{equation} \label{eqIII20(6)}
Y^{(6)lm}_{\alpha\beta}=
\frac{r^2}{\sqrt{2}}\left(
\begin{array}{c c c c}
0 & 0 & 0 & 0 \\
0 & 0 & 0 & 0 \\
0 & 0 & 1 & 0 \\
0 & 0 & 0 & s^2
\end{array}\right) Y^{lm},
\quad\quad
Y^{(7)lm}_{\alpha\beta}=
\frac{r^2}{\sqrt{2\lambda l(l+1)}}\left(
\begin{array}{c c c c}
0 & 0 & 0   & 0        \\
0 & 0 & 0   & 0        \\
0 & 0 & D_2 & D_1       \\
0 & 0 & D_1 & -s^2 D_2
\end{array}\right) Y^{lm},
\end{equation}
\end{mathletters}
\begin{mathletters} \label{eqIII25}
\begin{equation} \label{eqIII25(8)}
Y^{(8)lm}_{\alpha\beta}=\frac{r}{\sqrt{2l(l+1)}}\left(
\begin{array}{c c c c}
0 &        0            & s^{-1}\partial_{\varphi} & -s\,\partial_{\theta} \\
0 &        0            &          0               &                       \\
s^{-1}\partial_{\varphi}&          0               &     0    &    0       \\
-s\,\partial_{\theta}   &          0               &     0    &    0
\end{array}\right) Y^{lm},
\end{equation}
\begin{equation} \label{eqIII25(9)}
Y^{(9)lm}_{\alpha\beta}=\frac{rf^{-1}}{\sqrt{2l(l+1)}}\left(
\begin{array}{c c c c}
0 &        0                &        0                &           0         \\
0 &        0                & s^{-1}\partial_{\varphi} & -s\,\partial_{\theta} \\
0 & s^{-1}\partial_{\varphi}&        0                &          0          \\
0 & -s\,\partial_{\theta}   &        0                &          0
\end{array}\right) Y^{lm},
\end{equation}
\begin{equation} \label{eqIII25(10)}
Y^{(10)lm}_{\alpha\beta}=
\frac{r^2}{\sqrt{2\lambda l(l+1)}}\left(
\begin{array}{c c c c}
0 & 0 & 0          & 0            \\
0 & 0 & 0          & 0            \\
0 & 0 & s^{-1}D_1 & -s\,D_2       \\
0 & 0 & -s\,D_2     & -s\,D_1
\end{array}\right) Y^{lm},
\end{equation}
\end{mathletters}
where $f\equiv (1-2M/r)$,
$Y^{lm}(\theta,\varphi)$ are the standard scalar spherical harmonics,
$s\equiv\sin\theta$, $\lambda \equiv (l-1)(l+2)$, and the angular operators
$D_1$ and $D_2$ are given by
\begin{eqnarray} \label{D1D2}
D_1 \equiv 2(\partial_{\theta}-\cot\theta)\partial_{\varphi},
\quad\quad
D_2\equiv \partial_{\theta\theta}-\cot\theta\,\partial_{\theta}
-s^{-2} \partial_{\varphi\varphi}.
\end{eqnarray}
The radial factors involving $r$ and $f$ are introduced for dimensional
balance and for settling the horizon behavior.
The harmonics $Y^{(i)lm}_{\alpha\beta}$  constitute an orthonormal set,
in the sense expressed in  Eq.\ (\ref{II-60}). This can be readily
verified based on the identities
\begin{mathletters} \label{identities}
\begin{equation}
\int  Y^{l'm'} Y^{lm*} d\Omega = \delta_{ll'}\delta_{mm'},
\end{equation}
\begin{equation}
\int \left(Y^{l'm'}_{,\theta}Y^{lm*}_{,\theta}
+\sin^{-2}\theta\, Y^{l'm'}_{,\varphi} Y^{lm*}_{,\varphi}\right)
d\Omega = l(l+1)\delta_{ll'}\delta_{mm'},
\end{equation}
\begin{equation}
\int \left[(D_2Y^{l'm'})(D_2Y^{lm*})
+\sin^{-2}\theta (D_1Y^{l'm'})(D_1Y^{lm*})\right]d\Omega
= \lambda l(l+1)\delta_{ll'}\delta_{mm'},
\end{equation}
\end{mathletters}
where the integration is carried over a 2-sphere $r$=const,
an asterisk denotes complex conjugation, and $\delta_{nn'}$
is the Kronecker delta.

\section{separation of the field equations} \label{AppB}

The field equations (\ref{II-130}), which are fully separated with respect
to $l,m$, and uncoupled at their principle part with respect to $i$, are
obtained by substituting both expansions (\ref{II-70}) and (\ref{II-110})
into Eq.\ (\ref{II-30}), and then considering certain combinations of
the resulting equations and derivatives thereof. The necessary combinations
are given in table \ref{table:comb}.
\begin{table}[htb]
\centerline{$\begin{array}{c|c}\hline\hline
\text{To get equation for}\ldots  &
\text{use the combination}\ldots  \\ \hline\hline
i=1  & \{tt\}+f^2\{rr\}                  \\
i=2  & \{tr\}                 \\
i=3  & \{tt\}-f^2\{rr\}                  \\
i=4  & (s\{t\theta\})_{,\theta}+(s^{-1}\{t\varphi\})_{,\varphi} \\
i=5  & (s\{r\theta\})_{,\theta}+(s^{-1}\{r\varphi\})_{,\varphi} \\
i=6  & \{\theta\theta\}+s^{-2}\{\varphi\varphi\}                 \\
i=7  & s^{-2}\,D_1[s\{\theta\varphi\}]+
       s^{-1}D_2\left[s^2\{\theta\theta\}-\{\varphi\varphi\}\right]/2 \\
i=8  & \{t\theta\}_{,\varphi}-\{t\varphi\}_{,\theta}               \\
i=9  & \{r\theta\}_{,\varphi}-\{r\varphi\}_{,\theta}               \\
i=10  & s^{-2}D_1\left[s^2(\{\theta\theta\}-s^{-2}\{\varphi\varphi\})\right]/2
        -s^{-1}D_2[s\{\theta\varphi\}]                 \\
\hline\hline
\end{array}$}
\caption{\protect\footnotesize
Component combinations used in separating the perturbation equations.
Here $s\equiv \sin\theta$, and the angular differential operators
$D_1$ and $D_2$ are those defined in Eqs.\ (\ref{D1D2}). Curely brackets
represent Schwarzschild components of Eq.\ (\ref{II-30}); thus, ``$\{tt\}+f^2\{rr\}$'',
for example, stands for ``take the $tt$ component of Eq.\ (\ref{II-30}),
and add to it the $rr$ component of that equation, multiplied by $f^2$.''
}
\label{table:comb}
\end{table}

\section{Mode-decomposed linearized Einstein equations in
the Lorenz gauge} \label{AppC}

We give here the explicit expressions for the terms
${\tilde{\cal M}}^{(i)}_{\;(j)}$ appearing in the original form of
the separated field equations, Eqs.\ (\ref{II-130}).
For the numerical evolution in this paper we use a different form of the
separated equations [i.e., Eqs.\ (\ref{FE}), with Eqs.\ (\ref{II-150}), which
incorporate a constraint damping scheme], and so the original functions
${\tilde{\cal M}}^{(i)}_{\;(j)}$ are not needed in our analysis.
We nevertheless give them here, as they might be useful as a starting
point for anyone wishing to implement a different numerical scheme.
The terms ${\tilde{\cal M}}^{(i)}_{\;(j)}$ are given by
\begin{mathletters}\label{tildeM}
\begin{equation} \label{tildeM1}
{\tilde{\cal M}}^{(1)}_{\;(j)}\bar h^{(j)}=
\frac{1}{2}ff'\bar h^{(1)}_{,r}-\frac{1}{2}\,f'\bar h^{(2)}_{,t}
+\frac{f^2}{2r^2}\left(\bar h^{(1)}-\bar h^{(3)}-\bar h^{(5)}
-f\bar h^{(6)}\right),
\end{equation}
\begin{equation} \label{tildeM2}
{\tilde{\cal M}}^{(2)}_{\;(j)}\bar h^{(j)}=
\frac{1}{2}ff'\bar h^{(2)}_{,r}
-\frac{1}{2}\,f'\bar h^{(1)}_{,t}
+\frac{f^2}{2r^2}\left(\bar h^{(2)}-\bar h^{(4)}\right),
\end{equation}
\begin{eqnarray} \label{tildeM3}
{\tilde{\cal M}}^{(3)}_{\;(j)}\bar h^{(j)}=
\frac{1}{2}ff'\bar h^{(3)}_{,r}
+\frac{1}{2r^2}\left[1-8M/r+10(M/r)^2\right]\bar h^{(3)}
-\frac{f^2}{2r^2}\left[\bar h^{(1)}-\bar h^{(5)}-(1-4M/r)\bar h^{(6)}
\right],
\end{eqnarray}
\begin{equation} \label{tildeM4}
{\tilde{\cal M}}^{(4)}_{\;(j)}\bar h^{(j)}=
\frac{1}{4}f'f\bar h^{(4)}_{,r}
-\frac{1}{4}\,f'\bar h^{(5)}_{,t}
-\frac{3}{4}f'(f/r)\bar h^{(4)}
-\frac{1}{2}\,l(l+1)\,(f/r^2)\bar h^{(2)},
\end{equation}
\begin{eqnarray} \label{tildeM5}
{\tilde{\cal M}}^{(5)}_{\;(j)}\bar h^{(j)}=
\frac{1}{4}ff'\bar h^{(5)}_{,r}
-\frac{1}{4}f'\bar h^{(4)}_{,t}
+\frac{f}{r^2}(1-3.5M/r)\bar h^{(5)}
-\frac{f}{2r^2}\,l(l+1)\left(\bar h^{(1)}-\bar h^{(3)}-f\bar h^{(6)}\right)
-\frac{f^2}{2r^2}\bar h^{(7)},
\end{eqnarray}
\begin{equation} \label{tildeM6}
{\tilde{\cal M}}^{(6)}_{\;(j)}\bar h^{(j)}=
-\frac{f}{2r^2}\left[\bar h^{(1)}-\bar h^{(5)}
-(1-4M/r)\left(f^{-1}\bar h^{(3)}+\bar h^{(6)}\right)\right],
\end{equation}
\begin{equation} \label{tildeM7}
{\tilde{\cal M}}^{(7)}_{\;(j)}\bar h^{(j)}=
-\frac{f}{2r^2}\left(\bar h^{(7)}
+\lambda\,\bar h^{(5)}\right),
\end{equation}
\begin{equation} \label{tildeM8}
{\tilde{\cal M}}^{(8)}_{\;(j)}\bar h^{(j)}=
\frac{1}{4}f'f\bar h^{(8)}_{,r}
-\frac{1}{4}\,f'\bar h^{(9)}_{,t}
-\frac{3}{4}f'(f/r)\bar h^{(8)}
\end{equation}
\begin{equation} \label{tildeM9}
{\tilde{\cal M}}^{(9)}_{\;(j)}\bar h^{(j)}=
\frac{1}{4} f'\left(f\bar h^{(9)}_{,r}-\bar h^{(8)}_{,t}\right)
+\frac{f}{r^2}\left[(1-3.5M/r)\bar h^{(9)}-(f/2) \bar h^{(10)}
\right],
\end{equation}
\begin{equation} \label{tildeM10}
{\tilde{\cal M}}^{(10)}_{\;(j)}\bar h^{(j)}=
-\frac{f}{2r^2}\left(\bar h^{(10)}+\lambda\,\bar h^{(9)}\right),
\end{equation}
\end{mathletters}
where we have omitted all indices $l,m$ for brevity.


\end{document}